\documentclass[amsmath,amssymb,
 reprint,%
 superscriptaddress,%
 aps,%
 prb,%
longbibliography, %
]{revtex4-2}

\usepackage{graphicx}% Include figure files
\usepackage{dcolumn}% Align table columns on decimal point
\usepackage[export]{adjustbox}
\usepackage{siunitx}
\usepackage{bm}% bold math

\usepackage[utf8]{inputenc}
\usepackage[T1]{fontenc}
\usepackage{mathptmx}
\usepackage{etoolbox}
\usepackage{hyperref}
\usepackage{booktabs}
\usepackage{braket}

\DeclareSIUnit\angstrom{\text {Å}}

\newcommand*{\figref}[2][]{%
  \hyperref[{fig:#2}]{%
    Figure~\ref*{fig:#2}%
    \ifx\\#1\\%
    \else
      \,#1%
    \fi
  }%
}

\newcommand*{\tabref}[2][]{%
  \hyperref[{tab:#2}]{%
    Table~\ref*{tab:#2}%
    \ifx\\#1\\%
    \else
      \,#1%
    \fi
  }%
}

\newcommand*{\appref}[2][]{%
  \hyperref[{app:#2}]{%
    Appendix~\ref*{app:#2}%
    \ifx\\#1\\%
    \else
      \,#1%
    \fi
  }%
}

\newcommand*{\equationref}[2][]{%
  \hyperref[{eq:#2}]{%
    Equation~\eqref{eq:#2}%
    \ifx\\#1\\%
    \else
      \,#1%
    \fi
  }%
}

\def\SPSB#1#2{\rlap{\textsuperscript{#1}}\SB{#2}}

\def\SB#1{\textsubscript{#1}}
\def\VB{V\SPSB{--}{B}}

\makeatletter
\def\@email#1#2{%
 \endgroup
 \patchcmd{\titleblock@produce}
  {\frontmatter@RRAPformat}
  {\frontmatter@RRAPformat{\produce@RRAP{*#1\href{mailto:#2}{#2}}}\frontmatter@RRAPformat}
  {}{}
}%
\makeatother

\begin{document}

\title{Quantifying the creation of negatively charged boron vacancies\texorpdfstring{\\ in He-ion irradiated hexagonal boron nitride}{}}

\author{Amedeo Carbone}
\affiliation{Department of Electrical and Photonics Engineering, Technical University of Denmark, 2800 Kgs. Lyngby, Denmark}
\affiliation{NanoPhoton - Center for Nanophotonics, Technical University of Denmark, 2800 Kgs. Lyngby, Denmark}
\affiliation{Walter Schottky Institute, Technical University of Munich, 85748 Garching, Germany}

\author{Ilia D. Breev}
\affiliation{Center for Macroscopic Quantum States (bigQ), Department of Physics, Technical University of Denmark, 2800 Kgs. Lyngby, Denmark}

\author{Johannes Figueiredo}
\affiliation{Walter Schottky Institute, Technical University of Munich, 85748 Garching, Germany}
\affiliation{Munich Center for Quantum Science and Technology (MCQST), 80799 Munich, Germany}
\affiliation{TUM International Graduate School of Science and Engineering (IGSSE), 85748 Garching, Germany}

\author{Silvan Kretschmer}
\affiliation{Institute of Ion Beam Physics and Materials Research, Helmholtz-Zentrum Dresden-Rossendorf, 01328 Dresden, Germany}

\author{Leonard Geilen}
\affiliation{Walter Schottky Institute, Technical University of Munich, 85748 Garching, Germany}
\affiliation{Munich Center for Quantum Science and Technology (MCQST), 80799 Munich, Germany}
\affiliation{TUM International Graduate School of Science and Engineering (IGSSE), 85748 Garching, Germany}

\author{Amine Ben Mhenni}
\affiliation{Walter Schottky Institute, Technical University of Munich, 85748 Garching, Germany}
\affiliation{Munich Center for Quantum Science and Technology (MCQST), 80799 Munich, Germany}

\author{Johannes Arceri}
\affiliation{Fakultät für Physik, Ludwig-Maximilians-Universität München, 80799 Munich, Germany}
\affiliation{Munich Center for Quantum Science and Technology (MCQST), 80799 Munich, Germany}

\author{Arkady V. Krasheninnikov}
\affiliation{Institute of Ion Beam Physics and Materials Research, Helmholtz-Zentrum Dresden-Rossendorf, 01328 Dresden, Germany}

\author{Martijn Wubs}
\affiliation{Department of Electrical and Photonics Engineering, Technical University of Denmark, 2800 Kgs. Lyngby, Denmark}
\affiliation{NanoPhoton - Center for Nanophotonics, Technical University of Denmark, 2800 Kgs. Lyngby, Denmark}

\author{Alexander W. Holleitner}
\affiliation{Walter Schottky Institute, Technical University of Munich, 85748 Garching, Germany}
\affiliation{Munich Center for Quantum Science and Technology (MCQST), 80799 Munich, Germany}
\affiliation{TUM International Graduate School of Science and Engineering (IGSSE), 85748 Garching, Germany}

\author{Alexander Huck}
\affiliation{Center for Macroscopic Quantum States (bigQ), Department of Physics, Technical University of Denmark, 2800 Kgs. Lyngby, Denmark}

\author{Christoph Kastl}
\affiliation{Walter Schottky Institute, Technical University of Munich, 85748 Garching, Germany}
\affiliation{Munich Center for Quantum Science and Technology (MCQST), 80799 Munich, Germany}

\author{Nicolas Stenger}
 \email{niste@dtu.dk}
\affiliation{Department of Electrical and Photonics Engineering, Technical University of Denmark, 2800 Kgs. Lyngby, Denmark}
\affiliation{NanoPhoton - Center for Nanophotonics, Technical University of Denmark, 2800 Kgs. Lyngby, Denmark}

\date{\today}

\begin{abstract}

Hexagonal boron nitride (hBN) hosts luminescent defects possessing spin qualities compatible with quantum sensing protocols at room temperature. Vacancies, in particular, are readily obtained via exposure to high-energy ion beams. While the defect creation mechanism via such irradiation is well understood, the occurrence rate of optically active negatively charged vacancies (\VB) is still an open question. In this work, we exploit focused helium ions to systematically create optically active vacancy defects in hBN flakes at varying density. By comparing the density-dependent spin splitting measured by magnetic resonance to calculations based on a microscopic charge model, in which we introduce a correction term due to a constant background charge, we are able to quantify the number of {\VB} defects created by the ion irradiation. We find a lower bound for the fraction (0.2\%) of all vacancies in the optically active, negatively charged state. Our results provide a protocol for measuring the creation efficiency of {\VB}, which is necessary for understanding and optimizing luminescent centers in hBN.
\end{abstract}

%\keywords{quantum emitters, hexagonal boron nitride, photoluminescence, defects, ion beam}

\maketitle

\section{Introduction}

\begin{figure*}[hbpt]
  \centering
  \includegraphics[width=2\columnwidth]{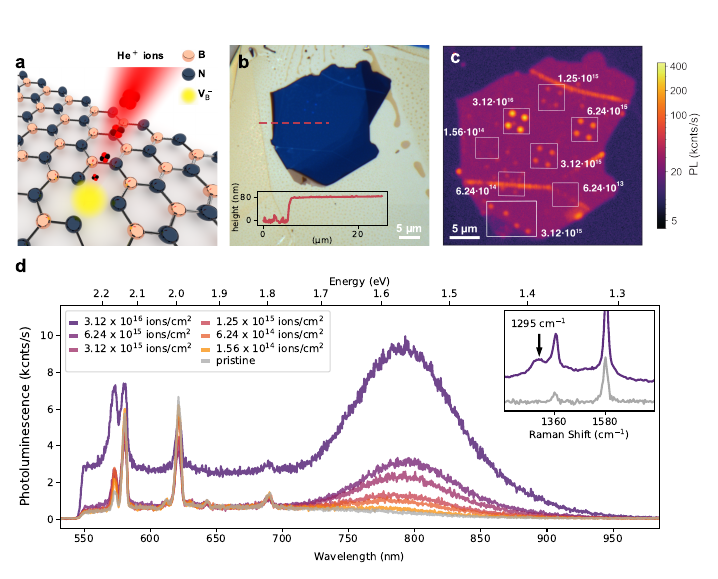}
  \caption[]{\textbf{Focused He-ion beam creation of {\VB} emitters in hexagonal boron nitride.} \textbf{a} Schematics of a focused He-ion beam creating (charged) boron vacancy defects in hBN. \textbf{b} Optical image of a 80 nm thick hBN crystal (blue) placed on a thick graphite flake. The inset shows a height profile along the dashed line. \textbf{c} Scanning confocal image of the photoluminescence (PL) after He-ion beam irradiation. The luminescence is filtered from 650 nm to 850 nm. The labels indicate the corresponding exposure fluences in units of ions/cm$^2$. \textbf{d} Representative PL spectra recorded at irradiated sites for varying fluences. For the highest irradiation fluence (purple) a broad, spectrally flat background PL appears and a defect-activated mode emerges in the Raman spectrum at \SI{1295}{cm^{-1}} (inset).}\label{fig:PL}
\end{figure*}

Luminescent centers in hBN emerged as bright and stable room temperature quantum emitters \cite{Tran2016, FischSciAdv}, rivaling established platforms in quantum communication and sensing applications \cite{Aharonovich2016, Gottscholl2021, Montblanch2023}. Emitters in hBN exhibit robust photoluminescence (PL) across a broad spectral range from the near-infrared \cite{Gottscholl2020} through the visible \cite{Tran2016, KumarAPL2023} up to the blue \cite{shevitski2019blue, Fournier2021, Gale2022, Horder2022, Chen2023Jul} and UV \cite{Bourrellier2016} regions, as well as magnetic response both as ensemble spin-1 systems \cite{Gottscholl2020} and as single spin-1 \cite{Stern2022, Stern2024} or spin-1/2 \cite{Gao2023} emitters. Among the most studied types of luminescent centers in hBN is the negatively charged boron vacancy (\VB), which consists of an electron localized at the position of the missing atom in the hexagonal lattice. These defects are typically addressed in PL ensemble measurements, where they exhibit a very broad spectrum centered around 800 nm \cite{Gottscholl2020, Hennessey2024, Healey2023, Gao2021, Kianinia2020, Guo2022, Glushkov2022, Ren2023, Liu2023, Baber2022, Ramsay2023, Toledo2018, Liang2023, Whitefield2023, Clua-Provost2024}. Importantly, {\VB} defects feature a spin-triplet ground state with a zero-field splitting of around 3.5 GHz \cite{Gottscholl2020}, a corresponding triplet excited state with a splitting of around 2.1 GHz \cite{Baber2022, Mu2022, Mathur2022, Yu2022}, and a metastable state acting as a non-radiative channel with a very short lifetime \cite{Whitefield2023, Clua-Provost2024}. The spin state can be optically read-out and coherently manipulated through optically detected magnetic resonance (ODMR) measurement schemes \cite{Gottscholl2020, Liang2023, Mendelson2022, Gao2022, Clua-Provost2024}. The spin sub-levels, and thus the two resonances in the ODMR spectrum, are sensitive to external magnetic fields \cite{Gottscholl2020, Baber2022}, which can be applied for quantum sensing schemes \cite{vaidya2023, huang2022, Kumar2022, Healey2023}, similar to nitrogen vacancy-centers in diamond \cite{Maze2008, Balasubramanian2009}. Because of their potential for applications in quantum technologies, the optical and magnetic properties of {\VB} defects have been a subject of intensive research, including their optical response at room \cite{Gottscholl2020, Baber2022, Sarkar2024, Liang2023, Clua-Provost2024} and cryogenic \cite{Kianinia2020, Sarkar2024} temperatures, as well as ODMR \cite{Gottscholl2020, Kianinia2020, Mu2022, Yu2022}, electron paramagnetic resonance (EPR) \cite{Gottscholl2020, Murzakhanov2021}, and spin-coherence measurements \cite{Gottscholl2021, Rizzato2023, Gong2023, Liu2022}. Furthermore, several studies discuss various sample preparation and optimization protocols, such as irradiation sputtering \cite{Glushkov2022}, ion beam fluence and energy tuning \cite{Guo2022, Sarkar2024}, high-temperature annealing \cite{Grosso2017, Ren2023, Liu2023}, water exposure \cite{Glushkov2022}, as well as the integration of defects into photonic circuits, such as nanocavities \cite{Froch2021, Finley2022}, plasmonic- \cite{Gao2021} and metasurfaces \cite{Sortino2024}. 

For photonic integration reliable, quantitative and deterministic methods for defect creation are desirable. In this context, different methods were successfully applied to create {\VB} in hBN, which include focused ion beam irradiation using H \cite{Hennessey2024}, He \cite{Grosso2017, Liang2023, Sarkar2024, Gao2021, Ren2023, Guo2022, Sasaki2023}, N \cite{Clua-Provost2024, Guo2022, Kianinia2020, Zabelotsky2023, Suzuki2023}, Ga \cite{Gottscholl2020, Mu2022}, C \cite{Guo2022, Baber2022, Ren2025}, Xe \cite{Kianinia2020} and Ar \cite{Guo2022, Kianinia2020}, neutron irradiation \cite{Toledo2018, Gottscholl2020, Kumar2022, Durand2023, Udvarhelyi2023, Haykal2022}, femtosecond laser writing \cite{Gao2021femto}, and electron beam irradiation \cite{Murzakhanov2021, Jin2009}. In particular, several works discussed the usage of a focused He-ion beam source to effectively create {\VB} defects in hBN \cite{Liang2023, Sarkar2024, Gao2021, Ren2023, Guo2022, Sasaki2023}. Focused He-ion beam exposure offers a nominal lateral defect positioning accuracy below 10 nm in very thin films \cite{Mitterreiter2020}, while preserving the local crystal structure of hBN at adequate doses. All such nanofabrication measures are crucial factors for an efficient integration of such luminescent centers in future photonic \cite{Grzeszczyk2024} and quantum imaging devices \cite{Healey2023}. In this context, several works \cite{Sarkar2024, Liang2023, Ren2023} already provided initial insights onto the He-ion creation process and the optical response of the created ensemble emitters, but a systematic understanding, in particular a reliable quantification of defect density and its impact on the magnetic and optical properties of the created emitters, is still lacking.

In the present work, we use a helium ion microscope (HIM) to irradiate hBN flakes, deposited on thick graphite flakes, to create {\VB} defects (\figref[a]{PL}). The thick graphite acts as both a mechanical and optical protection layer, reducing the impact of back-scattered ions from the SiO$_2$ substrate and suppressing unwanted background luminescence from the irradiation damage in the substrate. We systematically perform spatially- and time-resolved PL as well as ODMR measurements on defect ensembles for varying irradiation fluences, confirming the optical and magnetic activity of the created emitters. We show that the focused He-ion beam technique can site-selectively produce luminescent and background-free emitters, in good agreement with the existing literature \cite{Gottscholl2020, Liang2023, Sarkar2024, Gao2021, Ren2023, Guo2022, Sasaki2023}. Furthermore, we apply a microscopic charge model \cite{Gong2023, Udvarhelyi2023} to describe the defect density-dependent shift of the ODMR resonance frequencies. By introducing a background charge correction, we can quantify from the experimental data the density of the charged defects created by the ion exposure. By comparison to molecular dynamics (MD) simulations, which estimate the number of total vacancies created, we are able to quantify an overall efficiency of the charged defect creation process based on the focused He-ion beam irradiation. Our work provides a systematic study of defect production efficiency by comparing different irradiation fluences on homogeneous and uniform samples, offering a benchmark for future process optimization and device implementation.

\section{Methods}\label{sec:methods}

Thin layers of hBN and graphite are exfoliated using adhesive tape on Si/SiO$_2$ substrates. Subsequently, hBN/graphite heterostructures are assembled with a hot-pick-up technique \cite{Purdie2018} and cleaned by immersion in chloroform. The individual flake thickness is assessed with atomic force microscopy (AFM). To create luminescent defects in hBN, we use a focused He-ion beam (ion energy of \SI{30}{keV}) to pattern an array of circular patches with \SI{400}{nm} diameter and 2 µm pitch with ion fluences as defined in \tabref{doses}. Optical characterization is performed in a custom confocal microscope featuring a 532-nm laser and a grating spectrometer. Continuous-wave (cw) ODMR measurements, with the PL integrated from 650 nm to 850 nm, are performed with 532 nm laser excitation and a microwave excitation at 3 GHz -- 4 GHz. Molecular dynamics (MD) simulations are carried out using the LAMMPS package \cite{Plimpton1995}, to extract the statistics of defects produced by the impacts of He ions onto the hBN. More details about the methods can be found in \appref{methods} and the motivation for the simulation setup is discussed in Refs. \cite{Kretschmer2018, Ghaderzadeh2021, Lehtinen-2011}.

\section{Results}
\setlength{\tabcolsep}{.2cm}
\begin{table*}[htbp]
  \centering
  \begin{tabular}{ccccccc}
    \toprule
    \midrule
    \# & Dose $\left(\mu\text{C}/\text{cm}^2\right)$ & Fluence $\left(\text{ions}/\text{cm}^2\right)$ \ & $\rho_c$ $\left(e/\text{nm}^{-3}\right)$ \  & {\VB} density $\left(\text{nm}^{-3}\right)$ \ & V$_\text{B}^\text{(MD)}$ density $\left(\text{nm}^{-3}\right)$ \ & \VB/V$_\text{B}^\text{(MD)}$ \\
    \midrule
    1 & 5000    & \SI{3.12e16} \ & 0.0343(38) & 0.0059(51) & 7.67(9)   & 0.2 \%  \\
    2 & 1000    & \SI{6.24e15} \ & 0.0296(36) & 0.0012(49) & 1.53(2)   & 0.2 \%  \\
    3 & 500     & \SI{3.12e15} \ & 0.0291(40) & 0.0006(52) & 0.767(9)  & 0.2 \%  \\
    4 & 200     & \SI{1.25e15} \ & 0.0286(55) & 0.0002(65) & 0.307(4)  & 0.1 \%  \\
    5 & 100     & \SI{6.24e14} \ &     -      &     -      & 0.153(2)  &   -     \\
    6 & 25      & \SI{1.56e14} \ &     -      &     -      & 0.0384(5) &   -     \\
    7 & 10      & \SI{6.24e13} \ &     -      &     -      & 0.0153(2) &   -     \\
    \midrule
    \bottomrule
  \end{tabular}
  \caption{Summary of the different irradiation fluences (see \appref{he_irradiation} for details about the irradiation) and the corresponding charge density $\rho_c$ and {\VB} density extracted from the analysis in \figref[]{odmr}. The values are compared against the simulated density of (neutral) boron vacancies V$_\text{B}^\text{(MD)}$ from molecular dynamics simulations discussed in \figref[]{simulation}. The last column denotes the estimated creation yield of optically active defects defined as the ratio \VB/V$_\text{B}^\text{(MD)}$. Note that this value is a lower bound and the actual yield may be substantially higher due to defect annealing processes, as discussed in the section on molecular dynamics simulations. The errors of the least significant digits are given in parentheses.}
  \label{tab:doses}
\end{table*}

\subsection{Optical characterization of \texorpdfstring{\VB}~ centers}

\figref[b]{PL} depicts a typical hBN flake (80 nm thick) on graphite supported on Si/SiO$_2$. The sample is exposed with $2 \times 2$ arrays of circular patches of \SI{400}{nm} diameter and \SI{2}{\micro\meter} pitch. In those circular patches, the ion fluence is varied from \SI{6.24e13}{ions/cm^2} to \SI{3.12e16}{ions/cm^2} (see \tabref{doses}). To spatially map the defect emission, we use confocal fluorescence imaging. Here, the PL is filtered from 650 nm to 850 nm and recorded with an avalanche photodiode in photon counting mode while the excitation laser is scanned across the sample. The fluorescence mapping after He-ion irradiation (\figref[c]{PL}) clearly resolves the ion-exposed spots by their increased luminescence (\figref[a]{si_afm_pl}) as well as other bright spots, including folds and edges. A logarithmic scale is used for the color bar to visualize a larger dynamic range.

\begin{figure}[htb]
  \hspace{-1cm}\includegraphics[width=1.\columnwidth, valign=c]{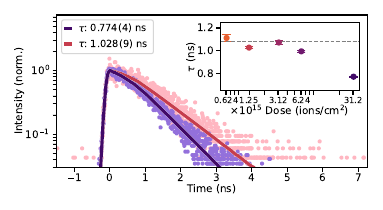}
  \caption[]{\textbf{Lifetime measurements on {\VB} emitters}. Time-resolved PL decays for two selected irradiation fluences of \SI{3.1e16}{ions \per cm^2} (purple) and \SI{1.25e15}{ions \per cm^2} (red). At the highest irradiation fluence, the decay times exhibit a significant shortening. The solid lines are fits with a double-exponential model with a short and long lifetime (the extracted decay times are shown in \appref{optical_characterization}). The inset shows the longer of both lifetimes $\tau$ extracted from the fit. Error bars are statistical errors provided by the fit routine.}\label{fig:lifetime}
\end{figure}

\figref[d]{PL} shows representative PL spectra recorded on the exposed spots for different ion fluences. The most prominent luminescence feature is a broad peak ($\text{FWHM} \approx \SI{100}{nm}$) centered around \SI{800}{nm} (\SI{1.57}{eV}), whose intensity increases monotonously with increasing He-ion exposure (\figref[a]{si_pl_dependency}). Consistent with earlier reports \cite{Kianinia2020, Gao2021, Clua-Provost2024}, the PL can be saturated at large powers, about 1.37 mW (measured at the entrance aperture of the objective) for spots with \SI{3.12e16}{ions/cm^2} (\figref[b]{si_pl_dependency}), which is indicative of emission from a finite number of localized defect centers \cite{Guo2022, Liang2023, Whitefield2023, Clua-Provost2024}. For comparison, the gray spectrum in \figref[d]{PL} was recorded in a location not exposed to He-ions, i.e. on pristine hBN, where this spectral feature is absent. The peak position and spectral shape of the ion-induced luminescence feature are consistent with other studies that demonstrated emission from {\VB} \cite{Gottscholl2020, Hennessey2024, Healey2023, Gao2021, Kianinia2020, Guo2022, Glushkov2022, Ren2023, Ivady2020}.  

The large broadening, which obscures the observation of a zero-phonon line, is generally attributed to dominant phonon-assisted emission processes from an ensemble of emitters \cite{Gottscholl2020, Gottscholl2021, Libbi2022}. Cavity-enhanced experiments put the zero-phonon line of {\VB} at around 770 nm \cite{Finley2022}, whereas simulations show that phonon-independent luminescence is extremely weak \cite{Libbi2022}. 
It should be noted that making a quantitative comparison of the line shape for such a broad luminescence feature across different studies is often challenging, unless meticulous calibration of the overall spectral efficiency of the optics used is performed. This accounts for slight variations between this work and other works \cite{Gottscholl2020, Reimers2020, Liang2023, Guo2022, Clua-Provost2024}. At the highest fluence (\SI{3.12e16}{ions/cm^2}, dark purple spectrum in \figref[d]{PL}), a background luminescence emerges across the full spectral range. This increased background correlates with the occurrence of an additional Raman mode at \SI{1295}{cm^{-1}}, which has previously been attributed to ion-beam induced damage in hBN \cite{Sarkar2024, Liang2023, Li2021} and which is absent in the pristine control spectrum (inset of \figref[d]{PL}, see \figref{si_raman} for the full Raman spectrum). The background luminescence and defect-activated Raman mode may hint towards the onset of ion-induced amorphization of the hBN lattice \cite{Liang2023} or the prevalent creation of larger defect clusters \cite{Li2021}.

Consistently, we find that the PL lifetime does not display any clear change with irradiation fluence (or equivalently the induced defect density) except for the highest irradiation fluence, where the lifetime drops significantly from about \SI{1.1}{ns} to \SI{0.7}{ns} (\figref{lifetime}). The lifetimes are in line with values reported by other studies \cite{Gottscholl2020, Baber2022, Mu2022, Sortino2024, Clua-Provost2024}. To fit the experimental data, a double-exponential model is applied, in which the first component captures the lifetime at longer time scales and the second component addresses the fine details at short time scales. The fit includes a convolution with the independently measured instrument response function. The time traces are shown in \figref{lifetime} exemplarily for \SI{1.25e15}{ion/cm^2} (red curve) and \SI{3.12e16}{ion/cm^2} (blue curve). \tabref{lifetimes} in \appref{optical_characterization} summarizes the extracted time constants (see \figref{si_lifetimes} for the full lifetime data). Note that for optical emission with low quantum yield, as is the case for {\VB}, the measured lifetimes in time-resolved PL reflect a combination of the radiative and non-radiative rates. Therefore, the shortening of the lifetime with the fluence indicates an increased non-radiative recombination, which is likely due to the significant damage imparted onto the hBN lattice at the highest fluence. So far, studies reported tuning of the lifetime via temperature \cite{Mu2022} or by exploiting spin-dependent non-radiative recombination using specific measurement pulse sequences \cite{Baber2022, Clua-Provost2024} (where first long optical pulses initialize the spin polarization in the state $\ket{m_s=0}$ and microwave $\pi$-pulses rotate the spin polarization in one of the states $\ket{m_s=\pm1}$), thus shortening the overall lifetime. In our case we are therefore measuring the PL decay dynamics of a population in which both the radiative and non-radiative recombination mechanisms play a role \cite{Clua-Provost2024}.

\subsection{Optically detected magnetic resonance and microscopic charge model}

\begin{figure*}[htbp]
  \centering
  \includegraphics[width=1.95\columnwidth, valign=c]{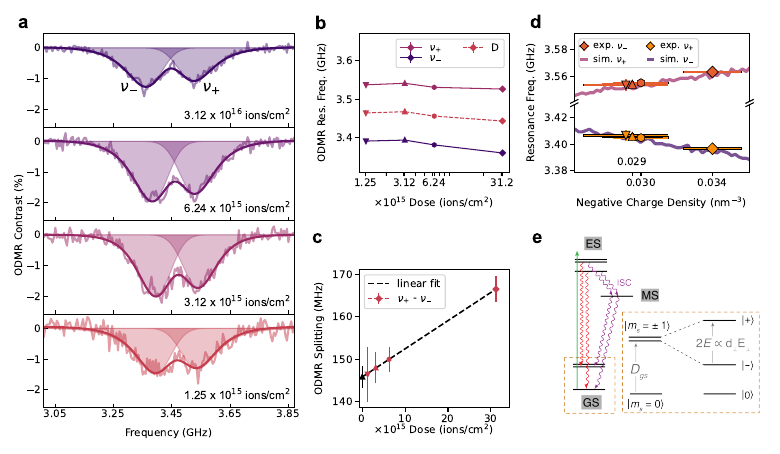}
  \caption{\textbf{ODMR characterization of the defects.} \textbf{a} ODMR contrast at zero magnetic field detected on different locations on the sample corresponding to different ion fluences. The solid lines are fits to the data with two exponentially modified Gaussians corresponding to the two spin-split resonances denoted $\nu_-$ and $\nu_+$. \textbf{b} Positions of the ODMR resonances $\nu_+$ and $\nu_-$ and their midpoint D. The error bars are smaller than the marker size.  \textbf{c} Magnitude of the splitting $\nu_+ - \nu_-$ extracted from the fit in \textbf{a}. The error bars denote the uncertainty of the parameters as given by the fit routine. The black data point shows the intercept extracted from the linear fit, with the relative fit error. \textbf{d} Level splitting calculated within a microscopic charge model for $\nu_+$ and $\nu_-$ (solid lines) as a function of negative charge density. The experimentally derived values of $\nu_+$ and $\nu_-$ (data points) are fit onto the data yielding an effective charge density for each data point. The symbols correspond to the ion fluences and associated symbols in \textbf{c}. \textbf{e} Schematic of the three-level structure of a {\VB} with ground (GS), excited (ES), and metastable (MS) state. The metastable state is populated via intersystem crossing (ISC). The ground state is split by sub-level mixing due to the in-plane $E_\perp$ field.}\label{fig:odmr}
\end{figure*}

The {\VB} defect is considered to be one of the few hBN defects whose atomic structure and level scheme are well known \cite{Gottscholl2020, Ivady2020, Reimers2020, Baber2022, Mu2022, Gong2023, Udvarhelyi2023, Durand2023}. \figref[e]{odmr} shows the level structure. The system exhibits a spin triplet ground state (GS) ${}^3A_2'$ and a corresponding spin triplet excited state (ES) ${}^3E''$. The ground state shows a so-called axial zero-field splitting $D_\text{gs}$ around \SI{3.46}{GHz} separating the $\ket{m_s=0}$ and $\ket{m_s=\pm1}$ spin sublevels. The term $m_s$ indicates the projection of electron spin along the $c$-axis of the hBN crystal, considered the quantization axis of the system \cite{Udvarhelyi2023}. An electric field $E_\perp$ perpendicular to the quantization axis mixes the $\ket{m_s=\pm1}$ sub-levels, resulting in new eigenstates $\ket{+}$ and $\ket{-}$ with a level splitting of $2E \propto d_\perp \cdot E_\perp$ (\figref[e]{odmr} inset), where $d_\perp \approx \SI{40}{Hz/(V\,cm^{-1})}$ denotes the corresponding perpendicular susceptibility. Transitions between the ground and excited states are optically addressable under green laser excitation (typical wavelength 532 nm). Additionally, a manifold of metastable singlet states (MS) is present, which can be populated from the excited state via intersystem crossing (ISC). The phonon-assisted radiative emission from excited state to ground state is rather weak since it is forbidden by symmetry \cite{Clua-Provost2024}, thus {\VB} emitters intrinsically have low radiative recombination rate and, hence, a low quantum efficiency. At the same time, the metastable state has a rather short lifetime (on the order of tens of nanoseconds) \cite{Whitefield2023, Clua-Provost2024}. Therefore, the systems relaxes overall rather rapidly, and the defect can be re-pumped efficiently, such that its PL can still be detected at sufficient intensity despite the intrinsically low quantum efficiency and the presence of the metastable state \cite{Clua-Provost2024}.

The ODMR spectra are measured at zero external magnetic field and on four different ion fluences (all on the same sample of \figref[c]{PL}) corresponding to four emitter densities (\figref[a]{odmr}). All ODMR spectra display two characteristic resonances. We verified that their splitting increases in frequency under a static external magnetic field \cite{Gottscholl2020, Baber2022} (\figref{si_odmr}). The slight difference in the overall magnitude of the contrast is due to a varying coupling efficiency of different positions in the sample to the nearby microwave antenna. To extract the splitting between the resonances, we fit the data with two exponentially modified Gaussians, which account for an asymmetric broadening of the resonances (see \figref[b]{si_sim} and the discussion in \appref{microscopic_charge_model}). \figref[b]{odmr} shows the resonance positions $\nu_+$ and $\nu_-$, extracted as the maxima of the individual resonance distribution, as a function of the irradiation fluence, as well as their midpoint $D$ defined as $(\nu_+ + \nu_-)/2$.

The splitting between the resonances increases with increasing ion fluence, or equivalently defect density (\figref[c]{odmr}). In literature, such splitting is attributed to the interaction of the {\VB} electronic spin with local fluctuations of the electric environment \cite{Gong2023, Durand2023, Udvarhelyi2023}. By contrast, variations of the mean $D$ are related to local strain fields \cite{Udvarhelyi2023} and temperature \cite{Liu2021}. By modeling the magnitude of the splitting, it is possible to estimate the local field strength and, therefore, the local charge density surrounding the emitters. Following the approach in Refs. \onlinecite{Gong2023, Durand2023, Udvarhelyi2023}, we apply a microscopic charge model to simulate the level splitting. Briefly, we consider level splitting due to a net electric field, which is generated by a discrete and random charge distribution in the vicinity of the {\VB} site. Importantly, the model still assumes overall charge neutrality, i.e. positive and negative point charges are placed around the {\VB} in equal numbers (see \appref{microscopic_charge_model} for details). The remaining net electric field arises solely from the locally net sum of the field components. To date, the microscopic origin of such a charge environment is not fully understood. A common assumption is that the charge fluctuations simply correspond to the number of charged {\VB} defects, which, as we discuss in the following, is not fully consistent with our experimental results.

\figref[d]{odmr} depicts the simulation results for a selected range of average charge densities, in which we fit the experimentally measured values for different ion fluences (orange data points) to the calculated values of $\nu_+$ and $\nu_-$ as a function of charge density (solid lines). In the fitting procedure, we match only the splitting $\nu_+ - \nu_-$ rather than the absolute values of $\nu_+$ and $\nu_-$, such that we eliminate strain effects affecting the mean $D$ \cite{Udvarhelyi2023} (see \figref{si_sim} for full data set and details of fitting procedure). The error bars take into account both the experimental uncertainty, i.e. the fitting error of the peak splitting (cf. \figref[c]{odmr}), and the theoretical uncertainty, i.e. the statistical spread of the eigenvalues in the simulation (cf. \figref[c]{si_sim}).

Generally, the extracted charge densities are in line with other reports \cite{Gong2023, Durand2023, Udvarhelyi2023}. This finding is consistent with the fact that we measure similar magnitudes of the splitting and incorporate the value of the out-of-plane susceptibility $d_\perp$ determined in Ref. \onlinecite{Gong2023} as the main parameter in the model. However, unlike Ref. \onlinecite{Udvarhelyi2023}, we do not observe a proportional scaling of charge density and PL intensity with ion fluence, whereby the latter can be taken as a good estimate of the relative number of optically active defects for a constant lifetime (cf. \figref[]{lifetime}) and for an excitation power below the saturation threshold (\figref[]{si_pl_dependency}). Specifically, the PL intensity of {\VB} emission increases by a factor of 10 (\figref[]{si_pl_dependency}) as the ion fluence is varied over more than one order of magnitude from \SI{1.25e15}{ions \per cm^2} to \SI{3.12e16}{ions \per cm^2} while, at the same time, the average charge density extracted from the model increases only slightly from \SI{0.029}{nm^{-3}} to \SI{0.034}{nm^{-3}}. Therefore, as a possible extension of the model, we propose that the overall charge density should include a constant positive and negative background charge, potentially due to other defect species. Charge transfer processes between such defects can then lead to local charge accumulation, while preserving overall charge neutrality. Such a background charge would lead to a finite level splitting already in the limit of small {\VB} defect density, as we clearly observe for \SI{1.25e15}{ions \per cm^2} (\figref[a]{odmr}). With these assumptions, we determine a background charge of \SI{0.0285(33)}{nm^{-3}}. From that, we can find the corrected values for the {\VB} density as given in \tabref{doses}. The origin of such a background charge may potentially be related to native defects, such as interstitials, impurities, or vacancies present after the crystal growth and, hence, before the irradiation process. Additionally, defects introduced already at very low irradiation fluences, below the rather large fluences needed to create enough optically active {\VB} sites for reliable ODMR characterization, may contribute substantially to the observed ODMR splitting (see below for a more detailed discussion on the impact of crystal quality).

\subsection{Molecular dynamics simulation of vacancy creation}\label{subsec:mdsim}

In a next step, we correlate the experimentally found production yield to molecular dynamics (MD) simulations of the ion matter interaction. We model the 30 keV He-ions impinging onto a \SI{72}{nm}-thick hBN sheet. The calculations are performed for a reference fluence (see \appref{md_sim}) and $0^{\circ}$ incidence angle. Contrary to the binary collision approximation (BCA), used in common SRIM (stopping range of ions in matter) simulations \cite{Guo2022, Sarkar2024, Gao2021, Liang2023, Hennessey2024}, MD simulations take into account the molecular structure of the target materials (an example is shown in \figref[c]{simulation}), which yields more precise estimates of the defect production. \figref[a]{simulation} depicts the depth profile of nitrogen (blue) and boron (red) vacancies at the reference fluence corresponding to \SI{0.8e16}{ions \per cm^2}. On average, we find densities of \SI{2.95}{vacancies\per nm^3} and \SI{2.46}{vacancies \per nm^3} for nitrogen and boron, respectively. \figref[b]{simulation} compares the MD approach to simulations based on BCA. For the latter, we find on average higher defect densities, which agrees with the general notion that these simplified methods tend to overestimate the defect creation.

\begin{figure}[htb]
  \centering
  \includegraphics[width=.95\columnwidth, valign=c]{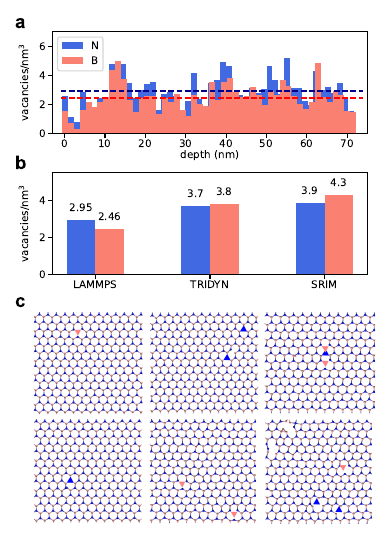}
  \caption[]{\textbf{Molecular dynamics (MD) and binary collision approximation (BCA) simulations for He-ion irradiated hBN.} \textbf{a} Density for both nitrogen (blue) and boron (red) vacancies along the depth of the hBN flake, under 30 keV He-ions irradiation (fluence \SI{0.8e16}{ions \per cm^2}) calculated within MD. The dashed lines indicate the averaged values of the vacancy densities. \textbf{b} Comparison of the vacancy densities averaged across an hBN film for different simulation algorithms LAMMPS (MD), TRIDYN (BCA), SRIM (BCA). \textbf{c} Six representative examples of the atomic structure of hBN sheets after He-ion irradiation simulated by MD. Boron and nitrogen vacancies are indicated in red and blue triangles, respectively.}\label{fig:simulation}
\end{figure}

We thus compare the results from the MD simulations and from the microscopic charge model to estimate the production efficiency of the irradiation process. The results are shown in the last column of \tabref{doses}. Overall, our analysis suggests that the production efficiency is rather low at about 0.2 \%, implying that, although with larger implantation fluences more vacancies are created, a large fraction of vacancies does not get promoted to {\VB} and remains optically inactive \cite{Abdi2018, Linderalv2021, Gong2023}. However, the actual fraction of defects that become optically active may be substantially better, as our simulations do not account for \textit{in situ} annealing of defects through annihilation of vacancies with interstitials, which are mobile at room temperature \cite{Weston2018, Strand2020}. The annealing should decrease the number of created vacancies, independent of their charge state. Unfortunately, direct simulation of annealing process using MD is challenging, as annealing happens at macroscopically long time scale, unachievable using MD simulations. In addition, charge transfer (e.g., between vacancies and interstitials) and mutual attraction of the defects through screened Coulomb interaction are possible, which should further affect the defect dynamics. The only way to model this process is to first perform extensive DFT calculations to assess all migration barriers, accounting for the interaction between the defects, and then perform kinetic Monte Carlo simulations. Such simulations will require substantial computational effort and are beyond the scope of this work. Overall, our values for the defect yield constitute a lower bound, since we cannot quantitatively account for the impact of defect annealing and since we included the potential effect of a constant background charge in the analysis of the ODMR splitting. In contrast, Ref. \onlinecite{Gong2023} reported a larger negatively charged {\VB} fraction of $\sim 1$\% for a fluence \SI{1e15}{ions \per cm^2}, which is comparable to our lowest irradiation fluence for which ODMR was measured. However, we note that the latter study concludes that the defect yield decreases steeply with ion fluence. With our background correction, the derived defect yield is constant within the experimental uncertainty. In other words, we tentatively state that the microscopic charge model without background correction tends to significantly overestimate the {\VB} density at low ion fluences. 

In this context, one may ask how the quality of the hBN crystal will impact the measured OMDR splitting and therefore the corresponding, extracted background charge. For this purpose, we compare our ODMR results with previously reported studies that explicitly observed {\VB} defects and provided sufficient information about the hBN quality, the irradiation parameters, and the ODMR results (Table \ref{tab:hbn_comparison}).  Overall, different studies, using different irradiation methods (both ions and neutrons) and different source crystals, report comparable values of the {\VB} ODMR splitting (about 100 - 150 MHz). Furthermore, a consistent observation throughout is that lower ion fluences result in smaller splitting, in line with our results. The lowest values (about 90 MHz) were achieved in high-pressure, high-temperature grown hBN (from NIMS, Japan) \cite{Sasaki2023, Suzuki2023}. Indeed, such high-pressure, high-temperature grown hBN is generally considered to be less defective, as evidenced for example by its lower intrinsic defect-related luminescence \cite{Schue2016}, compared to hBN crystals provided by commercial sources, such as HQgraphene (as used in the present work). Yet, we point out that based on the currently available data, we cannot make a conclusively statement on the relation between the hBN quality and the ODMR splitting in the limit of low irradiation fluence. Rather a key aspect is that ODMR splittings of irradiation created {\VB} close to 0 MHz were only reported in few-layer hBN \cite{Durand2023}. Here, it was argued that not the purity of the hBN, but rather the 2D geometry plays a decisive role in reducing the number of nearby background charges resulting in a clean ODMR spectrum with almost negligible peak splitting.

\section{Conclusions}
In summary, we quantified the defect creation within homogeneous hBN films using a helium ion microscope to site-selectively pattern of {\VB} defect ensembles. We analyzed the optical properties of the emitters as a function of ion fluence over more than two orders of magnitude. The spatially precise creation allowed us to study the magnetic response probed by ODMR contrast measurements in a consistent and systematic way. We simulated the level splitting obtained experimentally from the ODMR spectra employing a microscopic charge model, where the spatially random distribution of local charges results in a level splitting in ODMR. As a main finding, we observe a discrepancy between the scaling of the charge density, or equivalently the level splitting in ODMR, and the scaling of the defect PL, or equivalently the density of optically active defects, as function of ion fluence. Unlike other studies, which attributed the charge density in the microscopic model solely to {\VB} defects, we propose to include a background density in the model. With this modification, we were able to estimate the defect creation yield and compare it to molecular dynamics simulations. On one hand, further developments may involve the implementation of new theoretical models with the aim of addressing the fundamental question of the origin of the {\VB} charge \cite{PRM2023}, such as the nature of the charged state (dynamic or static) and the role of the local defect environment, as it has been done for N$_{\text{V}}$ centers \cite{Giri2023}. On the other hand, spatially precise creation of such emitters as demonstrated here may enable their integration in gated, light-emitting hetero-stacks, in which the PL emission can be tuned upon application of a bias voltage \cite{Fraunie2025} and a tunneling current can be coupled to the in-gap defect states by means of graphene electrodes \cite{Grzeszczyk2024, Park2024}.

\begin{acknowledgments}
We thank Moritz Fischer for helpful discussion and Dorte Rubæk Danielsen for help in the sample fabrication.
N. S. thanks the Novo Nordisk Foundation NERD Programme (project QuDec NNF23OC0082957).
Work at the Technical University of Munich was supported through the Deutsche Forschungsgemeinschaft (DFG, German Research Foundation) via the Munich Center for Quantum Science and Technology No. (MCQST)-EXC-2111-390814868. C.K., L.G., J.F. and A.W.H. acknowledge funding through the International Graduate School of Science and Engineering (TUM-IGSSE, project BrightQuanDTUM). N.S. and M.W. acknowledge the support from the Independent Research Fund Denmark, Natural Sciences (project no. 0135-00403B) and from the Danish National Research Foundation through NanoPhoton - Center for Nanophotonics (project no. DNRF147). A.H. and I.B. acknowledge the NNF Biomag project (NNF21OC0066526) and the Danish National Research Foundation through the center for macroscopic quantum states (bigQ, project no. DNRF0142). A.V.K. acknowledges funding from the German Research Foundation (DFG), projects KR 4866/9-1, and the collaborative research center “Chemistry of Synthetic 2D Materials” CRC-1415-417590517. Generous CPU time grants from the Technical University of Dresden computing cluster (TAURUS) and Gauss Centre for Supercomputing e.V., Supercomputer HAWK at Höchstleistungsrechenzentrum Stuttgart are greatly appreciated. 
\end{acknowledgments}

\section*{Conflict of Interest}
The authors have no conflicts to disclose.

\section*{Author Contributions}
N.S., A.H., A.W.H and C. K. conceived the project. N.S., C.K., A.H., M.W. and A.W.H. supervised the project. A.C. fabricated the sample. J.F. and A.C. performed the HIM irradiation. L.G. performed the AFM measurements. A.C. performed PL, Raman and lifetime measurements. I.B. and A.C performed PL and ODMR measurements. A.C. performed the microscopic charge model simulations. S.K. and A.V.K. performed the MD simulations. J.A. and A.B.M. performed measurements on further samples.

\section*{Data Availability}
The data that support the findings of this article are openly available \cite{Carbone2025}.

\renewcommand\thefigure{\thesection.\arabic{figure}}
\setcounter{figure}{0} 

\appendix
\section{Methods}\label{app:methods}

\subsection{Sample fabrication}\label{app:sample_fabrication}

\begin{figure}[b]
  \centering
  \includegraphics[width=.95\columnwidth, valign=c]{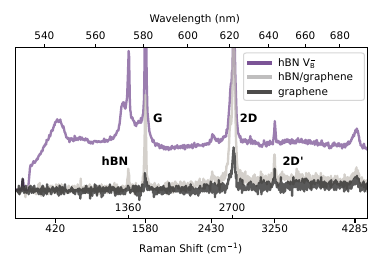}
  \caption[]{\textbf{Raman spectra taken on different spots of the irradiated hBN.} The purple spectrum is taken on hBN on graphite at an irradiated site with {VB} emitters (ion fluence \SI{3.12e16}{ions/cm^2}). The light grey spectrum is taken on pristine hBN on graphite. The dark grey spectrum is taken on graphite. The labels indicate the characteristic hBN phonon mode and the G, 2D, and 2D' modes of graphene.}\label{fig:si_raman}
\end{figure}

Thin layers of hBN (HQ Graphene) and graphite are exfoliated using adhesive tape (Nitto blue tape for hBN and Scotch Magic Tape for graphite) on Si/SiO$_2$ substrates (oxide thickness 90 nm, cleaned by sonication in acetone and subsequently isopropanol). For graphite exfoliation, we pre-heat the substrates, which were beforehand treated in a mild oxygen plasma for 10 minutes. Suitable hBN and graphite flakes are identified by optical microscopy based on size, contrast, and cleanliness. To assemble hBN/graphite heterostructures, we use a hot-pick-up technique based on poly-carbonate (PC)/polydimethylsiloxane (PDMS) stamps. A thin film of PC is placed on a droplet of PDMS on a glass plate. The stamp is brought in contact with the target hBN flake, and the substrate is heated to \SI{120}{\celsius} to pick-up the flake. The hBN is subsequently released on the target graphite flake by melting the PC film in contact with the target substrate at \SI{180}{\celsius}. The samples are cleaned by immersion in chloroform for few minutes. Finally, mild oxygen plasma cleaning (\SI{90}{\second}) is performed to remove any remaining polymer residues from the sample surface. The cleanliness and individual flake thickness of the prepared heterostructures are characterized by standard atomic force microscopy (AFM, neaSNOM Attocube Systems AG) in tapping mode.

\subsection{Optical Characterization}\label{app:optical_characterization}

\begin{figure}[bp]
  \centering
  \includegraphics[width=\columnwidth, valign=c]{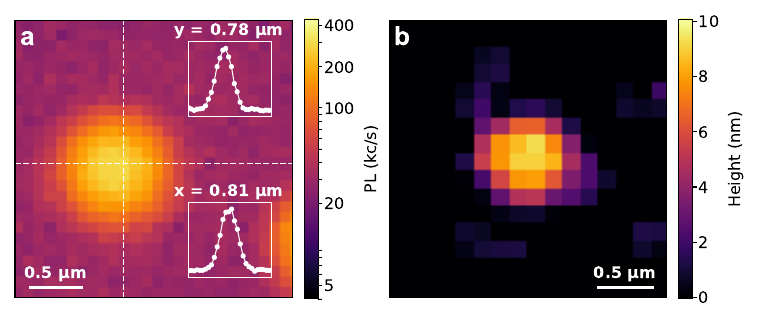}
  \caption[]{\textbf{Correlation of luminescence and local ion exposure.} \textbf{a} High-resolution PL mapping of a single HIM irradiation spot on hBN (circular patch of diameter \SI{0.4}{\micro m}). The insets are Gaussian fits of the PL intensity along the $x$ and $y$ directions. The corresponding FWHM (about 800 nm) is consistent with the width of the exposure pattern and the far-field optical resolution. \textbf{b} Topography scan of the same irradiated spot as measured by AFM. For the highest ion fluence (\SI{3.12e16}{ions/cm^2}), we find an elevated topography up to \SI{10}{nm}, presumably due to lattice swelling. Scale bars are 0.5 µm.}\label{fig:si_afm_pl}
\end{figure}

Optical properties are characterized by diffraction-limited confocal scanning photoluminescence (PL) mapping, Raman microscopy, and time-resolved photoluminescence (TRPL) under ambient conditions.  PL measurements are performed on a custom confocal microscope featuring a 532-nm laser (Cobolt, Hübner GmbH). The objective is a Nikon TU Plan Fluor $100\times$/0.9 NA EPI D. Typical laser powers used for the measurements are around 3 mW (0.122 mW measured at the entrance aperture of the objective). The signal is collected by a spectrometer (Andor-Solis SR-303i, 150 grooves/mm grating with blaze at 800 nm) equipped with a CCD (Andor Newton). Before the He-ion beam irradiation described below, we verify that the selected hBN flakes contain low background luminescence in the spectral range where emission from {\VB} is expected (\SI{700}{nm} -- \SI{950}{nm}) as well as that only a few intrinsic quantum emitters, with typical emission lines in the visible, are present. Raman spectroscopy is carried out in a commercial Raman microscope (WiTec Alpha, 532 nm laser, cw-excitation, 300 grooves/mm grating with blaze at 500 nm, Zeiss EC Epiplan-Neofluar objective lens $100\times$ / 0.9 NA).

\begin{figure}[hbt]
  \centering
  \includegraphics[width=.9\columnwidth, valign=c]{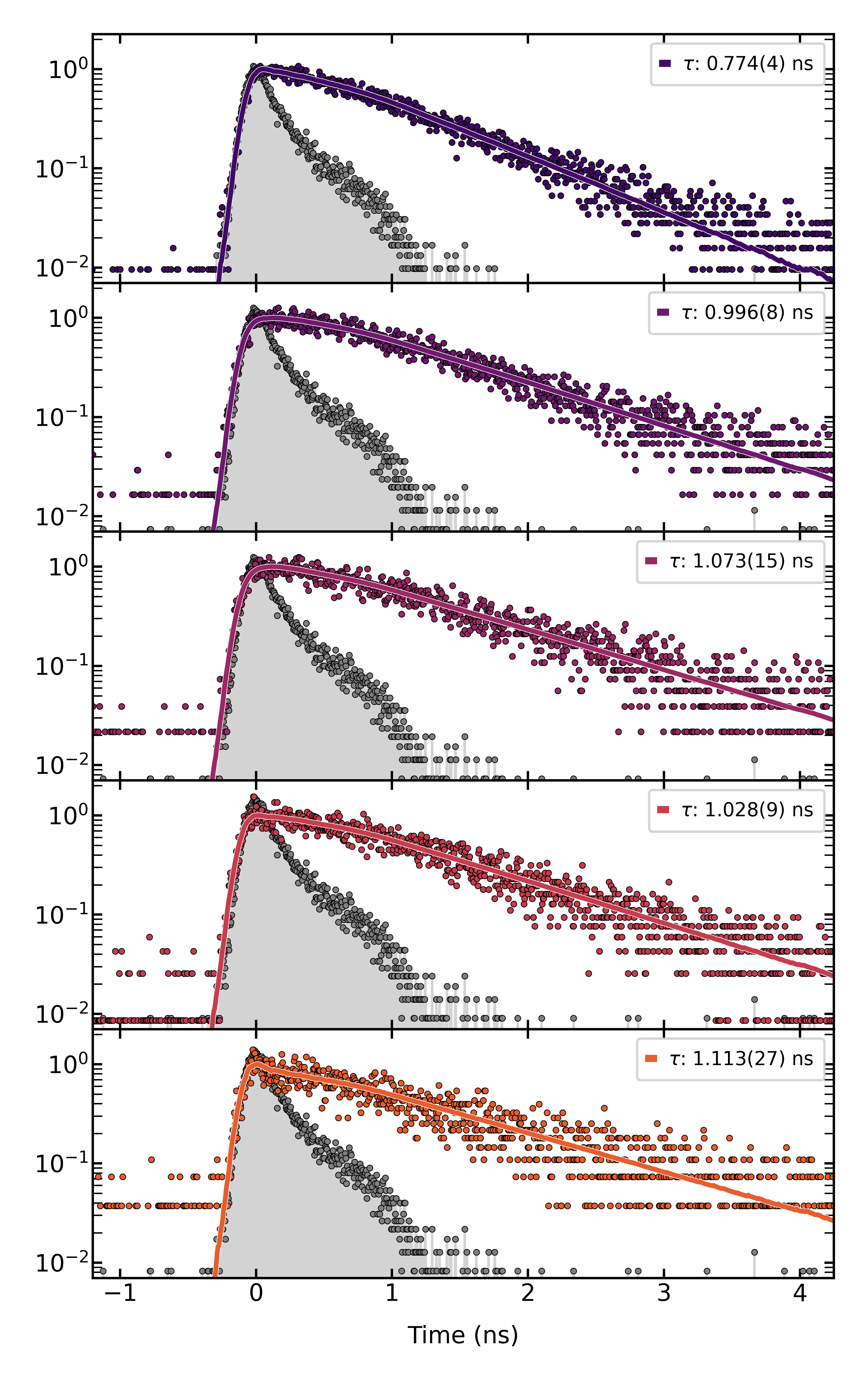}
  \caption[]{\textbf{Time-resolved photoluminescence and lifetime fitting of {\VB} emitters for different irradiation fluences}. The measured instrument response function (IRF) is shown along with fits for each decay curve. The fits are performed using a double exponential function to account for fine features at short timescales and improve the overall fitting at long time scales. The legends quote the longer one of both decay times.}\label{fig:si_lifetimes}
\end{figure}

\begin{table}[hbtp]
  \centering
  \begin{tabular}{ccc}
    \toprule
    \midrule
    \# & $\tau_1$ (ns) & $\tau_2$ (ns) \\
    \midrule
    1	& 0.774(1) 	& 0.031(1) \\
	2	& 0.996(3) 	& 0.070(4) \\
    3	& 1.073(5) 	& 0.138(8) \\
    4	& 1.028(3) 	& 0.031(2) \\
    5	& 1.113(9) 	& 0.017(3) \\
    \midrule
    \bottomrule
  \end{tabular}
  \caption{Time constants extracted from the double-exponential fit, as shown in \figref{lifetime} and \figref{si_lifetimes}.}
  \label{tab:lifetimes}
\end{table}

Time-resolved PL measurements (\figref[]{si_lifetimes}) are performed on the same setup as for PL experiments. For the excitation, we use a PicoQuant 515 nm laser diode in pulsed mode (200 µW of power, 10 MHz repetition rate), in combination with an avalanche photodiode (MPD) and a Time Tagger 20 (Swabian Instruments). The width of the instrument response function (measured on a reflective substrate) is around 250 ps. For fitting, we use the Python package \texttt{Lifefit} \cite{Steffen2016}, which uses an FFT-based, iterative reconvolution of exponential decays with an instrument response function to model the lifetime data. During fitting, the amplitude weights of the individual decays are optimized using a non-negative least squares algorithm, while the lifetime parameters are optimized by non-linear least squares regression. To account for shot noise, \texttt{Lifefit} uses $1/\sqrt{N+1}$ as weights in the least-square optimization, where $N$ is the number of photon counts in a time bin.

\subsection{Helium-ion irradiation}\label{app:he_irradiation}

\begin{figure}[bp]
  \centering
  \includegraphics[width=0.9\columnwidth]{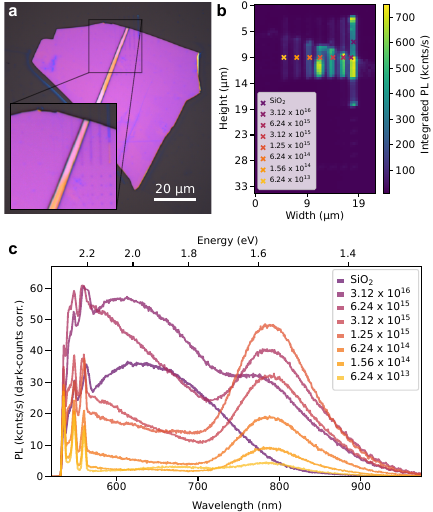}
  \caption[]{\textbf{Comparison of ion-induced luminescence from hBN and SiO$_2$.} \textbf{a} Optical image of a hBN film ($\simeq \SI{100}{nm}$ thickness). The hBN was placed directly on Si/SiO$_2$ and subsequently irradiated with 30 keV He-ions in a line pattern. The black rectangle highlights the area of the irradiation pattern, which includes positions on the bare Si/SiO$_2$. \textbf{b} PL map of the He-ion irradiated area resolving the defect PL. Importantly, also on bare Si/SiO$_2$ (purple cross) an increased PL is visible. \textbf{c} PL spectra of selected positions indicated \textbf{a}. The purple spectrum shows defect PL from bare Si/SiO$_2$, which is visible in the hBN spectra at high fluences as well. }\label{fig:SI_PL_SiO2}
\end{figure}

To create luminescent defects in the hBN films, we use a focused He-ion beam in a helium-ion microscope (Zeiss Orion). As irradiation patterns, we choose arrays of circular patches with \SI{400}{nm} diameter and 2 µm pitch. The former is on the order of the typical spot size in the confocal microscope, such that He-irradiated spots appear point-like in the PL images (\figref[]{si_afm_pl}), and the latter is much larger, so that the exposure spot can be well separated laterally. We use an acceleration voltage of \SI{30}{kV}, a beam spacing of 5 nm in x- and y-directions, a beam aperture of \SI{10}{\micro\meter}, dwell time of \SI{1}{\micro\second} and a beam current of \SI{0.1}{pA}. The values of the irradiation fluence are set in $\left(\mu\text{C}/\text{cm}^2\right)$ on the HIM (see \tabref{doses} in the main text for the fluences employed). To minimize unwanted irradiations, we use a large field of view and low imaging fluence for locating the heterostructures. For hBN placed directly on Si/SiO$_2$, we find significant background PL after ion exposure. The PL originates from both the Si/SiO$_2$ substrate and the hBN (\figref{SI_PL_SiO2}). To avoid such background PL, all the measurements in the main manuscript are conducted on hBN/graphite/SiO$_2$/Si heterostructures, resulting in spectra virtually free from unwanted background luminescence. 

\begin{figure}[tbp]
  \centering
  \includegraphics[width=.99\columnwidth, valign=c]{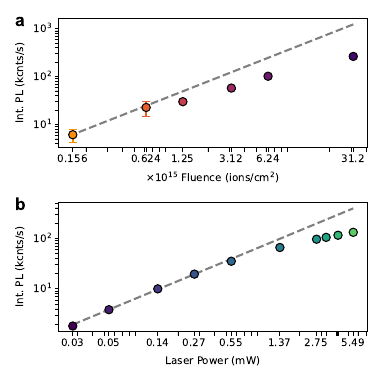}
  \caption[]{\textbf{Ion fluence and power dependence of the defect photoluminescence.} \textbf{a} Evolution of PL intensity (532 nm excitation, spectrum integrated from 650 nm to 850 nm) with irradiation fluence, measured on a HIM-irradiated hBN flake. Each data point is averaged over four irradiation spots for each fluence (cf. \figref[c]{PL}). The error bars denote the corresponding standard deviation. The PL intensity follows a sub-linear trend with irradiation fluence when compared to linear scaling (gray dashed line). \textbf{b} Evolution of PL intensity (532 nm excitation, spectrum integrated from 750 nm to 920 nm) with excitation power. The data is shown for the highest irradiation fluence of \SI{3.12e16}{ions/cm^2}. The  gray dashed line indicates linear behavior. The data points clearly show the onset of sub-linear saturation behavior at large power. The optical power is measured at the entrance aperture of the objective. Experimental parameters: 532 nm, objective Zeiss EC Epiplan-Neofluar 100x (NA = 0.9).}\label{fig:si_pl_dependency}
\end{figure}

The defect PL scales monotonously with ion fluence (\figref[a]{si_pl_dependency}). Furthermore, for a given defect density, the PL scales sub-linearly with laser power and shows saturation behavior as expected for a finite number of emitters (\figref[b]{si_pl_dependency}).

\subsection{Optically detected magnetic resonance (ODMR)} 
We perform  continuous-wave (CW) ODMR-measurements using a closed-loop coaxial antenna without electrical termination, positioned between the objective and the sample. This configuration allows the emitted light to pass through the loop. The microwave field is generated with an SRS SG384 signal generator and a +40 dBm microwave amplifier. For optical excitation, we use a laser at 532 nm, 6.0 mW with objective lens 60$\times$/0.7 NA. With the microwave excitation applied between 3 GHz -- 4 GHz, we detect the integrated PL (range 650 nm -- 850 nm). For control measurements, a magnetic field is applied through a permanent magnet placed in close proximity to the flake \figref[]{si_odmr}.

\begin{figure}[btp]
  \centering
  \includegraphics[width = .85\columnwidth]{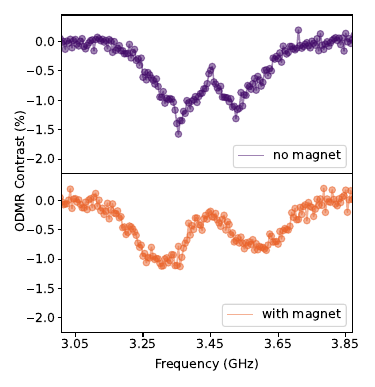}
  \caption[]{\textbf{ODMR with and without external magnetic field.} ODMR measurements performed on the highest irradiation fluence (\SI{3.12e16}{ions/cm^2}) with and without magnetic field $B$. The field is generated by a permanent magnet placed in close proximity to the sample. The application of the external $B$ field increases the splitting between the two resonances, as expected for a Zeeman effect.}\label{fig:si_odmr}
\end{figure}

\subsection{Microscopic charge model}\label{app:microscopic_charge_model}

\begin{figure}[tbp]
  \centering
  \includegraphics[width=\columnwidth, valign=c]{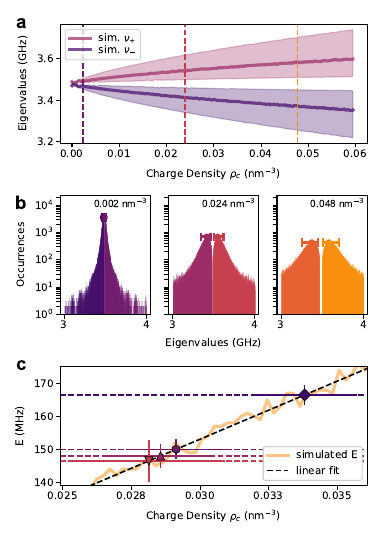}
  \caption[]{\textbf{Modeling of ODMR splitting.} \textbf{a} Simulated eigenvalues $\nu_{+, \text{sim.}}$ and $\nu_{-, \text{sim.}}$ as a result of the diagonalization of the Hamiltonian in \equationref{eq1} as a function of total charge density. The shaded areas are given by the half width at half maximum estimated for each side of each eigenvalues distribution. The three dashed vertical lines indicate the densities for which the distributions displayed in \textbf{b} were calculated. \textbf{b} Semi-log plot of the statistical distribution of the simulated eigenvalues for three selected charge densities. A Gaussian filter was applied to smoothen statistical variations in the histogram arising from the finite number of simulations. We define the splitting as the difference between the maxima of the distributions of the two eigenvalues, as indicated by the circles. The distributions have an asymmetric peak shape falling off exponentially for large detuning. We quantify the asymmetry by the half-width at half-maximum, as indicated by the error bars. These values are then used to plot the statistical variation of the eigenvalue distribution (shaded areas in \textbf{a}). \textbf{c} In the experimentally relevant range for the charge density, we linearize the model and match the result to the experimental data points.}
  \label{fig:si_sim}
\end{figure}

We model the {\VB} defect as a net negative charge situated at one lattice site surrounded by a number of $N_-$ negative and $N_+$ positive charges. The charges are localized on random sites of the hexagonal lattice and positive and negative charges are assumed to occur in equal numbers, $N_+ = N_-$, to preserve overall charge neutrality within a fixed shell volume $V$ with radius \SI{10}{nm}. We exclude charges on sites which are closer than two times the inter-atomic distance (a=\SI{0.1446}{nm} \cite{Hod2012}) effectively neglecting multi-vacancies or defect complexes and cutting off strong fields from very close charges that would otherwise lead to divergent behavior.

Such a random charge distribution creates a net electric field on the {\VB} lattice site, whose in-plane components $\mathcal{E}_{x, y}$, perpendicular to the quantization axis, mix the spin sublevels $\ket{m_s=\pm1}$. The mixing leads to new eigenstates $\ket{\pm}$ with energies $\nu_{\pm}$ separated by a splitting $2E\propto d_{\perp}\mathcal{E}_{x, y}$. The out-of-plane field components are not considered because the $d_{\parallel}$ component of the susceptibility vanishes due to mirror symmetry \cite{Gong2023, Udvarhelyi2023}. The {\VB} ground state Hamiltonian can then be written as \cite{Gong2023, Durand2023, Udvarhelyi2023}:
\begin{multline}
\mathcal{H} = D_\text{gs} S_z^2 + d_{\perp} \left(\mathcal{E}_x \left( S_y^2 - S_x^2 \right) +
\mathcal{E}_y \left( S_x S_y + S_y S_x \right)\right) + \\ \sum_{i=1}^3 A_{zz} I_z^i S_z
\label{eq:eq1}
\end{multline}
with $D_\text{gs}\approx \SI{3.48}{GHz}$, $d_{\perp}\approx \SI{40}{Hz \per (V~cm^{-1})}$ according to Ref. \onlinecite{Gong2023}, $S_i$ being the spin-1 operator for {\VB} electronic spin, $I^i$ being the spin-1 operator for the closest three ${}^{14}\text{N}$ nuclear spins, and $A_{zz} \approx \SI{47}{MHz}$ being the hyperfine coupling strength \cite{Gottscholl2020}. The nuclear spins are treated as random perturbations not correlated to the state of {\VB}, and we simply set each as $I^i = \{-1, 0, 1\}$ with uniform probability, and average over different spin configurations to extract their contribution \cite{Gong2023}. 

Diagonalizing $\mathcal{H}$, we obtain the corresponding eigenvalues $\nu_{\pm}$ as a function of $\rho_c = |e| \cdot N_{+/-} /V$. These are then averaged over 10$^4$ possible configurations to take into account the fact that each {\VB} in the ensemble sees a different charge environment. The full results are shown in \figref[a]{si_sim}. Note that, at each charge density, the calculated eigenvalues show a statistical distribution which falls off approximately exponentially with increasing detuning from the zero-field splitting resulting in an asymmetric distribution around the peak maxima (\figref[b]{si_sim}). 

For such data, calculating simply the mean of the eigenvalues would skew the result towards large detuning and overestimate the corresponding peak splitting. Therefore, we rather quote the position of the peak maxima as the level splitting. To account for such an asymmetric broadening also in the evaluation of the experimental spectra, we use an exponentially modified Gaussian for fitting the ODMR data (see \figref[a]{odmr} in the main text). 
To fit the experimentally derived splitting $\nu_+ - \nu_-$ (in \figref[c]{odmr}) to the simulations we apply the procedure outlined in \figref[c]{si_sim}. Briefly, we linearly approximate the simulation results in the experimentally relevant range of charge densities. Then, we use this linear fit to match the experimentally measured splitting to a charge density. Lastly, we consider both the variance of the model results, i.e. the error from the linear fit, and the uncertainty of the experiment, i.e. the error bar from fitting the ODMR data, to calculate the error bar of the charge density by Gaussian error propagation.

\subsection{Molecular Dynamics (MD) simulations}\label{app:md_sim}
MD simulations are carried out using the LAMMPS package \cite{Plimpton1995} to extract the statistics of defect creation produced by the interaction of He ions with hBN. The interaction of the ion beam with the hBN structure is modeled via an extended Tersoff potential for boron nitride (BN-ExTeP) \cite{Los2017}, with a smooth transition to the Ziegler–Biersack–Littmark (ZBL) potential \cite{Ziegler1985-book} for small distances. The simulation spans a material thickness of \SI{720}{\angstrom} and is performed for a reference fluence, given by a number of 990 ions and a covered surface of $\sim$35 Å $\times$ 35 Å. The MD approach is compared to other Monte Carlo methods, which are the TRIDYN code \cite{Moller1984} and the SRIM software \cite{ZIEGLER20101818}.

\subsection{Comparison of ODMR splitting across different studies}\label{app:zfs}

In Table \ref{tab:hbn_comparison} we compare our ODMR results with previously reported studies that explicitly observed {\VB} defects and provided sufficient information about the hBN quality, the irradiation parameters and the ODMR results.% A more detailed discussion regarding this table is presented in section \ref{subsec:mdsim} in the main text.
\null
\vfill
\begin{table*}[htb]
    \centering
    \begin{tabular}{lllll}
        \toprule
        \midrule
        \textbf{Reference}                                          & \textbf{Splitting}    & \textbf{hBN source}        & \textbf{Irradiation} & \textbf{Fluence}                  \\
                                                                    & \textbf{[MHz]}        & \textbf{(thickness [nm])}  & \textbf{Method}      & \textbf{[× 10$^{15}$ cm$^{-2}$]}  \\
        \midrule
        This work                                                   & $145 - 165$           & HQgraphene (80)            & FIB He$^+$ (30 keV)  & 1 – 30                            \\
        Gottscholl, Nat. Mater. 2020 \cite{Gottscholl2020}          & $\sim 100$            & HQgraphene (80000)         & Therm. neutrons      & 2300                              \\
        Guo, ACS Omega 2022 \cite{Guo2022}                          & $\sim 120 - 160$      & HQgraphene (10 - 100)      & IB N$^+$ (30 keV)    & 0.01 - 1                          \\
                                                                    & $115$                 & HQgraphene (10 - 100)      & IB He$^+$ (30 keV)   & 0.1                               \\
        Ren, J. Lumin. 2023 \cite{Ren2023}                          & $\sim 120$            & HQgraphene (>1000)         & FIB He$^+$ (50 keV)  & 0.1                               \\
        Gong, Nat. Commun. 2023 \cite{Gong2023}                     & $\sim 120 - 160$      & Commerc. avail. hBN        & IB He$^+$ (3 keV)    & 0.03 - 1                          \\
        Gao, Nano Lett. 2021 \cite{Gao2021}                        & $\sim 100 - 120$       & 2D Semiconductors         & IB He$^+$ (0.3 - 2.5 keV)    & 0.05                      \\
        Zabelotsky, Appl. Nan. Mat. 2023 \cite{Zabelotsky2023}      & 140                   & NIMS (100/150)             & FIB N$^+$ (12 keV)   & 0.625                             \\
        Sasaki, Appl. Phys. Lett. 2023 \cite{Sasaki2023}            & $\sim 90 - 100$       & NIMS (66)                  & FIB He$^+$ (30 keV)  & 1                                 \\
        Suzuki, Appl. Phys. Expr. 2023 \cite{Suzuki2023}            & 150                   & NIMS (>100)                & FIB N$^+$ (40 keV)   & 1                                 \\
                                                                    & 90 (high temp. irr.)  & NIMS (>100)                & FIB N$^+$ (40 keV)   & 1                                 \\
        Liang, Adv. Opt. Mat. 2023 \cite{Liang2023}                 & $\sim 90 - 125$       & NIMS (>100)                & FIB He$^+$ (500 keV) & 1 – 100                           \\
        Udvarhelyi, npj Comput. Mater. 2023 \cite{Udvarhelyi2023}   & $\sim 75 - 120$       & Own prod. (bulk)           & Therm. neutrons      & 26 - 260                          \\
        Durand, Phys. Rev. Lett. 2023 \cite{Durand2023}             & $\sim 120$            & Own prod. (15)          & Therm. neutrons      & 26                                \\
                                                                     & $\sim 0$               & Own prod. (few layer)          & Therm. neutrons      & 26                                \\
        Haykal, Nat. Commun. 2022 \cite{Haykal2022}                 & 120                   & Own prod. (bulk)           & Therm. neutrons      & 26 - 260                          \\
        \bottomrule
    \end{tabular}
    \caption{Summary of works involving measurement of ODMR on {\VB} defects in hBN and their parameters. The ODMR splitting is quoted as values given directly in the cited sources or as values extracted from a plot cited sources (indicated with a $\sim$ symbol). In ref. \onlinecite{Suzuki2023}, ODMR values for ion irradiation under elevated temperatures can be found as well (denoted as high.temp. irr. in the table). The hBN source is mentioned with reference to the supplier, namely HQgraphene, NIMS (that is, high-pressure high-temperature synthesized high-quality hBN \cite{Taniguchi2007, Kubota2007}), and hBN from own production. For the irradiation methods, we quote focused ion beam (FIB), ion implanters or parallel ion beams (IB), and thermal neutrons (from a reactor source). The fluence is given in units of ions/cm$^{-2}$ throughout for consistency and provided as a range when available in the source publication.}
    \label{tab:hbn_comparison}
\end{table*}

\bibliography{02_bibliography}

\providecommand{\noopsort}[1]{}\providecommand{\singleletter}[1]{#1}%
\begin{thebibliography}{88}%
\makeatletter
\providecommand \@ifxundefined [1]{%
 \@ifx{#1\undefined}
}%
\providecommand \@ifnum [1]{%
 \ifnum #1\expandafter \@firstoftwo
 \else \expandafter \@secondoftwo
 \fi
}%
\providecommand \@ifx [1]{%
 \ifx #1\expandafter \@firstoftwo
 \else \expandafter \@secondoftwo
 \fi
}%
\providecommand \natexlab [1]{#1}%
\providecommand \enquote  [1]{``#1''}%
\providecommand \bibnamefont  [1]{#1}%
\providecommand \bibfnamefont [1]{#1}%
\providecommand \citenamefont [1]{#1}%
\providecommand \href@noop [0]{\@secondoftwo}%
\providecommand \href [0]{\begingroup \@sanitize@url \@href}%
\providecommand \@href[1]{\@@startlink{#1}\@@href}%
\providecommand \@@href[1]{\endgroup#1\@@endlink}%
\providecommand \@sanitize@url [0]{\catcode `\\12\catcode `\$12\catcode `\&12\catcode `\#12\catcode `\^12\catcode `\_12\catcode `\%12\relax}%
\providecommand \@@startlink[1]{}%
\providecommand \@@endlink[0]{}%
\providecommand \url  [0]{\begingroup\@sanitize@url \@url }%
\providecommand \@url [1]{\endgroup\@href {#1}{\urlprefix }}%
\providecommand \urlprefix  [0]{URL }%
\providecommand \Eprint [0]{\href }%
\providecommand \doibase [0]{https://doi.org/}%
\providecommand \selectlanguage [0]{\@gobble}%
\providecommand \bibinfo  [0]{\@secondoftwo}%
\providecommand \bibfield  [0]{\@secondoftwo}%
\providecommand \translation [1]{[#1]}%
\providecommand \BibitemOpen [0]{}%
\providecommand \bibitemStop [0]{}%
\providecommand \bibitemNoStop [0]{.\EOS\space}%
\providecommand \EOS [0]{\spacefactor3000\relax}%
\providecommand \BibitemShut  [1]{\csname bibitem#1\endcsname}%
\let\auto@bib@innerbib\@empty
%</preamble>
\bibitem [{\citenamefont {Tran}\ \emph {et~al.}(2016)\citenamefont {Tran}, \citenamefont {Bray}, \citenamefont {Ford}, \citenamefont {Toth},\ and\ \citenamefont {Aharonovich}}]{Tran2016}%
  \BibitemOpen
  \bibfield  {author} {\bibinfo {author} {\bibfnamefont {T.~T.}\ \bibnamefont {Tran}}, \bibinfo {author} {\bibfnamefont {K.}~\bibnamefont {Bray}}, \bibinfo {author} {\bibfnamefont {M.~J.}\ \bibnamefont {Ford}}, \bibinfo {author} {\bibfnamefont {M.}~\bibnamefont {Toth}},\ and\ \bibinfo {author} {\bibfnamefont {I.}~\bibnamefont {Aharonovich}},\ }\bibfield  {title} {\bibinfo {title} {Quantum emission from hexagonal boron nitride monolayers},\ }\href {https://doi.org/10.1038/nnano.2015.242} {\bibfield  {journal} {\bibinfo  {journal} {Nature Nanotechnology}\ }\textbf {\bibinfo {volume} {11}},\ \bibinfo {pages} {37} (\bibinfo {year} {2016})}\BibitemShut {NoStop}%
\bibitem [{\citenamefont {Fischer}\ \emph {et~al.}(2021)\citenamefont {Fischer}, \citenamefont {Caridad}, \citenamefont {Sajid}, \citenamefont {Ghaderzadeh}, \citenamefont {Ghorbani-Asl}, \citenamefont {Gammelgaard}, \citenamefont {B{\o}ggild}, \citenamefont {Thygesen}, \citenamefont {Krasheninnikov}, \citenamefont {Xiao}, \citenamefont {Wubs},\ and\ \citenamefont {Stenger}}]{FischSciAdv}%
  \BibitemOpen
  \bibfield  {author} {\bibinfo {author} {\bibfnamefont {M.}~\bibnamefont {Fischer}}, \bibinfo {author} {\bibfnamefont {J.~M.}\ \bibnamefont {Caridad}}, \bibinfo {author} {\bibfnamefont {A.}~\bibnamefont {Sajid}}, \bibinfo {author} {\bibfnamefont {S.}~\bibnamefont {Ghaderzadeh}}, \bibinfo {author} {\bibfnamefont {M.}~\bibnamefont {Ghorbani-Asl}}, \bibinfo {author} {\bibfnamefont {L.}~\bibnamefont {Gammelgaard}}, \bibinfo {author} {\bibfnamefont {P.}~\bibnamefont {B{\o}ggild}}, \bibinfo {author} {\bibfnamefont {K.~S.}\ \bibnamefont {Thygesen}}, \bibinfo {author} {\bibfnamefont {A.~V.}\ \bibnamefont {Krasheninnikov}}, \bibinfo {author} {\bibfnamefont {S.}~\bibnamefont {Xiao}}, \bibinfo {author} {\bibfnamefont {M.}~\bibnamefont {Wubs}},\ and\ \bibinfo {author} {\bibfnamefont {N.}~\bibnamefont {Stenger}},\ }\bibfield  {title} {\bibinfo {title} {{Controlled generation of luminescent centers in hexagonal boron nitride by irradiation engineering}},\ }\href {https://doi.org/10.1126/sciadv.abe7138} {\bibfield
  {journal} {\bibinfo  {journal} {Science Advances}\ }\textbf {\bibinfo {volume} {7}},\ \bibinfo {pages} {7138} (\bibinfo {year} {2021})}\BibitemShut {NoStop}%
\bibitem [{\citenamefont {Aharonovich}\ \emph {et~al.}(2016)\citenamefont {Aharonovich}, \citenamefont {Englund},\ and\ \citenamefont {Toth}}]{Aharonovich2016}%
  \BibitemOpen
  \bibfield  {author} {\bibinfo {author} {\bibfnamefont {I.}~\bibnamefont {Aharonovich}}, \bibinfo {author} {\bibfnamefont {D.}~\bibnamefont {Englund}},\ and\ \bibinfo {author} {\bibfnamefont {M.}~\bibnamefont {Toth}},\ }\bibfield  {title} {\bibinfo {title} {Solid-state single-photon emitters},\ }\href {https://doi.org/10.1038/nphoton.2016.186} {\bibfield  {journal} {\bibinfo  {journal} {Nature Photonics}\ }\textbf {\bibinfo {volume} {10}},\ \bibinfo {pages} {631} (\bibinfo {year} {2016})}\BibitemShut {NoStop}%
\bibitem [{\citenamefont {Gottscholl}\ \emph {et~al.}(2021)\citenamefont {Gottscholl}, \citenamefont {Diez}, \citenamefont {Soltamov}, \citenamefont {Kasper}, \citenamefont {Sperlich}, \citenamefont {Kianinia}, \citenamefont {Bradac}, \citenamefont {Aharonovich},\ and\ \citenamefont {Dyakonov}}]{Gottscholl2021}%
  \BibitemOpen
  \bibfield  {author} {\bibinfo {author} {\bibfnamefont {A.}~\bibnamefont {Gottscholl}}, \bibinfo {author} {\bibfnamefont {M.}~\bibnamefont {Diez}}, \bibinfo {author} {\bibfnamefont {V.}~\bibnamefont {Soltamov}}, \bibinfo {author} {\bibfnamefont {C.}~\bibnamefont {Kasper}}, \bibinfo {author} {\bibfnamefont {A.}~\bibnamefont {Sperlich}}, \bibinfo {author} {\bibfnamefont {M.}~\bibnamefont {Kianinia}}, \bibinfo {author} {\bibfnamefont {C.}~\bibnamefont {Bradac}}, \bibinfo {author} {\bibfnamefont {I.}~\bibnamefont {Aharonovich}},\ and\ \bibinfo {author} {\bibfnamefont {V.}~\bibnamefont {Dyakonov}},\ }\bibfield  {title} {\bibinfo {title} {{Room temperature coherent control of spin defects in hexagonal boron nitride}},\ }\href {https://doi.org/10.1126/sciadv.abf3630} {\bibfield  {journal} {\bibinfo  {journal} {Science Advances}\ }\textbf {\bibinfo {volume} {7}},\ \bibinfo {pages} {eabf3630} (\bibinfo {year} {2021})}\BibitemShut {NoStop}%
\bibitem [{\citenamefont {Montblanch}\ \emph {et~al.}(2023)\citenamefont {Montblanch}, \citenamefont {Barbone}, \citenamefont {Aharonovich}, \citenamefont {Atat{\"{u}}re},\ and\ \citenamefont {Ferrari}}]{Montblanch2023}%
  \BibitemOpen
  \bibfield  {author} {\bibinfo {author} {\bibfnamefont {A.~R.}\ \bibnamefont {Montblanch}}, \bibinfo {author} {\bibfnamefont {M.}~\bibnamefont {Barbone}}, \bibinfo {author} {\bibfnamefont {I.}~\bibnamefont {Aharonovich}}, \bibinfo {author} {\bibfnamefont {M.}~\bibnamefont {Atat{\"{u}}re}},\ and\ \bibinfo {author} {\bibfnamefont {A.~C.}\ \bibnamefont {Ferrari}},\ }\bibfield  {title} {\bibinfo {title} {{Layered materials as a platform for quantum technologies}},\ }\href {https://doi.org/10.1038/s41565-023-01354-x} {\bibfield  {journal} {\bibinfo  {journal} {Nature Nanotechnology}\ }\textbf {\bibinfo {volume} {18}},\ \bibinfo {pages} {555} (\bibinfo {year} {2023})}\BibitemShut {NoStop}%
\bibitem [{\citenamefont {Gottscholl}\ \emph {et~al.}(2020)\citenamefont {Gottscholl}, \citenamefont {Kianinia}, \citenamefont {Soltamov}, \citenamefont {Orlinskii}, \citenamefont {Mamin}, \citenamefont {Bradac}, \citenamefont {Kasper}, \citenamefont {Krambrock}, \citenamefont {Sperlich}, \citenamefont {Toth}, \citenamefont {Aharonovich},\ and\ \citenamefont {Dyakonov}}]{Gottscholl2020}%
  \BibitemOpen
  \bibfield  {author} {\bibinfo {author} {\bibfnamefont {A.}~\bibnamefont {Gottscholl}}, \bibinfo {author} {\bibfnamefont {M.}~\bibnamefont {Kianinia}}, \bibinfo {author} {\bibfnamefont {V.}~\bibnamefont {Soltamov}}, \bibinfo {author} {\bibfnamefont {S.}~\bibnamefont {Orlinskii}}, \bibinfo {author} {\bibfnamefont {G.}~\bibnamefont {Mamin}}, \bibinfo {author} {\bibfnamefont {C.}~\bibnamefont {Bradac}}, \bibinfo {author} {\bibfnamefont {C.}~\bibnamefont {Kasper}}, \bibinfo {author} {\bibfnamefont {K.}~\bibnamefont {Krambrock}}, \bibinfo {author} {\bibfnamefont {A.}~\bibnamefont {Sperlich}}, \bibinfo {author} {\bibfnamefont {M.}~\bibnamefont {Toth}}, \bibinfo {author} {\bibfnamefont {I.}~\bibnamefont {Aharonovich}},\ and\ \bibinfo {author} {\bibfnamefont {V.}~\bibnamefont {Dyakonov}},\ }\bibfield  {title} {\bibinfo {title} {Initialization and read-out of intrinsic spin defects in a van der waals crystal at room temperature},\ }\href {https://doi.org/10.1038/s41563-020-0619-6} {\bibfield  {journal} {\bibinfo
  {journal} {Nature Materials}\ }\textbf {\bibinfo {volume} {19}},\ \bibinfo {pages} {540} (\bibinfo {year} {2020})}\BibitemShut {NoStop}%
\bibitem [{\citenamefont {Kumar}\ \emph {et~al.}(2023)\citenamefont {Kumar}, \citenamefont {Cholsuk}, \citenamefont {Zand}, \citenamefont {Mishuk}, \citenamefont {Matthes}, \citenamefont {Eilenberger}, \citenamefont {Suwanna},\ and\ \citenamefont {Vogl}}]{KumarAPL2023}%
  \BibitemOpen
  \bibfield  {author} {\bibinfo {author} {\bibfnamefont {A.}~\bibnamefont {Kumar}}, \bibinfo {author} {\bibfnamefont {C.}~\bibnamefont {Cholsuk}}, \bibinfo {author} {\bibfnamefont {A.}~\bibnamefont {Zand}}, \bibinfo {author} {\bibfnamefont {M.~N.}\ \bibnamefont {Mishuk}}, \bibinfo {author} {\bibfnamefont {T.}~\bibnamefont {Matthes}}, \bibinfo {author} {\bibfnamefont {F.}~\bibnamefont {Eilenberger}}, \bibinfo {author} {\bibfnamefont {S.}~\bibnamefont {Suwanna}},\ and\ \bibinfo {author} {\bibfnamefont {T.}~\bibnamefont {Vogl}},\ }\bibfield  {title} {\bibinfo {title} {{Localized creation of yellow single photon emitting carbon complexes in hexagonal boron nitride}},\ }\href {https://doi.org/10.1063/5.0147560} {\bibfield  {journal} {\bibinfo  {journal} {APL Materials}\ }\textbf {\bibinfo {volume} {11}},\ \bibinfo {pages} {071108} (\bibinfo {year} {2023})}\BibitemShut {NoStop}%
\bibitem [{\citenamefont {Shevitski}\ \emph {et~al.}(2019)\citenamefont {Shevitski}, \citenamefont {Gilbert}, \citenamefont {Chen}, \citenamefont {Kastl}, \citenamefont {Barnard}, \citenamefont {Wong}, \citenamefont {Ogletree}, \citenamefont {Watanabe}, \citenamefont {Taniguchi}, \citenamefont {Zettl} \emph {et~al.}}]{shevitski2019blue}%
  \BibitemOpen
  \bibfield  {author} {\bibinfo {author} {\bibfnamefont {B.}~\bibnamefont {Shevitski}}, \bibinfo {author} {\bibfnamefont {S.~M.}\ \bibnamefont {Gilbert}}, \bibinfo {author} {\bibfnamefont {C.~T.}\ \bibnamefont {Chen}}, \bibinfo {author} {\bibfnamefont {C.}~\bibnamefont {Kastl}}, \bibinfo {author} {\bibfnamefont {E.~S.}\ \bibnamefont {Barnard}}, \bibinfo {author} {\bibfnamefont {E.}~\bibnamefont {Wong}}, \bibinfo {author} {\bibfnamefont {D.~F.}\ \bibnamefont {Ogletree}}, \bibinfo {author} {\bibfnamefont {K.}~\bibnamefont {Watanabe}}, \bibinfo {author} {\bibfnamefont {T.}~\bibnamefont {Taniguchi}}, \bibinfo {author} {\bibfnamefont {A.}~\bibnamefont {Zettl}}, \emph {et~al.},\ }\bibfield  {title} {\bibinfo {title} {Blue-light-emitting color centers in high-quality hexagonal boron nitride},\ }\href@noop {} {\bibfield  {journal} {\bibinfo  {journal} {Physical Review B}\ }\textbf {\bibinfo {volume} {100}},\ \bibinfo {pages} {155419} (\bibinfo {year} {2019})}\BibitemShut {NoStop}%
\bibitem [{\citenamefont {Fournier}\ \emph {et~al.}(2021)\citenamefont {Fournier}, \citenamefont {Plaud}, \citenamefont {Roux}, \citenamefont {Pierret}, \citenamefont {Rosticher}, \citenamefont {Watanabe}, \citenamefont {Taniguchi}, \citenamefont {Buil}, \citenamefont {Qu{\'e}lin}, \citenamefont {Barjon}, \citenamefont {Hermier},\ and\ \citenamefont {Delteil}}]{Fournier2021}%
  \BibitemOpen
  \bibfield  {author} {\bibinfo {author} {\bibfnamefont {C.}~\bibnamefont {Fournier}}, \bibinfo {author} {\bibfnamefont {A.}~\bibnamefont {Plaud}}, \bibinfo {author} {\bibfnamefont {S.}~\bibnamefont {Roux}}, \bibinfo {author} {\bibfnamefont {A.}~\bibnamefont {Pierret}}, \bibinfo {author} {\bibfnamefont {M.}~\bibnamefont {Rosticher}}, \bibinfo {author} {\bibfnamefont {K.}~\bibnamefont {Watanabe}}, \bibinfo {author} {\bibfnamefont {T.}~\bibnamefont {Taniguchi}}, \bibinfo {author} {\bibfnamefont {S.}~\bibnamefont {Buil}}, \bibinfo {author} {\bibfnamefont {X.}~\bibnamefont {Qu{\'e}lin}}, \bibinfo {author} {\bibfnamefont {J.}~\bibnamefont {Barjon}}, \bibinfo {author} {\bibfnamefont {J.-P.}\ \bibnamefont {Hermier}},\ and\ \bibinfo {author} {\bibfnamefont {A.}~\bibnamefont {Delteil}},\ }\bibfield  {title} {\bibinfo {title} {Position-controlled quantum emitters with reproducible emission wavelength in hexagonal boron nitride},\ }\href {https://doi.org/10.1038/s41467-021-24019-6} {\bibfield  {journal} {\bibinfo
  {journal} {Nature Communications}\ }\textbf {\bibinfo {volume} {12}},\ \bibinfo {pages} {3779} (\bibinfo {year} {2021})}\BibitemShut {NoStop}%
\bibitem [{\citenamefont {Gale}\ \emph {et~al.}(2022)\citenamefont {Gale}, \citenamefont {Li}, \citenamefont {Chen}, \citenamefont {Watanabe}, \citenamefont {Taniguchi}, \citenamefont {Aharonovich},\ and\ \citenamefont {Toth}}]{Gale2022}%
  \BibitemOpen
  \bibfield  {author} {\bibinfo {author} {\bibfnamefont {A.}~\bibnamefont {Gale}}, \bibinfo {author} {\bibfnamefont {C.}~\bibnamefont {Li}}, \bibinfo {author} {\bibfnamefont {Y.}~\bibnamefont {Chen}}, \bibinfo {author} {\bibfnamefont {K.}~\bibnamefont {Watanabe}}, \bibinfo {author} {\bibfnamefont {T.}~\bibnamefont {Taniguchi}}, \bibinfo {author} {\bibfnamefont {I.}~\bibnamefont {Aharonovich}},\ and\ \bibinfo {author} {\bibfnamefont {M.}~\bibnamefont {Toth}},\ }\bibfield  {title} {\bibinfo {title} {{Site-Specific Fabrication of Blue Quantum Emitters in Hexagonal Boron Nitride}},\ }\href {https://doi.org/10.1021/acsphotonics.2c00631} {\bibfield  {journal} {\bibinfo  {journal} {ACS Photonics}\ }\textbf {\bibinfo {volume} {9}},\ \bibinfo {pages} {2170} (\bibinfo {year} {2022})}\BibitemShut {NoStop}%
\bibitem [{\citenamefont {Horder}\ \emph {et~al.}(2022)\citenamefont {Horder}, \citenamefont {White}, \citenamefont {Gale}, \citenamefont {Li}, \citenamefont {Watanabe}, \citenamefont {Taniguchi}, \citenamefont {Kianinia}, \citenamefont {Aharonovich},\ and\ \citenamefont {Toth}}]{Horder2022}%
  \BibitemOpen
  \bibfield  {author} {\bibinfo {author} {\bibfnamefont {J.}~\bibnamefont {Horder}}, \bibinfo {author} {\bibfnamefont {S.~J.}\ \bibnamefont {White}}, \bibinfo {author} {\bibfnamefont {A.}~\bibnamefont {Gale}}, \bibinfo {author} {\bibfnamefont {C.}~\bibnamefont {Li}}, \bibinfo {author} {\bibfnamefont {K.}~\bibnamefont {Watanabe}}, \bibinfo {author} {\bibfnamefont {T.}~\bibnamefont {Taniguchi}}, \bibinfo {author} {\bibfnamefont {M.}~\bibnamefont {Kianinia}}, \bibinfo {author} {\bibfnamefont {I.}~\bibnamefont {Aharonovich}},\ and\ \bibinfo {author} {\bibfnamefont {M.}~\bibnamefont {Toth}},\ }\bibfield  {title} {\bibinfo {title} {{Coherence Properties of Electron-Beam-Activated Emitters in Hexagonal Boron Nitride Under Resonant Excitation}},\ }\href {https://doi.org/10.1103/PhysRevApplied.18.064021} {\bibfield  {journal} {\bibinfo  {journal} {Physical Review Applied}\ }\textbf {\bibinfo {volume} {18}},\ \bibinfo {pages} {064021} (\bibinfo {year} {2022})}\BibitemShut {NoStop}%
\bibitem [{\citenamefont {Chen}\ \emph {et~al.}(2023)\citenamefont {Chen}, \citenamefont {Gale}, \citenamefont {Yamamura}, \citenamefont {Horder}, \citenamefont {Condos}, \citenamefont {Watanabe}, \citenamefont {Taniguchi}, \citenamefont {Toth},\ and\ \citenamefont {Aharonovich}}]{Chen2023Jul}%
  \BibitemOpen
  \bibfield  {author} {\bibinfo {author} {\bibfnamefont {Y.}~\bibnamefont {Chen}}, \bibinfo {author} {\bibfnamefont {A.}~\bibnamefont {Gale}}, \bibinfo {author} {\bibfnamefont {K.}~\bibnamefont {Yamamura}}, \bibinfo {author} {\bibfnamefont {J.}~\bibnamefont {Horder}}, \bibinfo {author} {\bibfnamefont {A.}~\bibnamefont {Condos}}, \bibinfo {author} {\bibfnamefont {K.}~\bibnamefont {Watanabe}}, \bibinfo {author} {\bibfnamefont {T.}~\bibnamefont {Taniguchi}}, \bibinfo {author} {\bibfnamefont {M.}~\bibnamefont {Toth}},\ and\ \bibinfo {author} {\bibfnamefont {I.}~\bibnamefont {Aharonovich}},\ }\bibfield  {title} {\bibinfo {title} {{Annealing of blue quantum emitters in carbon-doped hexagonal boron nitride}},\ }\href {https://doi.org/10.1063/5.0155311} {\bibfield  {journal} {\bibinfo  {journal} {Applied Physics Letters}\ }\textbf {\bibinfo {volume} {123}},\ \bibinfo {pages} {041902} (\bibinfo {year} {2023})}\BibitemShut {NoStop}%
\bibitem [{\citenamefont {Bourrellier}\ \emph {et~al.}(2016)\citenamefont {Bourrellier}, \citenamefont {Meuret}, \citenamefont {Tararan}, \citenamefont {St{\'{e}}phan}, \citenamefont {Kociak}, \citenamefont {Tizei},\ and\ \citenamefont {Zobelli}}]{Bourrellier2016}%
  \BibitemOpen
  \bibfield  {author} {\bibinfo {author} {\bibfnamefont {R.}~\bibnamefont {Bourrellier}}, \bibinfo {author} {\bibfnamefont {S.}~\bibnamefont {Meuret}}, \bibinfo {author} {\bibfnamefont {A.}~\bibnamefont {Tararan}}, \bibinfo {author} {\bibfnamefont {O.}~\bibnamefont {St{\'{e}}phan}}, \bibinfo {author} {\bibfnamefont {M.}~\bibnamefont {Kociak}}, \bibinfo {author} {\bibfnamefont {L.~H.}\ \bibnamefont {Tizei}},\ and\ \bibinfo {author} {\bibfnamefont {A.}~\bibnamefont {Zobelli}},\ }\bibfield  {title} {\bibinfo {title} {{Bright UV single photon emission at point defects in h-BN}},\ }\href {https://doi.org/10.1021/acs.nanolett.6b01368} {\bibfield  {journal} {\bibinfo  {journal} {Nano Letters}\ }\textbf {\bibinfo {volume} {16}},\ \bibinfo {pages} {4317} (\bibinfo {year} {2016})}\BibitemShut {NoStop}%
\bibitem [{\citenamefont {Stern}\ \emph {et~al.}(2022)\citenamefont {Stern}, \citenamefont {Gu}, \citenamefont {Jarman}, \citenamefont {Eizagirre~Barker}, \citenamefont {Mendelson}, \citenamefont {Chugh}, \citenamefont {Schott}, \citenamefont {Tan}, \citenamefont {Sirringhaus}, \citenamefont {Aharonovich},\ and\ \citenamefont {Atat{\"u}re}}]{Stern2022}%
  \BibitemOpen
  \bibfield  {author} {\bibinfo {author} {\bibfnamefont {H.~L.}\ \bibnamefont {Stern}}, \bibinfo {author} {\bibfnamefont {Q.}~\bibnamefont {Gu}}, \bibinfo {author} {\bibfnamefont {J.}~\bibnamefont {Jarman}}, \bibinfo {author} {\bibfnamefont {S.}~\bibnamefont {Eizagirre~Barker}}, \bibinfo {author} {\bibfnamefont {N.}~\bibnamefont {Mendelson}}, \bibinfo {author} {\bibfnamefont {D.}~\bibnamefont {Chugh}}, \bibinfo {author} {\bibfnamefont {S.}~\bibnamefont {Schott}}, \bibinfo {author} {\bibfnamefont {H.~H.}\ \bibnamefont {Tan}}, \bibinfo {author} {\bibfnamefont {H.}~\bibnamefont {Sirringhaus}}, \bibinfo {author} {\bibfnamefont {I.}~\bibnamefont {Aharonovich}},\ and\ \bibinfo {author} {\bibfnamefont {M.}~\bibnamefont {Atat{\"u}re}},\ }\bibfield  {title} {\bibinfo {title} {Room-temperature optically detected magnetic resonance of single defects in hexagonal boron nitride},\ }\href {https://doi.org/10.1038/s41467-022-28169-z} {\bibfield  {journal} {\bibinfo  {journal} {Nature Communications}\ }\textbf {\bibinfo
  {volume} {13}},\ \bibinfo {pages} {618} (\bibinfo {year} {2022})}\BibitemShut {NoStop}%
\bibitem [{\citenamefont {Stern}\ \emph {et~al.}(2024)\citenamefont {Stern}, \citenamefont {{M. Gilardoni}}, \citenamefont {Gu}, \citenamefont {{Eizagirre Barker}}, \citenamefont {Powell}, \citenamefont {Deng}, \citenamefont {Fraser}, \citenamefont {Follet}, \citenamefont {Li}, \citenamefont {Ramsay}, \citenamefont {Tan}, \citenamefont {Aharonovich},\ and\ \citenamefont {Atat{\"{u}}re}}]{Stern2024}%
  \BibitemOpen
  \bibfield  {author} {\bibinfo {author} {\bibfnamefont {H.~L.}\ \bibnamefont {Stern}}, \bibinfo {author} {\bibfnamefont {C.}~\bibnamefont {{M. Gilardoni}}}, \bibinfo {author} {\bibfnamefont {Q.}~\bibnamefont {Gu}}, \bibinfo {author} {\bibfnamefont {S.}~\bibnamefont {{Eizagirre Barker}}}, \bibinfo {author} {\bibfnamefont {O.~F.~J.}\ \bibnamefont {Powell}}, \bibinfo {author} {\bibfnamefont {X.}~\bibnamefont {Deng}}, \bibinfo {author} {\bibfnamefont {S.~A.}\ \bibnamefont {Fraser}}, \bibinfo {author} {\bibfnamefont {L.}~\bibnamefont {Follet}}, \bibinfo {author} {\bibfnamefont {C.}~\bibnamefont {Li}}, \bibinfo {author} {\bibfnamefont {A.~J.}\ \bibnamefont {Ramsay}}, \bibinfo {author} {\bibfnamefont {H.~H.}\ \bibnamefont {Tan}}, \bibinfo {author} {\bibfnamefont {I.}~\bibnamefont {Aharonovich}},\ and\ \bibinfo {author} {\bibfnamefont {M.}~\bibnamefont {Atat{\"{u}}re}},\ }\bibfield  {title} {\bibinfo {title} {{A quantum coherent spin in hexagonal boron nitride at ambient conditions}},\ }\href
  {https://doi.org/10.1038/s41563-024-01887-z} {\bibfield  {journal} {\bibinfo  {journal} {Nature Materials}\ }\textbf {\bibinfo {volume} {23}},\ \bibinfo {pages} {1379} (\bibinfo {year} {2024})}\BibitemShut {NoStop}%
\bibitem [{\citenamefont {Gao}\ \emph {et~al.}(2024)\citenamefont {Gao}, \citenamefont {Vaidya}, \citenamefont {Dikshit}, \citenamefont {Ju}, \citenamefont {Shen}, \citenamefont {Jin}, \citenamefont {Zhang},\ and\ \citenamefont {Li}}]{Gao2023}%
  \BibitemOpen
  \bibfield  {author} {\bibinfo {author} {\bibfnamefont {X.}~\bibnamefont {Gao}}, \bibinfo {author} {\bibfnamefont {S.}~\bibnamefont {Vaidya}}, \bibinfo {author} {\bibfnamefont {S.}~\bibnamefont {Dikshit}}, \bibinfo {author} {\bibfnamefont {P.}~\bibnamefont {Ju}}, \bibinfo {author} {\bibfnamefont {K.}~\bibnamefont {Shen}}, \bibinfo {author} {\bibfnamefont {Y.}~\bibnamefont {Jin}}, \bibinfo {author} {\bibfnamefont {S.}~\bibnamefont {Zhang}},\ and\ \bibinfo {author} {\bibfnamefont {T.}~\bibnamefont {Li}},\ }\bibfield  {title} {\bibinfo {title} {{Nanotube spin defects for omnidirectional magnetic field sensing}},\ }\href {https://doi.org/10.1038/s41467-024-51941-2} {\bibfield  {journal} {\bibinfo  {journal} {Nature Communications}\ }\textbf {\bibinfo {volume} {15}},\ \bibinfo {pages} {7697} (\bibinfo {year} {2024})}\BibitemShut {NoStop}%
\bibitem [{\citenamefont {Hennessey}\ \emph {et~al.}(2024)\citenamefont {Hennessey}, \citenamefont {Whitefield}, \citenamefont {Gale}, \citenamefont {Kianinia}, \citenamefont {Scott}, \citenamefont {Aharonovich},\ and\ \citenamefont {Toth}}]{Hennessey2024}%
  \BibitemOpen
  \bibfield  {author} {\bibinfo {author} {\bibfnamefont {M.}~\bibnamefont {Hennessey}}, \bibinfo {author} {\bibfnamefont {B.}~\bibnamefont {Whitefield}}, \bibinfo {author} {\bibfnamefont {A.}~\bibnamefont {Gale}}, \bibinfo {author} {\bibfnamefont {M.}~\bibnamefont {Kianinia}}, \bibinfo {author} {\bibfnamefont {J.~A.}\ \bibnamefont {Scott}}, \bibinfo {author} {\bibfnamefont {I.}~\bibnamefont {Aharonovich}},\ and\ \bibinfo {author} {\bibfnamefont {M.}~\bibnamefont {Toth}},\ }\bibfield  {title} {\bibinfo {title} {{Framework for Engineering of Spin Defects in Hexagonal Boron Nitride by Focused Ion Beams}},\ }\href {https://doi.org/10.1002/qute.202300459} {\bibfield  {journal} {\bibinfo  {journal} {Advanced Quantum Technologies}\ }\textbf {\bibinfo {volume} {2300459}},\ \bibinfo {pages} {2300118} (\bibinfo {year} {2024})}\BibitemShut {NoStop}%
\bibitem [{\citenamefont {Healey}\ \emph {et~al.}(2023)\citenamefont {Healey}, \citenamefont {Scholten}, \citenamefont {Yang}, \citenamefont {Scott}, \citenamefont {Abrahams}, \citenamefont {Robertson}, \citenamefont {Hou}, \citenamefont {Guo}, \citenamefont {Rahman}, \citenamefont {Lu}, \citenamefont {Kianinia}, \citenamefont {Aharonovich},\ and\ \citenamefont {Tetienne}}]{Healey2023}%
  \BibitemOpen
  \bibfield  {author} {\bibinfo {author} {\bibfnamefont {A.~J.}\ \bibnamefont {Healey}}, \bibinfo {author} {\bibfnamefont {S.~C.}\ \bibnamefont {Scholten}}, \bibinfo {author} {\bibfnamefont {T.}~\bibnamefont {Yang}}, \bibinfo {author} {\bibfnamefont {J.~A.}\ \bibnamefont {Scott}}, \bibinfo {author} {\bibfnamefont {G.~J.}\ \bibnamefont {Abrahams}}, \bibinfo {author} {\bibfnamefont {I.~O.}\ \bibnamefont {Robertson}}, \bibinfo {author} {\bibfnamefont {X.~F.}\ \bibnamefont {Hou}}, \bibinfo {author} {\bibfnamefont {Y.~F.}\ \bibnamefont {Guo}}, \bibinfo {author} {\bibfnamefont {S.}~\bibnamefont {Rahman}}, \bibinfo {author} {\bibfnamefont {Y.}~\bibnamefont {Lu}}, \bibinfo {author} {\bibfnamefont {M.}~\bibnamefont {Kianinia}}, \bibinfo {author} {\bibfnamefont {I.}~\bibnamefont {Aharonovich}},\ and\ \bibinfo {author} {\bibfnamefont {J.~P.}\ \bibnamefont {Tetienne}},\ }\bibfield  {title} {\bibinfo {title} {{Quantum microscopy with van der Waals heterostructures}},\ }\href {https://doi.org/10.1038/s41567-022-01815-5}
  {\bibfield  {journal} {\bibinfo  {journal} {Nature Physics}\ }\textbf {\bibinfo {volume} {19}},\ \bibinfo {pages} {87} (\bibinfo {year} {2023})}\BibitemShut {NoStop}%
\bibitem [{\citenamefont {Gao}\ \emph {et~al.}(2021{\natexlab{a}})\citenamefont {Gao}, \citenamefont {Jiang}, \citenamefont {Llacsahuanga~Allcca}, \citenamefont {Shen}, \citenamefont {Sadi}, \citenamefont {Solanki}, \citenamefont {Ju}, \citenamefont {Xu}, \citenamefont {Upadhyaya}, \citenamefont {Chen}, \citenamefont {Bhave},\ and\ \citenamefont {Li}}]{Gao2021}%
  \BibitemOpen
  \bibfield  {author} {\bibinfo {author} {\bibfnamefont {X.}~\bibnamefont {Gao}}, \bibinfo {author} {\bibfnamefont {B.}~\bibnamefont {Jiang}}, \bibinfo {author} {\bibfnamefont {A.~E.}\ \bibnamefont {Llacsahuanga~Allcca}}, \bibinfo {author} {\bibfnamefont {K.}~\bibnamefont {Shen}}, \bibinfo {author} {\bibfnamefont {M.~A.}\ \bibnamefont {Sadi}}, \bibinfo {author} {\bibfnamefont {A.~B.}\ \bibnamefont {Solanki}}, \bibinfo {author} {\bibfnamefont {P.}~\bibnamefont {Ju}}, \bibinfo {author} {\bibfnamefont {Z.}~\bibnamefont {Xu}}, \bibinfo {author} {\bibfnamefont {P.}~\bibnamefont {Upadhyaya}}, \bibinfo {author} {\bibfnamefont {Y.~P.}\ \bibnamefont {Chen}}, \bibinfo {author} {\bibfnamefont {S.~A.}\ \bibnamefont {Bhave}},\ and\ \bibinfo {author} {\bibfnamefont {T.}~\bibnamefont {Li}},\ }\bibfield  {title} {\bibinfo {title} {High-contrast plasmonic-enhanced shallow spin defects in hexagonal boron nitride for quantum sensing},\ }\href {https://doi.org/10.1021/acs.nanolett.1c02495} {\bibfield  {journal} {\bibinfo
  {journal} {Nano Letters}\ }\textbf {\bibinfo {volume} {21}},\ \bibinfo {pages} {7708} (\bibinfo {year} {2021}{\natexlab{a}})}\BibitemShut {NoStop}%
\bibitem [{\citenamefont {Kianinia}\ \emph {et~al.}(2020)\citenamefont {Kianinia}, \citenamefont {White}, \citenamefont {Fröch}, \citenamefont {Bradac},\ and\ \citenamefont {Aharonovich}}]{Kianinia2020}%
  \BibitemOpen
  \bibfield  {author} {\bibinfo {author} {\bibfnamefont {M.}~\bibnamefont {Kianinia}}, \bibinfo {author} {\bibfnamefont {S.}~\bibnamefont {White}}, \bibinfo {author} {\bibfnamefont {J.~E.}\ \bibnamefont {Fröch}}, \bibinfo {author} {\bibfnamefont {C.}~\bibnamefont {Bradac}},\ and\ \bibinfo {author} {\bibfnamefont {I.}~\bibnamefont {Aharonovich}},\ }\bibfield  {title} {\bibinfo {title} {Generation of spin defects in hexagonal boron nitride},\ }\href {https://doi.org/10.1021/acsphotonics.0c00614} {\bibfield  {journal} {\bibinfo  {journal} {ACS Photonics}\ }\textbf {\bibinfo {volume} {7}},\ \bibinfo {pages} {2147} (\bibinfo {year} {2020})}\BibitemShut {NoStop}%
\bibitem [{\citenamefont {Guo}\ \emph {et~al.}(2022)\citenamefont {Guo}, \citenamefont {Liu}, \citenamefont {Li}, \citenamefont {Yang}, \citenamefont {Yu}, \citenamefont {Meng}, \citenamefont {Wang}, \citenamefont {Zeng}, \citenamefont {Yan}, \citenamefont {Li}, \citenamefont {Wang}, \citenamefont {Xu}, \citenamefont {Wang}, \citenamefont {Tang}, \citenamefont {Li},\ and\ \citenamefont {Guo}}]{Guo2022}%
  \BibitemOpen
  \bibfield  {author} {\bibinfo {author} {\bibfnamefont {N.~J.}\ \bibnamefont {Guo}}, \bibinfo {author} {\bibfnamefont {W.}~\bibnamefont {Liu}}, \bibinfo {author} {\bibfnamefont {Z.~P.}\ \bibnamefont {Li}}, \bibinfo {author} {\bibfnamefont {Y.~Z.}\ \bibnamefont {Yang}}, \bibinfo {author} {\bibfnamefont {S.}~\bibnamefont {Yu}}, \bibinfo {author} {\bibfnamefont {Y.}~\bibnamefont {Meng}}, \bibinfo {author} {\bibfnamefont {Z.~A.}\ \bibnamefont {Wang}}, \bibinfo {author} {\bibfnamefont {X.~D.}\ \bibnamefont {Zeng}}, \bibinfo {author} {\bibfnamefont {F.~F.}\ \bibnamefont {Yan}}, \bibinfo {author} {\bibfnamefont {Q.}~\bibnamefont {Li}}, \bibinfo {author} {\bibfnamefont {J.~F.}\ \bibnamefont {Wang}}, \bibinfo {author} {\bibfnamefont {J.~S.}\ \bibnamefont {Xu}}, \bibinfo {author} {\bibfnamefont {Y.~T.}\ \bibnamefont {Wang}}, \bibinfo {author} {\bibfnamefont {J.~S.}\ \bibnamefont {Tang}}, \bibinfo {author} {\bibfnamefont {C.~F.}\ \bibnamefont {Li}},\ and\ \bibinfo {author} {\bibfnamefont {G.~C.}\ \bibnamefont {Guo}},\
  }\bibfield  {title} {\bibinfo {title} {{Generation of Spin Defects by Ion Implantation in Hexagonal Boron Nitride}},\ }\href {https://doi.org/10.1021/acsomega.1c04564} {\bibfield  {journal} {\bibinfo  {journal} {ACS Omega}\ }\textbf {\bibinfo {volume} {7}},\ \bibinfo {pages} {1733} (\bibinfo {year} {2022})}\BibitemShut {NoStop}%
\bibitem [{\citenamefont {Glushkov}\ \emph {et~al.}(2022)\citenamefont {Glushkov}, \citenamefont {Macha}, \citenamefont {Räth}, \citenamefont {Navikas}, \citenamefont {Ronceray}, \citenamefont {Cheon}, \citenamefont {Ahmed}, \citenamefont {Avsar}, \citenamefont {Watanabe}, \citenamefont {Taniguchi}, \citenamefont {Shorubalko}, \citenamefont {Kis}, \citenamefont {Fantner},\ and\ \citenamefont {Radenovic}}]{Glushkov2022}%
  \BibitemOpen
  \bibfield  {author} {\bibinfo {author} {\bibfnamefont {E.}~\bibnamefont {Glushkov}}, \bibinfo {author} {\bibfnamefont {M.}~\bibnamefont {Macha}}, \bibinfo {author} {\bibfnamefont {E.}~\bibnamefont {Räth}}, \bibinfo {author} {\bibfnamefont {V.}~\bibnamefont {Navikas}}, \bibinfo {author} {\bibfnamefont {N.}~\bibnamefont {Ronceray}}, \bibinfo {author} {\bibfnamefont {C.~Y.}\ \bibnamefont {Cheon}}, \bibinfo {author} {\bibfnamefont {A.}~\bibnamefont {Ahmed}}, \bibinfo {author} {\bibfnamefont {A.}~\bibnamefont {Avsar}}, \bibinfo {author} {\bibfnamefont {K.}~\bibnamefont {Watanabe}}, \bibinfo {author} {\bibfnamefont {T.}~\bibnamefont {Taniguchi}}, \bibinfo {author} {\bibfnamefont {I.}~\bibnamefont {Shorubalko}}, \bibinfo {author} {\bibfnamefont {A.}~\bibnamefont {Kis}}, \bibinfo {author} {\bibfnamefont {G.}~\bibnamefont {Fantner}},\ and\ \bibinfo {author} {\bibfnamefont {A.}~\bibnamefont {Radenovic}},\ }\bibfield  {title} {\bibinfo {title} {Engineering optically active defects in hexagonal boron nitride using
  focused ion beam and water},\ }\href {https://doi.org/10.1021/acsnano.1c07086} {\bibfield  {journal} {\bibinfo  {journal} {ACS Nano}\ }\textbf {\bibinfo {volume} {16}},\ \bibinfo {pages} {3695} (\bibinfo {year} {2022})}\BibitemShut {NoStop}%
\bibitem [{\citenamefont {Ren}\ \emph {et~al.}(2023)\citenamefont {Ren}, \citenamefont {Wu},\ and\ \citenamefont {Xu}}]{Ren2023}%
  \BibitemOpen
  \bibfield  {author} {\bibinfo {author} {\bibfnamefont {F.}~\bibnamefont {Ren}}, \bibinfo {author} {\bibfnamefont {Y.}~\bibnamefont {Wu}},\ and\ \bibinfo {author} {\bibfnamefont {Z.}~\bibnamefont {Xu}},\ }\bibfield  {title} {\bibinfo {title} {{Creation and repair of luminescence defects in hexagonal boron nitride by irradiation and annealing for optical neutron detection}},\ }\href {https://doi.org/10.1016/j.jlumin.2023.119911} {\bibfield  {journal} {\bibinfo  {journal} {Journal of Luminescence}\ }\textbf {\bibinfo {volume} {261}},\ \bibinfo {pages} {119911} (\bibinfo {year} {2023})}\BibitemShut {NoStop}%
\bibitem [{\citenamefont {Liu}\ \emph {et~al.}(2023)\citenamefont {Liu}, \citenamefont {Wu}, \citenamefont {Jing}, \citenamefont {Cheng}, \citenamefont {Zhan}, \citenamefont {Bao}, \citenamefont {Yan}, \citenamefont {Xu}, \citenamefont {Zhang}, \citenamefont {Li}, \citenamefont {Liu}, \citenamefont {Liu},\ and\ \citenamefont {Shen}}]{Liu2023}%
  \BibitemOpen
  \bibfield  {author} {\bibinfo {author} {\bibfnamefont {G.~L.}\ \bibnamefont {Liu}}, \bibinfo {author} {\bibfnamefont {X.~Y.}\ \bibnamefont {Wu}}, \bibinfo {author} {\bibfnamefont {P.~T.}\ \bibnamefont {Jing}}, \bibinfo {author} {\bibfnamefont {Z.}~\bibnamefont {Cheng}}, \bibinfo {author} {\bibfnamefont {D.}~\bibnamefont {Zhan}}, \bibinfo {author} {\bibfnamefont {Y.}~\bibnamefont {Bao}}, \bibinfo {author} {\bibfnamefont {J.~X.}\ \bibnamefont {Yan}}, \bibinfo {author} {\bibfnamefont {H.}~\bibnamefont {Xu}}, \bibinfo {author} {\bibfnamefont {L.~G.}\ \bibnamefont {Zhang}}, \bibinfo {author} {\bibfnamefont {B.~H.}\ \bibnamefont {Li}}, \bibinfo {author} {\bibfnamefont {K.~W.}\ \bibnamefont {Liu}}, \bibinfo {author} {\bibfnamefont {L.}~\bibnamefont {Liu}},\ and\ \bibinfo {author} {\bibfnamefont {D.~Z.}\ \bibnamefont {Shen}},\ }\bibfield  {title} {\bibinfo {title} {{Single Photon Emitters in Hexagonal Boron Nitride Fabricated by Focused Helium Ion Beam}},\ }\href {https://doi.org/10.1002/adom.202302083} {\bibfield
  {journal} {\bibinfo  {journal} {Advanced Optical Materials}\ }\textbf {\bibinfo {volume} {2302083}},\ \bibinfo {pages} {1} (\bibinfo {year} {2023})}\BibitemShut {NoStop}%
\bibitem [{\citenamefont {Baber}\ \emph {et~al.}(2022)\citenamefont {Baber}, \citenamefont {Malein}, \citenamefont {Khatri}, \citenamefont {Keatley}, \citenamefont {Guo}, \citenamefont {Withers}, \citenamefont {Ramsay},\ and\ \citenamefont {Luxmoore}}]{Baber2022}%
  \BibitemOpen
  \bibfield  {author} {\bibinfo {author} {\bibfnamefont {S.}~\bibnamefont {Baber}}, \bibinfo {author} {\bibfnamefont {R.~N.~E.}\ \bibnamefont {Malein}}, \bibinfo {author} {\bibfnamefont {P.}~\bibnamefont {Khatri}}, \bibinfo {author} {\bibfnamefont {P.~S.}\ \bibnamefont {Keatley}}, \bibinfo {author} {\bibfnamefont {S.}~\bibnamefont {Guo}}, \bibinfo {author} {\bibfnamefont {F.}~\bibnamefont {Withers}}, \bibinfo {author} {\bibfnamefont {A.~J.}\ \bibnamefont {Ramsay}},\ and\ \bibinfo {author} {\bibfnamefont {I.~J.}\ \bibnamefont {Luxmoore}},\ }\bibfield  {title} {\bibinfo {title} {Excited state spectroscopy of boron vacancy defects in hexagonal boron nitride using time-resolved optically detected magnetic resonance},\ }\href {https://doi.org/10.1021/acs.nanolett.1c04366} {\bibfield  {journal} {\bibinfo  {journal} {Nano Letters}\ }\textbf {\bibinfo {volume} {22}},\ \bibinfo {pages} {461} (\bibinfo {year} {2022})}\BibitemShut {NoStop}%
\bibitem [{\citenamefont {Ramsay}\ \emph {et~al.}(2023)\citenamefont {Ramsay}, \citenamefont {Hekmati}, \citenamefont {Patrickson}, \citenamefont {Baber}, \citenamefont {Arvidsson-Shukur}, \citenamefont {Bennett},\ and\ \citenamefont {Luxmoore}}]{Ramsay2023}%
  \BibitemOpen
  \bibfield  {author} {\bibinfo {author} {\bibfnamefont {A.~J.}\ \bibnamefont {Ramsay}}, \bibinfo {author} {\bibfnamefont {R.}~\bibnamefont {Hekmati}}, \bibinfo {author} {\bibfnamefont {C.~J.}\ \bibnamefont {Patrickson}}, \bibinfo {author} {\bibfnamefont {S.}~\bibnamefont {Baber}}, \bibinfo {author} {\bibfnamefont {D.~R.~M.}\ \bibnamefont {Arvidsson-Shukur}}, \bibinfo {author} {\bibfnamefont {A.~J.}\ \bibnamefont {Bennett}},\ and\ \bibinfo {author} {\bibfnamefont {I.~J.}\ \bibnamefont {Luxmoore}},\ }\bibfield  {title} {\bibinfo {title} {{Coherence protection of spin qubits in hexagonal boron nitride}},\ }\href {https://doi.org/10.1038/s41467-023-36196-7} {\bibfield  {journal} {\bibinfo  {journal} {Nature Communications}\ }\textbf {\bibinfo {volume} {14}},\ \bibinfo {pages} {461} (\bibinfo {year} {2023})}\BibitemShut {NoStop}%
\bibitem [{\citenamefont {Toledo}\ \emph {et~al.}(2018)\citenamefont {Toledo}, \citenamefont {de~Jesus}, \citenamefont {Kianinia}, \citenamefont {Leal}, \citenamefont {Fantini}, \citenamefont {Cury}, \citenamefont {S{\'{a}}far}, \citenamefont {Aharonovich},\ and\ \citenamefont {Krambrock}}]{Toledo2018}%
  \BibitemOpen
  \bibfield  {author} {\bibinfo {author} {\bibfnamefont {J.~R.}\ \bibnamefont {Toledo}}, \bibinfo {author} {\bibfnamefont {D.~B.}\ \bibnamefont {de~Jesus}}, \bibinfo {author} {\bibfnamefont {M.}~\bibnamefont {Kianinia}}, \bibinfo {author} {\bibfnamefont {A.~S.}\ \bibnamefont {Leal}}, \bibinfo {author} {\bibfnamefont {C.}~\bibnamefont {Fantini}}, \bibinfo {author} {\bibfnamefont {L.~A.}\ \bibnamefont {Cury}}, \bibinfo {author} {\bibfnamefont {G.~A.~M.}\ \bibnamefont {S{\'{a}}far}}, \bibinfo {author} {\bibfnamefont {I.}~\bibnamefont {Aharonovich}},\ and\ \bibinfo {author} {\bibfnamefont {K.}~\bibnamefont {Krambrock}},\ }\bibfield  {title} {\bibinfo {title} {{Electron paramagnetic resonance signature of point defects in neutron-irradiated hexagonal boron nitride}},\ }\href {https://doi.org/10.1103/PhysRevB.98.155203} {\bibfield  {journal} {\bibinfo  {journal} {Physical Review B}\ }\textbf {\bibinfo {volume} {98}},\ \bibinfo {pages} {155203} (\bibinfo {year} {2018})}\BibitemShut {NoStop}%
\bibitem [{\citenamefont {Liang}\ \emph {et~al.}(2023)\citenamefont {Liang}, \citenamefont {Chen}, \citenamefont {Yang}, \citenamefont {Watanabe}, \citenamefont {Taniguchi}, \citenamefont {Eda},\ and\ \citenamefont {Bettiol}}]{Liang2023}%
  \BibitemOpen
  \bibfield  {author} {\bibinfo {author} {\bibfnamefont {H.}~\bibnamefont {Liang}}, \bibinfo {author} {\bibfnamefont {Y.}~\bibnamefont {Chen}}, \bibinfo {author} {\bibfnamefont {C.}~\bibnamefont {Yang}}, \bibinfo {author} {\bibfnamefont {K.}~\bibnamefont {Watanabe}}, \bibinfo {author} {\bibfnamefont {T.}~\bibnamefont {Taniguchi}}, \bibinfo {author} {\bibfnamefont {G.}~\bibnamefont {Eda}},\ and\ \bibinfo {author} {\bibfnamefont {A.~A.}\ \bibnamefont {Bettiol}},\ }\bibfield  {title} {\bibinfo {title} {{High Sensitivity Spin Defects in hBN Created by High‐Energy He Beam Irradiation}},\ }\href {https://doi.org/10.1002/adom.202201941} {\bibfield  {journal} {\bibinfo  {journal} {Advanced Optical Materials}\ }\textbf {\bibinfo {volume} {11}},\ \bibinfo {pages} {2201941} (\bibinfo {year} {2023})}\BibitemShut {NoStop}%
\bibitem [{\citenamefont {Whitefield}\ \emph {et~al.}(2023)\citenamefont {Whitefield}, \citenamefont {Toth}, \citenamefont {Aharonovich}, \citenamefont {Tetienne},\ and\ \citenamefont {Kianinia}}]{Whitefield2023}%
  \BibitemOpen
  \bibfield  {author} {\bibinfo {author} {\bibfnamefont {B.}~\bibnamefont {Whitefield}}, \bibinfo {author} {\bibfnamefont {M.}~\bibnamefont {Toth}}, \bibinfo {author} {\bibfnamefont {I.}~\bibnamefont {Aharonovich}}, \bibinfo {author} {\bibfnamefont {J.~P.}\ \bibnamefont {Tetienne}},\ and\ \bibinfo {author} {\bibfnamefont {M.}~\bibnamefont {Kianinia}},\ }\bibfield  {title} {\bibinfo {title} {{Magnetic Field Sensitivity Optimization of Negatively Charged Boron Vacancy Defects in hBN.}},\ }\href {https://doi.org/10.1002/qute.202300118} {\bibfield  {journal} {\bibinfo  {journal} {Advanced Quantum Technologies}\ }\textbf {\bibinfo {volume} {2300118}},\ \bibinfo {pages} {1} (\bibinfo {year} {2023})}\BibitemShut {NoStop}%
\bibitem [{\citenamefont {Clua-Provost}\ \emph {et~al.}(2024)\citenamefont {Clua-Provost}, \citenamefont {Mu}, \citenamefont {Durand}, \citenamefont {Schrader}, \citenamefont {Happacher}, \citenamefont {Bocquel}, \citenamefont {Maletinsky}, \citenamefont {Frauni{\'{e}}}, \citenamefont {Marie}, \citenamefont {Robert}, \citenamefont {Seine}, \citenamefont {Janzen}, \citenamefont {Edgar}, \citenamefont {Gil}, \citenamefont {Cassabois},\ and\ \citenamefont {Jacques}}]{Clua-Provost2024}%
  \BibitemOpen
  \bibfield  {author} {\bibinfo {author} {\bibfnamefont {T.}~\bibnamefont {Clua-Provost}}, \bibinfo {author} {\bibfnamefont {Z.}~\bibnamefont {Mu}}, \bibinfo {author} {\bibfnamefont {A.}~\bibnamefont {Durand}}, \bibinfo {author} {\bibfnamefont {C.}~\bibnamefont {Schrader}}, \bibinfo {author} {\bibfnamefont {J.}~\bibnamefont {Happacher}}, \bibinfo {author} {\bibfnamefont {J.}~\bibnamefont {Bocquel}}, \bibinfo {author} {\bibfnamefont {P.}~\bibnamefont {Maletinsky}}, \bibinfo {author} {\bibfnamefont {J.}~\bibnamefont {Frauni{\'{e}}}}, \bibinfo {author} {\bibfnamefont {X.}~\bibnamefont {Marie}}, \bibinfo {author} {\bibfnamefont {C.}~\bibnamefont {Robert}}, \bibinfo {author} {\bibfnamefont {G.}~\bibnamefont {Seine}}, \bibinfo {author} {\bibfnamefont {E.}~\bibnamefont {Janzen}}, \bibinfo {author} {\bibfnamefont {J.~H.}\ \bibnamefont {Edgar}}, \bibinfo {author} {\bibfnamefont {B.}~\bibnamefont {Gil}}, \bibinfo {author} {\bibfnamefont {G.}~\bibnamefont {Cassabois}},\ and\ \bibinfo {author} {\bibfnamefont
  {V.}~\bibnamefont {Jacques}},\ }\bibfield  {title} {\bibinfo {title} {{Spin-dependent photodynamics of boron-vacancy centers in hexagonal boron nitride}},\ }\href {https://doi.org/10.1103/PhysRevB.110.014104} {\bibfield  {journal} {\bibinfo  {journal} {Physical Review B}\ }\textbf {\bibinfo {volume} {110}},\ \bibinfo {pages} {014104} (\bibinfo {year} {2024})}\BibitemShut {NoStop}%
\bibitem [{\citenamefont {Mu}\ \emph {et~al.}(2022)\citenamefont {Mu}, \citenamefont {Cai}, \citenamefont {Chen}, \citenamefont {Kenny}, \citenamefont {Jiang}, \citenamefont {Ru}, \citenamefont {Lyu}, \citenamefont {Koh}, \citenamefont {Liu}, \citenamefont {Aharonovich},\ and\ \citenamefont {Gao}}]{Mu2022}%
  \BibitemOpen
  \bibfield  {author} {\bibinfo {author} {\bibfnamefont {Z.}~\bibnamefont {Mu}}, \bibinfo {author} {\bibfnamefont {H.}~\bibnamefont {Cai}}, \bibinfo {author} {\bibfnamefont {D.}~\bibnamefont {Chen}}, \bibinfo {author} {\bibfnamefont {J.}~\bibnamefont {Kenny}}, \bibinfo {author} {\bibfnamefont {Z.}~\bibnamefont {Jiang}}, \bibinfo {author} {\bibfnamefont {S.}~\bibnamefont {Ru}}, \bibinfo {author} {\bibfnamefont {X.}~\bibnamefont {Lyu}}, \bibinfo {author} {\bibfnamefont {T.~S.}\ \bibnamefont {Koh}}, \bibinfo {author} {\bibfnamefont {X.}~\bibnamefont {Liu}}, \bibinfo {author} {\bibfnamefont {I.}~\bibnamefont {Aharonovich}},\ and\ \bibinfo {author} {\bibfnamefont {W.}~\bibnamefont {Gao}},\ }\bibfield  {title} {\bibinfo {title} {{Excited-State Optically Detected Magnetic Resonance of Spin Defects in Hexagonal Boron Nitride}},\ }\href {https://doi.org/10.1103/PhysRevLett.128.216402} {\bibfield  {journal} {\bibinfo  {journal} {Physical Review Letters}\ }\textbf {\bibinfo {volume} {128}},\ \bibinfo {pages} {216402}
  (\bibinfo {year} {2022})}\BibitemShut {NoStop}%
\bibitem [{\citenamefont {Mathur}\ \emph {et~al.}(2022)\citenamefont {Mathur}, \citenamefont {Mukherjee}, \citenamefont {Gao}, \citenamefont {Luo}, \citenamefont {McCullian}, \citenamefont {Li}, \citenamefont {Vamivakas},\ and\ \citenamefont {Fuchs}}]{Mathur2022}%
  \BibitemOpen
  \bibfield  {author} {\bibinfo {author} {\bibfnamefont {N.}~\bibnamefont {Mathur}}, \bibinfo {author} {\bibfnamefont {A.}~\bibnamefont {Mukherjee}}, \bibinfo {author} {\bibfnamefont {X.}~\bibnamefont {Gao}}, \bibinfo {author} {\bibfnamefont {J.}~\bibnamefont {Luo}}, \bibinfo {author} {\bibfnamefont {B.~A.}\ \bibnamefont {McCullian}}, \bibinfo {author} {\bibfnamefont {T.}~\bibnamefont {Li}}, \bibinfo {author} {\bibfnamefont {A.~N.}\ \bibnamefont {Vamivakas}},\ and\ \bibinfo {author} {\bibfnamefont {G.~D.}\ \bibnamefont {Fuchs}},\ }\bibfield  {title} {\bibinfo {title} {{Excited-state spin-resonance spectroscopy of {\VB} defect centers in hexagonal boron nitride}},\ }\href {https://doi.org/10.1038/s41467-022-30772-z} {\bibfield  {journal} {\bibinfo  {journal} {Nature Communications}\ }\textbf {\bibinfo {volume} {13}},\ \bibinfo {pages} {3233} (\bibinfo {year} {2022})}\BibitemShut {NoStop}%
\bibitem [{\citenamefont {Yu}\ \emph {et~al.}(2022)\citenamefont {Yu}, \citenamefont {Sun}, \citenamefont {Wang}, \citenamefont {Zhang}, \citenamefont {Ye}, \citenamefont {Zhou}, \citenamefont {Liu}, \citenamefont {Wang}, \citenamefont {Shi}, \citenamefont {Wang},\ and\ \citenamefont {Du}}]{Yu2022}%
  \BibitemOpen
  \bibfield  {author} {\bibinfo {author} {\bibfnamefont {P.}~\bibnamefont {Yu}}, \bibinfo {author} {\bibfnamefont {H.}~\bibnamefont {Sun}}, \bibinfo {author} {\bibfnamefont {M.}~\bibnamefont {Wang}}, \bibinfo {author} {\bibfnamefont {T.}~\bibnamefont {Zhang}}, \bibinfo {author} {\bibfnamefont {X.}~\bibnamefont {Ye}}, \bibinfo {author} {\bibfnamefont {J.}~\bibnamefont {Zhou}}, \bibinfo {author} {\bibfnamefont {H.}~\bibnamefont {Liu}}, \bibinfo {author} {\bibfnamefont {C.~J.}\ \bibnamefont {Wang}}, \bibinfo {author} {\bibfnamefont {F.}~\bibnamefont {Shi}}, \bibinfo {author} {\bibfnamefont {Y.}~\bibnamefont {Wang}},\ and\ \bibinfo {author} {\bibfnamefont {J.}~\bibnamefont {Du}},\ }\bibfield  {title} {\bibinfo {title} {{Excited-State Spectroscopy of Spin Defects in Hexagonal Boron Nitride}},\ }\href {https://doi.org/10.1021/acs.nanolett.1c04841} {\bibfield  {journal} {\bibinfo  {journal} {Nano Letters}\ }\textbf {\bibinfo {volume} {22}},\ \bibinfo {pages} {3545} (\bibinfo {year} {2022})}\BibitemShut {NoStop}%
\bibitem [{\citenamefont {Mendelson}\ \emph {et~al.}(2022)\citenamefont {Mendelson}, \citenamefont {Ritika}, \citenamefont {Kianinia}, \citenamefont {Scott}, \citenamefont {Kim}, \citenamefont {Fröch}, \citenamefont {Gazzana}, \citenamefont {Westerhausen}, \citenamefont {Xiao}, \citenamefont {Mohajerani}, \citenamefont {Strauf}, \citenamefont {Toth}, \citenamefont {Aharonovich},\ and\ \citenamefont {Xu}}]{Mendelson2022}%
  \BibitemOpen
  \bibfield  {author} {\bibinfo {author} {\bibfnamefont {N.}~\bibnamefont {Mendelson}}, \bibinfo {author} {\bibfnamefont {R.}~\bibnamefont {Ritika}}, \bibinfo {author} {\bibfnamefont {M.}~\bibnamefont {Kianinia}}, \bibinfo {author} {\bibfnamefont {J.}~\bibnamefont {Scott}}, \bibinfo {author} {\bibfnamefont {S.}~\bibnamefont {Kim}}, \bibinfo {author} {\bibfnamefont {J.~E.}\ \bibnamefont {Fröch}}, \bibinfo {author} {\bibfnamefont {C.}~\bibnamefont {Gazzana}}, \bibinfo {author} {\bibfnamefont {M.}~\bibnamefont {Westerhausen}}, \bibinfo {author} {\bibfnamefont {L.}~\bibnamefont {Xiao}}, \bibinfo {author} {\bibfnamefont {S.~S.}\ \bibnamefont {Mohajerani}}, \bibinfo {author} {\bibfnamefont {S.}~\bibnamefont {Strauf}}, \bibinfo {author} {\bibfnamefont {M.}~\bibnamefont {Toth}}, \bibinfo {author} {\bibfnamefont {I.}~\bibnamefont {Aharonovich}},\ and\ \bibinfo {author} {\bibfnamefont {Z.-Q.}\ \bibnamefont {Xu}},\ }\bibfield  {title} {\bibinfo {title} {Coupling spin defects in a layered material to nanoscale plasmonic
  cavities},\ }\href {https://doi.org/https://doi.org/10.1002/adma.202106046} {\bibfield  {journal} {\bibinfo  {journal} {Advanced Materials}\ }\textbf {\bibinfo {volume} {34}},\ \bibinfo {pages} {2106046} (\bibinfo {year} {2022})}\BibitemShut {NoStop}%
\bibitem [{\citenamefont {Gao}\ \emph {et~al.}(2022)\citenamefont {Gao}, \citenamefont {Vaidya}, \citenamefont {Li}, \citenamefont {Ju}, \citenamefont {Jiang}, \citenamefont {Xu}, \citenamefont {Allcca}, \citenamefont {Shen}, \citenamefont {Taniguchi}, \citenamefont {Watanabe}, \citenamefont {Bhave}, \citenamefont {Chen}, \citenamefont {Ping},\ and\ \citenamefont {Li}}]{Gao2022}%
  \BibitemOpen
  \bibfield  {author} {\bibinfo {author} {\bibfnamefont {X.}~\bibnamefont {Gao}}, \bibinfo {author} {\bibfnamefont {S.}~\bibnamefont {Vaidya}}, \bibinfo {author} {\bibfnamefont {K.}~\bibnamefont {Li}}, \bibinfo {author} {\bibfnamefont {P.}~\bibnamefont {Ju}}, \bibinfo {author} {\bibfnamefont {B.}~\bibnamefont {Jiang}}, \bibinfo {author} {\bibfnamefont {Z.}~\bibnamefont {Xu}}, \bibinfo {author} {\bibfnamefont {A.~E.~L.}\ \bibnamefont {Allcca}}, \bibinfo {author} {\bibfnamefont {K.}~\bibnamefont {Shen}}, \bibinfo {author} {\bibfnamefont {T.}~\bibnamefont {Taniguchi}}, \bibinfo {author} {\bibfnamefont {K.}~\bibnamefont {Watanabe}}, \bibinfo {author} {\bibfnamefont {S.~A.}\ \bibnamefont {Bhave}}, \bibinfo {author} {\bibfnamefont {Y.~P.}\ \bibnamefont {Chen}}, \bibinfo {author} {\bibfnamefont {Y.}~\bibnamefont {Ping}},\ and\ \bibinfo {author} {\bibfnamefont {T.}~\bibnamefont {Li}},\ }\bibfield  {title} {\bibinfo {title} {{Nuclear spin polarization and control in hexagonal boron nitride}},\ }\href
  {https://doi.org/10.1038/s41563-022-01329-8} {\bibfield  {journal} {\bibinfo  {journal} {Nature Materials}\ }\textbf {\bibinfo {volume} {21}},\ \bibinfo {pages} {1024} (\bibinfo {year} {2022})}\BibitemShut {NoStop}%
\bibitem [{\citenamefont {Vaidya}\ \emph {et~al.}(2023)\citenamefont {Vaidya}, \citenamefont {Gao}, \citenamefont {Dikshit},\ and\ \citenamefont {Tongcang}}]{vaidya2023}%
  \BibitemOpen
  \bibfield  {author} {\bibinfo {author} {\bibfnamefont {S.}~\bibnamefont {Vaidya}}, \bibinfo {author} {\bibfnamefont {X.}~\bibnamefont {Gao}}, \bibinfo {author} {\bibfnamefont {I.}~\bibnamefont {Dikshit}, \bibfnamefont {Saakshi~Aharonovich}},\ and\ \bibinfo {author} {\bibfnamefont {L.}~\bibnamefont {Tongcang}},\ }\bibfield  {title} {\bibinfo {title} {Quantum sensing and imaging with spin defects in hexagonal boron nitride},\ }\href {https://doi.org/10.1080/23746149.2023.2206049} {\bibfield  {journal} {\bibinfo  {journal} {Advances in Physics X}\ }\textbf {\bibinfo {volume} {8}},\ \bibinfo {pages} {2206049} (\bibinfo {year} {2023})}\BibitemShut {NoStop}%
\bibitem [{\citenamefont {Huang}\ \emph {et~al.}(2022)\citenamefont {Huang}, \citenamefont {Zhou}, \citenamefont {Chen}, \citenamefont {Lu}, \citenamefont {McLaughlin}, \citenamefont {Li}, \citenamefont {Alghamdi}, \citenamefont {Djugba}, \citenamefont {Shi}, \citenamefont {Wang},\ and\ \citenamefont {Du}}]{huang2022}%
  \BibitemOpen
  \bibfield  {author} {\bibinfo {author} {\bibfnamefont {M.}~\bibnamefont {Huang}}, \bibinfo {author} {\bibfnamefont {J.}~\bibnamefont {Zhou}}, \bibinfo {author} {\bibfnamefont {D.}~\bibnamefont {Chen}}, \bibinfo {author} {\bibfnamefont {H.}~\bibnamefont {Lu}}, \bibinfo {author} {\bibfnamefont {N.~J.}\ \bibnamefont {McLaughlin}}, \bibinfo {author} {\bibfnamefont {S.}~\bibnamefont {Li}}, \bibinfo {author} {\bibfnamefont {M.}~\bibnamefont {Alghamdi}}, \bibinfo {author} {\bibfnamefont {D.}~\bibnamefont {Djugba}}, \bibinfo {author} {\bibfnamefont {J.}~\bibnamefont {Shi}}, \bibinfo {author} {\bibfnamefont {H.}~\bibnamefont {Wang}},\ and\ \bibinfo {author} {\bibfnamefont {C.~R.}\ \bibnamefont {Du}},\ }\bibfield  {title} {\bibinfo {title} {{Wide field imaging of van der Waals ferromagnet {F}e$_3${G}e{T}e$_2$ by spin defects in hexagonal boron nitride}},\ }\href {https://doi.org/10.1038/s41467-022-33016-2} {\bibfield  {journal} {\bibinfo  {journal} {Nature Communications}\ }\textbf {\bibinfo {volume} {13}},\ \bibinfo
  {pages} {5369} (\bibinfo {year} {2022})}\BibitemShut {NoStop}%
\bibitem [{\citenamefont {Kumar}\ \emph {et~al.}(2022)\citenamefont {Kumar}, \citenamefont {Fabre}, \citenamefont {Durand}, \citenamefont {Clua-Provost}, \citenamefont {Li}, \citenamefont {Edgar}, \citenamefont {Rougemaille}, \citenamefont {Coraux}, \citenamefont {Marie}, \citenamefont {Renucci}, \citenamefont {Robert}, \citenamefont {Robert-Philip}, \citenamefont {Gil}, \citenamefont {Cassabois}, \citenamefont {Finco},\ and\ \citenamefont {Jacques}}]{Kumar2022}%
  \BibitemOpen
  \bibfield  {author} {\bibinfo {author} {\bibfnamefont {P.}~\bibnamefont {Kumar}}, \bibinfo {author} {\bibfnamefont {F.}~\bibnamefont {Fabre}}, \bibinfo {author} {\bibfnamefont {A.}~\bibnamefont {Durand}}, \bibinfo {author} {\bibfnamefont {T.}~\bibnamefont {Clua-Provost}}, \bibinfo {author} {\bibfnamefont {J.}~\bibnamefont {Li}}, \bibinfo {author} {\bibfnamefont {J.}~\bibnamefont {Edgar}}, \bibinfo {author} {\bibfnamefont {N.}~\bibnamefont {Rougemaille}}, \bibinfo {author} {\bibfnamefont {J.}~\bibnamefont {Coraux}}, \bibinfo {author} {\bibfnamefont {X.}~\bibnamefont {Marie}}, \bibinfo {author} {\bibfnamefont {P.}~\bibnamefont {Renucci}}, \bibinfo {author} {\bibfnamefont {C.}~\bibnamefont {Robert}}, \bibinfo {author} {\bibfnamefont {I.}~\bibnamefont {Robert-Philip}}, \bibinfo {author} {\bibfnamefont {B.}~\bibnamefont {Gil}}, \bibinfo {author} {\bibfnamefont {G.}~\bibnamefont {Cassabois}}, \bibinfo {author} {\bibfnamefont {A.}~\bibnamefont {Finco}},\ and\ \bibinfo {author} {\bibfnamefont {V.}~\bibnamefont
  {Jacques}},\ }\bibfield  {title} {\bibinfo {title} {{Magnetic Imaging with Spin Defects in Hexagonal Boron Nitride}},\ }\href {https://doi.org/10.1103/PhysRevApplied.18.L061002} {\bibfield  {journal} {\bibinfo  {journal} {Physical Review Applied}\ }\textbf {\bibinfo {volume} {18}},\ \bibinfo {pages} {L061002} (\bibinfo {year} {2022})}\BibitemShut {NoStop}%
\bibitem [{\citenamefont {Maze}\ \emph {et~al.}(2008)\citenamefont {Maze}, \citenamefont {Stanwix}, \citenamefont {Hodges}, \citenamefont {Hong}, \citenamefont {Taylor}, \citenamefont {Cappellaro}, \citenamefont {Jiang}, \citenamefont {Dutt}, \citenamefont {Togan}, \citenamefont {Zibrov}, \citenamefont {Yacoby}, \citenamefont {Walsworth},\ and\ \citenamefont {Lukin}}]{Maze2008}%
  \BibitemOpen
  \bibfield  {author} {\bibinfo {author} {\bibfnamefont {J.~R.}\ \bibnamefont {Maze}}, \bibinfo {author} {\bibfnamefont {P.~L.}\ \bibnamefont {Stanwix}}, \bibinfo {author} {\bibfnamefont {J.~S.}\ \bibnamefont {Hodges}}, \bibinfo {author} {\bibfnamefont {S.}~\bibnamefont {Hong}}, \bibinfo {author} {\bibfnamefont {J.~M.}\ \bibnamefont {Taylor}}, \bibinfo {author} {\bibfnamefont {P.}~\bibnamefont {Cappellaro}}, \bibinfo {author} {\bibfnamefont {L.}~\bibnamefont {Jiang}}, \bibinfo {author} {\bibfnamefont {M.~V.}\ \bibnamefont {Dutt}}, \bibinfo {author} {\bibfnamefont {E.}~\bibnamefont {Togan}}, \bibinfo {author} {\bibfnamefont {A.~S.}\ \bibnamefont {Zibrov}}, \bibinfo {author} {\bibfnamefont {A.}~\bibnamefont {Yacoby}}, \bibinfo {author} {\bibfnamefont {R.~L.}\ \bibnamefont {Walsworth}},\ and\ \bibinfo {author} {\bibfnamefont {M.~D.}\ \bibnamefont {Lukin}},\ }\bibfield  {title} {\bibinfo {title} {{Nanoscale magnetic sensing with an individual electronic spin in diamond}},\ }\href
  {https://doi.org/10.1038/nature07279} {\bibfield  {journal} {\bibinfo  {journal} {Nature}\ }\textbf {\bibinfo {volume} {455}},\ \bibinfo {pages} {644} (\bibinfo {year} {2008})}\BibitemShut {NoStop}%
\bibitem [{\citenamefont {Balasubramanian}\ \emph {et~al.}(2009)\citenamefont {Balasubramanian}, \citenamefont {Neumann}, \citenamefont {Twitchen}, \citenamefont {Markham}, \citenamefont {Kolesov}, \citenamefont {Mizuochi}, \citenamefont {Isoya}, \citenamefont {Achard}, \citenamefont {Beck}, \citenamefont {Tissler}, \citenamefont {Jacques}, \citenamefont {Hemmer}, \citenamefont {Jelezko},\ and\ \citenamefont {Wrachtrup}}]{Balasubramanian2009}%
  \BibitemOpen
  \bibfield  {author} {\bibinfo {author} {\bibfnamefont {G.}~\bibnamefont {Balasubramanian}}, \bibinfo {author} {\bibfnamefont {P.}~\bibnamefont {Neumann}}, \bibinfo {author} {\bibfnamefont {D.}~\bibnamefont {Twitchen}}, \bibinfo {author} {\bibfnamefont {M.}~\bibnamefont {Markham}}, \bibinfo {author} {\bibfnamefont {R.}~\bibnamefont {Kolesov}}, \bibinfo {author} {\bibfnamefont {N.}~\bibnamefont {Mizuochi}}, \bibinfo {author} {\bibfnamefont {J.}~\bibnamefont {Isoya}}, \bibinfo {author} {\bibfnamefont {J.}~\bibnamefont {Achard}}, \bibinfo {author} {\bibfnamefont {J.}~\bibnamefont {Beck}}, \bibinfo {author} {\bibfnamefont {J.}~\bibnamefont {Tissler}}, \bibinfo {author} {\bibfnamefont {V.}~\bibnamefont {Jacques}}, \bibinfo {author} {\bibfnamefont {P.~R.}\ \bibnamefont {Hemmer}}, \bibinfo {author} {\bibfnamefont {F.}~\bibnamefont {Jelezko}},\ and\ \bibinfo {author} {\bibfnamefont {J.}~\bibnamefont {Wrachtrup}},\ }\bibfield  {title} {\bibinfo {title} {{Ultralong spin coherence time in isotopically engineered
  diamond}},\ }\href {https://doi.org/10.1038/nmat2420} {\bibfield  {journal} {\bibinfo  {journal} {Nature Materials}\ }\textbf {\bibinfo {volume} {8}},\ \bibinfo {pages} {383} (\bibinfo {year} {2009})}\BibitemShut {NoStop}%
\bibitem [{\citenamefont {Sarkar}\ \emph {et~al.}(2024)\citenamefont {Sarkar}, \citenamefont {Xu}, \citenamefont {Mathew}, \citenamefont {Lal}, \citenamefont {Chung}, \citenamefont {Lee}, \citenamefont {Watanabe}, \citenamefont {Taniguchi}, \citenamefont {Venkatesan},\ and\ \citenamefont {Grade{\v{c}}ak}}]{Sarkar2024}%
  \BibitemOpen
  \bibfield  {author} {\bibinfo {author} {\bibfnamefont {S.}~\bibnamefont {Sarkar}}, \bibinfo {author} {\bibfnamefont {Y.}~\bibnamefont {Xu}}, \bibinfo {author} {\bibfnamefont {S.}~\bibnamefont {Mathew}}, \bibinfo {author} {\bibfnamefont {M.}~\bibnamefont {Lal}}, \bibinfo {author} {\bibfnamefont {J.-Y.}\ \bibnamefont {Chung}}, \bibinfo {author} {\bibfnamefont {H.~Y.}\ \bibnamefont {Lee}}, \bibinfo {author} {\bibfnamefont {K.}~\bibnamefont {Watanabe}}, \bibinfo {author} {\bibfnamefont {T.}~\bibnamefont {Taniguchi}}, \bibinfo {author} {\bibfnamefont {T.}~\bibnamefont {Venkatesan}},\ and\ \bibinfo {author} {\bibfnamefont {S.}~\bibnamefont {Grade{\v{c}}ak}},\ }\bibfield  {title} {\bibinfo {title} {{Identifying Luminescent Boron Vacancies in h-BN Generated Using Controlled He$^+$ Ion Irradiation}},\ }\href {https://doi.org/10.1021/acs.nanolett.3c03113} {\bibfield  {journal} {\bibinfo  {journal} {Nano Letters}\ }\textbf {\bibinfo {volume} {24}},\ \bibinfo {pages} {43} (\bibinfo {year} {2024})}\BibitemShut {NoStop}%
\bibitem [{\citenamefont {Murzakhanov}\ \emph {et~al.}(2021)\citenamefont {Murzakhanov}, \citenamefont {Yavkin}, \citenamefont {Mamin}, \citenamefont {Orlinskii}, \citenamefont {Mumdzhi}, \citenamefont {Gracheva}, \citenamefont {Gabbasov}, \citenamefont {Smirnov}, \citenamefont {Davydov},\ and\ \citenamefont {Soltamov}}]{Murzakhanov2021}%
  \BibitemOpen
  \bibfield  {author} {\bibinfo {author} {\bibfnamefont {F.}~\bibnamefont {Murzakhanov}}, \bibinfo {author} {\bibfnamefont {B.}~\bibnamefont {Yavkin}}, \bibinfo {author} {\bibfnamefont {G.}~\bibnamefont {Mamin}}, \bibinfo {author} {\bibfnamefont {S.}~\bibnamefont {Orlinskii}}, \bibinfo {author} {\bibfnamefont {I.}~\bibnamefont {Mumdzhi}}, \bibinfo {author} {\bibfnamefont {I.}~\bibnamefont {Gracheva}}, \bibinfo {author} {\bibfnamefont {B.}~\bibnamefont {Gabbasov}}, \bibinfo {author} {\bibfnamefont {A.}~\bibnamefont {Smirnov}}, \bibinfo {author} {\bibfnamefont {V.}~\bibnamefont {Davydov}},\ and\ \bibinfo {author} {\bibfnamefont {V.}~\bibnamefont {Soltamov}},\ }\bibfield  {title} {\bibinfo {title} {{Creation of Negatively Charged Boron Vacancies in Hexagonal Boron Nitride Crystal by Electron Irradiation and Mechanism of Inhomogeneous Broadening of Boron Vacancy-Related Spin Resonance Lines}},\ }\href {https://doi.org/10.3390/nano11061373} {\bibfield  {journal} {\bibinfo  {journal} {Nanomaterials}\ }\textbf
  {\bibinfo {volume} {11}},\ \bibinfo {pages} {1373} (\bibinfo {year} {2021})}\BibitemShut {NoStop}%
\bibitem [{\citenamefont {Rizzato}\ \emph {et~al.}(2023)\citenamefont {Rizzato}, \citenamefont {Schalk}, \citenamefont {Mohr}, \citenamefont {Hermann}, \citenamefont {Leibold}, \citenamefont {Bruckmaier}, \citenamefont {Salvitti}, \citenamefont {Qian}, \citenamefont {Ji}, \citenamefont {Astakhov}, \citenamefont {Kentsch}, \citenamefont {Helm}, \citenamefont {Stier}, \citenamefont {Finley},\ and\ \citenamefont {Bucher}}]{Rizzato2023}%
  \BibitemOpen
  \bibfield  {author} {\bibinfo {author} {\bibfnamefont {R.}~\bibnamefont {Rizzato}}, \bibinfo {author} {\bibfnamefont {M.}~\bibnamefont {Schalk}}, \bibinfo {author} {\bibfnamefont {S.}~\bibnamefont {Mohr}}, \bibinfo {author} {\bibfnamefont {J.~C.}\ \bibnamefont {Hermann}}, \bibinfo {author} {\bibfnamefont {J.~P.}\ \bibnamefont {Leibold}}, \bibinfo {author} {\bibfnamefont {F.}~\bibnamefont {Bruckmaier}}, \bibinfo {author} {\bibfnamefont {G.}~\bibnamefont {Salvitti}}, \bibinfo {author} {\bibfnamefont {C.}~\bibnamefont {Qian}}, \bibinfo {author} {\bibfnamefont {P.}~\bibnamefont {Ji}}, \bibinfo {author} {\bibfnamefont {G.~V.}\ \bibnamefont {Astakhov}}, \bibinfo {author} {\bibfnamefont {U.}~\bibnamefont {Kentsch}}, \bibinfo {author} {\bibfnamefont {M.}~\bibnamefont {Helm}}, \bibinfo {author} {\bibfnamefont {A.~V.}\ \bibnamefont {Stier}}, \bibinfo {author} {\bibfnamefont {J.~J.}\ \bibnamefont {Finley}},\ and\ \bibinfo {author} {\bibfnamefont {D.~B.}\ \bibnamefont {Bucher}},\ }\bibfield  {title} {\bibinfo {title}
  {{Extending the coherence of spin defects in hBN enables advanced qubit control and quantum sensing}},\ }\href {https://doi.org/10.1038/s41467-023-40473-w} {\bibfield  {journal} {\bibinfo  {journal} {Nature Communications}\ }\textbf {\bibinfo {volume} {14}},\ \bibinfo {pages} {5089} (\bibinfo {year} {2023})}\BibitemShut {NoStop}%
\bibitem [{\citenamefont {Gong}\ \emph {et~al.}(2023)\citenamefont {Gong}, \citenamefont {He}, \citenamefont {Gao}, \citenamefont {Ju}, \citenamefont {Liu}, \citenamefont {Ye}, \citenamefont {Henriksen}, \citenamefont {Li},\ and\ \citenamefont {Zu}}]{Gong2023}%
  \BibitemOpen
  \bibfield  {author} {\bibinfo {author} {\bibfnamefont {R.}~\bibnamefont {Gong}}, \bibinfo {author} {\bibfnamefont {G.}~\bibnamefont {He}}, \bibinfo {author} {\bibfnamefont {X.}~\bibnamefont {Gao}}, \bibinfo {author} {\bibfnamefont {P.}~\bibnamefont {Ju}}, \bibinfo {author} {\bibfnamefont {Z.}~\bibnamefont {Liu}}, \bibinfo {author} {\bibfnamefont {B.}~\bibnamefont {Ye}}, \bibinfo {author} {\bibfnamefont {E.~A.}\ \bibnamefont {Henriksen}}, \bibinfo {author} {\bibfnamefont {T.}~\bibnamefont {Li}},\ and\ \bibinfo {author} {\bibfnamefont {C.}~\bibnamefont {Zu}},\ }\bibfield  {title} {\bibinfo {title} {{Coherent dynamics of strongly interacting electronic spin defects in hexagonal boron nitride}},\ }\href {https://doi.org/10.1038/s41467-023-39115-y} {\bibfield  {journal} {\bibinfo  {journal} {Nature Communications}\ }\textbf {\bibinfo {volume} {14}},\ \bibinfo {pages} {3299} (\bibinfo {year} {2023})}\BibitemShut {NoStop}%
\bibitem [{\citenamefont {Liu}\ \emph {et~al.}(2022)\citenamefont {Liu}, \citenamefont {Iv{\'{a}}dy}, \citenamefont {Li}, \citenamefont {Yang}, \citenamefont {Yu}, \citenamefont {Meng}, \citenamefont {Wang}, \citenamefont {Guo}, \citenamefont {Yan}, \citenamefont {Li}, \citenamefont {Wang}, \citenamefont {Xu}, \citenamefont {Liu}, \citenamefont {Zhou}, \citenamefont {Dong}, \citenamefont {Chen}, \citenamefont {Sun}, \citenamefont {Wang}, \citenamefont {Tang}, \citenamefont {Gali}, \citenamefont {Li},\ and\ \citenamefont {Guo}}]{Liu2022}%
  \BibitemOpen
  \bibfield  {author} {\bibinfo {author} {\bibfnamefont {W.}~\bibnamefont {Liu}}, \bibinfo {author} {\bibfnamefont {V.}~\bibnamefont {Iv{\'{a}}dy}}, \bibinfo {author} {\bibfnamefont {Z.-P.}\ \bibnamefont {Li}}, \bibinfo {author} {\bibfnamefont {Y.-Z.}\ \bibnamefont {Yang}}, \bibinfo {author} {\bibfnamefont {S.}~\bibnamefont {Yu}}, \bibinfo {author} {\bibfnamefont {Y.}~\bibnamefont {Meng}}, \bibinfo {author} {\bibfnamefont {Z.-A.}\ \bibnamefont {Wang}}, \bibinfo {author} {\bibfnamefont {N.-J.}\ \bibnamefont {Guo}}, \bibinfo {author} {\bibfnamefont {F.-F.}\ \bibnamefont {Yan}}, \bibinfo {author} {\bibfnamefont {Q.}~\bibnamefont {Li}}, \bibinfo {author} {\bibfnamefont {J.-F.}\ \bibnamefont {Wang}}, \bibinfo {author} {\bibfnamefont {J.-S.}\ \bibnamefont {Xu}}, \bibinfo {author} {\bibfnamefont {X.}~\bibnamefont {Liu}}, \bibinfo {author} {\bibfnamefont {Z.-Q.}\ \bibnamefont {Zhou}}, \bibinfo {author} {\bibfnamefont {Y.}~\bibnamefont {Dong}}, \bibinfo {author} {\bibfnamefont {X.-D.}\ \bibnamefont {Chen}}, \bibinfo
  {author} {\bibfnamefont {F.-W.}\ \bibnamefont {Sun}}, \bibinfo {author} {\bibfnamefont {Y.-T.}\ \bibnamefont {Wang}}, \bibinfo {author} {\bibfnamefont {J.-S.}\ \bibnamefont {Tang}}, \bibinfo {author} {\bibfnamefont {A.}~\bibnamefont {Gali}}, \bibinfo {author} {\bibfnamefont {C.-F.}\ \bibnamefont {Li}},\ and\ \bibinfo {author} {\bibfnamefont {G.-C.}\ \bibnamefont {Guo}},\ }\bibfield  {title} {\bibinfo {title} {{Coherent dynamics of multi-spin {\VB} center in hexagonal boron nitride}},\ }\href {https://doi.org/10.1038/s41467-022-33399-2} {\bibfield  {journal} {\bibinfo  {journal} {Nature Communications}\ }\textbf {\bibinfo {volume} {13}},\ \bibinfo {pages} {5713} (\bibinfo {year} {2022})}\BibitemShut {NoStop}%
\bibitem [{\citenamefont {Grosso}\ \emph {et~al.}(2017)\citenamefont {Grosso}, \citenamefont {Moon}, \citenamefont {Lienhard}, \citenamefont {Ali}, \citenamefont {Efetov}, \citenamefont {Furchi}, \citenamefont {Jarillo-Herrero}, \citenamefont {Ford}, \citenamefont {Aharonovich},\ and\ \citenamefont {Englund}}]{Grosso2017}%
  \BibitemOpen
  \bibfield  {author} {\bibinfo {author} {\bibfnamefont {G.}~\bibnamefont {Grosso}}, \bibinfo {author} {\bibfnamefont {H.}~\bibnamefont {Moon}}, \bibinfo {author} {\bibfnamefont {B.}~\bibnamefont {Lienhard}}, \bibinfo {author} {\bibfnamefont {S.}~\bibnamefont {Ali}}, \bibinfo {author} {\bibfnamefont {D.~K.}\ \bibnamefont {Efetov}}, \bibinfo {author} {\bibfnamefont {M.~M.}\ \bibnamefont {Furchi}}, \bibinfo {author} {\bibfnamefont {P.}~\bibnamefont {Jarillo-Herrero}}, \bibinfo {author} {\bibfnamefont {M.~J.}\ \bibnamefont {Ford}}, \bibinfo {author} {\bibfnamefont {I.}~\bibnamefont {Aharonovich}},\ and\ \bibinfo {author} {\bibfnamefont {D.}~\bibnamefont {Englund}},\ }\bibfield  {title} {\bibinfo {title} {Tunable and high-purity room temperature single-photon emission from atomic defects in hexagonal boron nitride},\ }\href {https://doi.org/10.1038/s41467-017-00810-2} {\bibfield  {journal} {\bibinfo  {journal} {Nature Communications}\ }\textbf {\bibinfo {volume} {8}},\ \bibinfo {pages} {705} (\bibinfo {year}
  {2017})}\BibitemShut {NoStop}%
\bibitem [{\citenamefont {Fr{\"{o}}ch}\ \emph {et~al.}(2021)\citenamefont {Fr{\"{o}}ch}, \citenamefont {Spencer}, \citenamefont {Kianinia}, \citenamefont {Totonjian}, \citenamefont {Nguyen}, \citenamefont {Gottscholl}, \citenamefont {Dyakonov}, \citenamefont {Toth}, \citenamefont {Kim},\ and\ \citenamefont {Aharonovich}}]{Froch2021}%
  \BibitemOpen
  \bibfield  {author} {\bibinfo {author} {\bibfnamefont {J.~E.}\ \bibnamefont {Fr{\"{o}}ch}}, \bibinfo {author} {\bibfnamefont {L.~P.}\ \bibnamefont {Spencer}}, \bibinfo {author} {\bibfnamefont {M.}~\bibnamefont {Kianinia}}, \bibinfo {author} {\bibfnamefont {D.~D.}\ \bibnamefont {Totonjian}}, \bibinfo {author} {\bibfnamefont {M.}~\bibnamefont {Nguyen}}, \bibinfo {author} {\bibfnamefont {A.}~\bibnamefont {Gottscholl}}, \bibinfo {author} {\bibfnamefont {V.}~\bibnamefont {Dyakonov}}, \bibinfo {author} {\bibfnamefont {M.}~\bibnamefont {Toth}}, \bibinfo {author} {\bibfnamefont {S.}~\bibnamefont {Kim}},\ and\ \bibinfo {author} {\bibfnamefont {I.}~\bibnamefont {Aharonovich}},\ }\bibfield  {title} {\bibinfo {title} {{Coupling Spin Defects in Hexagonal Boron Nitride to Monolithic Bullseye Cavities}},\ }\href {https://doi.org/10.1021/acs.nanolett.1c01843} {\bibfield  {journal} {\bibinfo  {journal} {Nano Letters}\ }\textbf {\bibinfo {volume} {21}},\ \bibinfo {pages} {6549} (\bibinfo {year} {2021})}\BibitemShut {NoStop}%
\bibitem [{\citenamefont {Qian}\ \emph {et~al.}(2022)\citenamefont {Qian}, \citenamefont {Villafañe}, \citenamefont {Schalk}, \citenamefont {Astakhov}, \citenamefont {Kentsch}, \citenamefont {Helm}, \citenamefont {Soubelet}, \citenamefont {Wilson}, \citenamefont {Rizzato}, \citenamefont {Mohr}, \citenamefont {Holleitner}, \citenamefont {Bucher}, \citenamefont {Stier},\ and\ \citenamefont {Finley}}]{Finley2022}%
  \BibitemOpen
  \bibfield  {author} {\bibinfo {author} {\bibfnamefont {C.}~\bibnamefont {Qian}}, \bibinfo {author} {\bibfnamefont {V.}~\bibnamefont {Villafañe}}, \bibinfo {author} {\bibfnamefont {M.}~\bibnamefont {Schalk}}, \bibinfo {author} {\bibfnamefont {G.~V.}\ \bibnamefont {Astakhov}}, \bibinfo {author} {\bibfnamefont {U.}~\bibnamefont {Kentsch}}, \bibinfo {author} {\bibfnamefont {M.}~\bibnamefont {Helm}}, \bibinfo {author} {\bibfnamefont {P.}~\bibnamefont {Soubelet}}, \bibinfo {author} {\bibfnamefont {N.~P.}\ \bibnamefont {Wilson}}, \bibinfo {author} {\bibfnamefont {R.}~\bibnamefont {Rizzato}}, \bibinfo {author} {\bibfnamefont {S.}~\bibnamefont {Mohr}}, \bibinfo {author} {\bibfnamefont {A.~W.}\ \bibnamefont {Holleitner}}, \bibinfo {author} {\bibfnamefont {D.~B.}\ \bibnamefont {Bucher}}, \bibinfo {author} {\bibfnamefont {A.~V.}\ \bibnamefont {Stier}},\ and\ \bibinfo {author} {\bibfnamefont {J.~J.}\ \bibnamefont {Finley}},\ }\bibfield  {title} {\bibinfo {title} {Unveiling the zero-phonon line of the boron vacancy
  center by cavity-enhanced emission},\ }\href {https://doi.org/10.1021/acs.nanolett.2c00739} {\bibfield  {journal} {\bibinfo  {journal} {Nano Letters}\ }\textbf {\bibinfo {volume} {22}},\ \bibinfo {pages} {5137} (\bibinfo {year} {2022})}\BibitemShut {NoStop}%
\bibitem [{\citenamefont {Sortino}\ \emph {et~al.}(2024)\citenamefont {Sortino}, \citenamefont {Gale}, \citenamefont {K{\"{u}}hner}, \citenamefont {Li}, \citenamefont {Biechteler}, \citenamefont {Wendisch}, \citenamefont {Kianinia}, \citenamefont {Ren}, \citenamefont {Toth}, \citenamefont {Maier}, \citenamefont {Aharonovich},\ and\ \citenamefont {Tittl}}]{Sortino2024}%
  \BibitemOpen
  \bibfield  {author} {\bibinfo {author} {\bibfnamefont {L.}~\bibnamefont {Sortino}}, \bibinfo {author} {\bibfnamefont {A.}~\bibnamefont {Gale}}, \bibinfo {author} {\bibfnamefont {L.}~\bibnamefont {K{\"{u}}hner}}, \bibinfo {author} {\bibfnamefont {C.}~\bibnamefont {Li}}, \bibinfo {author} {\bibfnamefont {J.}~\bibnamefont {Biechteler}}, \bibinfo {author} {\bibfnamefont {F.~J.}\ \bibnamefont {Wendisch}}, \bibinfo {author} {\bibfnamefont {M.}~\bibnamefont {Kianinia}}, \bibinfo {author} {\bibfnamefont {H.}~\bibnamefont {Ren}}, \bibinfo {author} {\bibfnamefont {M.}~\bibnamefont {Toth}}, \bibinfo {author} {\bibfnamefont {S.~A.}\ \bibnamefont {Maier}}, \bibinfo {author} {\bibfnamefont {I.}~\bibnamefont {Aharonovich}},\ and\ \bibinfo {author} {\bibfnamefont {A.}~\bibnamefont {Tittl}},\ }\bibfield  {title} {\bibinfo {title} {{Optically addressable spin defects coupled to bound states in the continuum metasurfaces}},\ }\href {https://doi.org/10.1038/s41467-024-46272-1} {\bibfield  {journal} {\bibinfo  {journal} {Nature
  Communications}\ }\textbf {\bibinfo {volume} {15}},\ \bibinfo {pages} {2008} (\bibinfo {year} {2024})}\BibitemShut {NoStop}%
\bibitem [{\citenamefont {Sasaki}\ \emph {et~al.}(2023)\citenamefont {Sasaki}, \citenamefont {Nakamura}, \citenamefont {Gu}, \citenamefont {Tsukamoto}, \citenamefont {Nakaharai}, \citenamefont {Iwasaki}, \citenamefont {Watanabe}, \citenamefont {Taniguchi}, \citenamefont {Ogawa}, \citenamefont {Morita},\ and\ \citenamefont {Kobayashi}}]{Sasaki2023}%
  \BibitemOpen
  \bibfield  {author} {\bibinfo {author} {\bibfnamefont {K.}~\bibnamefont {Sasaki}}, \bibinfo {author} {\bibfnamefont {Y.}~\bibnamefont {Nakamura}}, \bibinfo {author} {\bibfnamefont {H.}~\bibnamefont {Gu}}, \bibinfo {author} {\bibfnamefont {M.}~\bibnamefont {Tsukamoto}}, \bibinfo {author} {\bibfnamefont {S.}~\bibnamefont {Nakaharai}}, \bibinfo {author} {\bibfnamefont {T.}~\bibnamefont {Iwasaki}}, \bibinfo {author} {\bibfnamefont {K.}~\bibnamefont {Watanabe}}, \bibinfo {author} {\bibfnamefont {T.}~\bibnamefont {Taniguchi}}, \bibinfo {author} {\bibfnamefont {S.}~\bibnamefont {Ogawa}}, \bibinfo {author} {\bibfnamefont {Y.}~\bibnamefont {Morita}},\ and\ \bibinfo {author} {\bibfnamefont {K.}~\bibnamefont {Kobayashi}},\ }\bibfield  {title} {\bibinfo {title} {{Magnetic field imaging by hBN quantum sensor nanoarray}},\ }\href {https://doi.org/10.1063/5.0147072} {\bibfield  {journal} {\bibinfo  {journal} {Applied Physics Letters}\ }\textbf {\bibinfo {volume} {122}} (\bibinfo {year} {2023})}\BibitemShut {NoStop}%
\bibitem [{\citenamefont {Zabelotsky}\ \emph {et~al.}(2023)\citenamefont {Zabelotsky}, \citenamefont {Singh}, \citenamefont {Haim}, \citenamefont {Malkinson}, \citenamefont {Kadkhodazadeh}, \citenamefont {Radko}, \citenamefont {Aharonovich}, \citenamefont {Steinberg}, \citenamefont {Berg-S{\o}rensen}, \citenamefont {Huck}, \citenamefont {Taniguchi}, \citenamefont {Watanabe},\ and\ \citenamefont {Bar-Gill}}]{Zabelotsky2023}%
  \BibitemOpen
  \bibfield  {author} {\bibinfo {author} {\bibfnamefont {T.}~\bibnamefont {Zabelotsky}}, \bibinfo {author} {\bibfnamefont {S.}~\bibnamefont {Singh}}, \bibinfo {author} {\bibfnamefont {G.}~\bibnamefont {Haim}}, \bibinfo {author} {\bibfnamefont {R.}~\bibnamefont {Malkinson}}, \bibinfo {author} {\bibfnamefont {S.}~\bibnamefont {Kadkhodazadeh}}, \bibinfo {author} {\bibfnamefont {I.~P.}\ \bibnamefont {Radko}}, \bibinfo {author} {\bibfnamefont {I.}~\bibnamefont {Aharonovich}}, \bibinfo {author} {\bibfnamefont {H.}~\bibnamefont {Steinberg}}, \bibinfo {author} {\bibfnamefont {K.}~\bibnamefont {Berg-S{\o}rensen}}, \bibinfo {author} {\bibfnamefont {A.}~\bibnamefont {Huck}}, \bibinfo {author} {\bibfnamefont {T.}~\bibnamefont {Taniguchi}}, \bibinfo {author} {\bibfnamefont {K.}~\bibnamefont {Watanabe}},\ and\ \bibinfo {author} {\bibfnamefont {N.}~\bibnamefont {Bar-Gill}},\ }\bibfield  {title} {\bibinfo {title} {{Creation of Boron Vacancies in Hexagonal Boron Nitride Exfoliated from Bulk Crystals for Quantum Sensing}},\
  }\href {https://doi.org/10.1021/acsanm.3c03395} {\bibfield  {journal} {\bibinfo  {journal} {ACS Applied Nano Materials}\ }\textbf {\bibinfo {volume} {6}},\ \bibinfo {pages} {21671} (\bibinfo {year} {2023})}\BibitemShut {NoStop}%
\bibitem [{\citenamefont {Suzuki}\ \emph {et~al.}(2023)\citenamefont {Suzuki}, \citenamefont {Yamazaki}, \citenamefont {Taniguchi}, \citenamefont {Watanabe}, \citenamefont {Nishiya}, \citenamefont {Matsushita}, \citenamefont {Harii}, \citenamefont {Masuyama}, \citenamefont {Hijikata},\ and\ \citenamefont {Ohshima}}]{Suzuki2023}%
  \BibitemOpen
  \bibfield  {author} {\bibinfo {author} {\bibfnamefont {T.}~\bibnamefont {Suzuki}}, \bibinfo {author} {\bibfnamefont {Y.}~\bibnamefont {Yamazaki}}, \bibinfo {author} {\bibfnamefont {T.}~\bibnamefont {Taniguchi}}, \bibinfo {author} {\bibfnamefont {K.}~\bibnamefont {Watanabe}}, \bibinfo {author} {\bibfnamefont {Y.}~\bibnamefont {Nishiya}}, \bibinfo {author} {\bibfnamefont {Y.-i.}\ \bibnamefont {Matsushita}}, \bibinfo {author} {\bibfnamefont {K.}~\bibnamefont {Harii}}, \bibinfo {author} {\bibfnamefont {Y.}~\bibnamefont {Masuyama}}, \bibinfo {author} {\bibfnamefont {Y.}~\bibnamefont {Hijikata}},\ and\ \bibinfo {author} {\bibfnamefont {T.}~\bibnamefont {Ohshima}},\ }\bibfield  {title} {\bibinfo {title} {{Spin property improvement of boron vacancy defect in hexagonal boron nitride by thermal treatment}},\ }\href {https://doi.org/10.35848/1882-0786/acc442} {\bibfield  {journal} {\bibinfo  {journal} {Applied Physics Express}\ }\textbf {\bibinfo {volume} {16}},\ \bibinfo {pages} {032006} (\bibinfo {year}
  {2023})}\BibitemShut {NoStop}%
\bibitem [{\citenamefont {Ren}\ \emph {et~al.}(2025)\citenamefont {Ren}, \citenamefont {Xu},\ and\ \citenamefont {Wu}}]{Ren2025}%
  \BibitemOpen
  \bibfield  {author} {\bibinfo {author} {\bibfnamefont {F.}~\bibnamefont {Ren}}, \bibinfo {author} {\bibfnamefont {Z.}~\bibnamefont {Xu}},\ and\ \bibinfo {author} {\bibfnamefont {Y.}~\bibnamefont {Wu}},\ }\bibfield  {title} {\bibinfo {title} {{Optimization of carbon irradiation parameters for creating spin defects in hexagonal boron nitride}},\ }\bibfield  {journal} {\bibinfo  {journal} {Nanotechnology and Precision Engineering}\ }\textbf {\bibinfo {volume} {8}},\ \href {https://doi.org/10.1063/10.0036119} {10.1063/10.0036119} (\bibinfo {year} {2025})\BibitemShut {NoStop}%
\bibitem [{\citenamefont {Durand}\ \emph {et~al.}(2023)\citenamefont {Durand}, \citenamefont {Clua-Provost}, \citenamefont {Fabre}, \citenamefont {Kumar}, \citenamefont {Li}, \citenamefont {Edgar}, \citenamefont {Udvarhelyi}, \citenamefont {Gali}, \citenamefont {Marie}, \citenamefont {Robert}, \citenamefont {G\'erard}, \citenamefont {Gil}, \citenamefont {Cassabois},\ and\ \citenamefont {Jacques}}]{Durand2023}%
  \BibitemOpen
  \bibfield  {author} {\bibinfo {author} {\bibfnamefont {A.}~\bibnamefont {Durand}}, \bibinfo {author} {\bibfnamefont {T.}~\bibnamefont {Clua-Provost}}, \bibinfo {author} {\bibfnamefont {F.}~\bibnamefont {Fabre}}, \bibinfo {author} {\bibfnamefont {P.}~\bibnamefont {Kumar}}, \bibinfo {author} {\bibfnamefont {J.}~\bibnamefont {Li}}, \bibinfo {author} {\bibfnamefont {J.~H.}\ \bibnamefont {Edgar}}, \bibinfo {author} {\bibfnamefont {P.}~\bibnamefont {Udvarhelyi}}, \bibinfo {author} {\bibfnamefont {A.}~\bibnamefont {Gali}}, \bibinfo {author} {\bibfnamefont {X.}~\bibnamefont {Marie}}, \bibinfo {author} {\bibfnamefont {C.}~\bibnamefont {Robert}}, \bibinfo {author} {\bibfnamefont {J.~M.}\ \bibnamefont {G\'erard}}, \bibinfo {author} {\bibfnamefont {B.}~\bibnamefont {Gil}}, \bibinfo {author} {\bibfnamefont {G.}~\bibnamefont {Cassabois}},\ and\ \bibinfo {author} {\bibfnamefont {V.}~\bibnamefont {Jacques}},\ }\bibfield  {title} {\bibinfo {title} {Optically active spin defects in few-layer thick hexagonal boron nitride},\
  }\href {https://doi.org/10.1103/PhysRevLett.131.116902} {\bibfield  {journal} {\bibinfo  {journal} {Phys. Rev. Lett.}\ }\textbf {\bibinfo {volume} {131}},\ \bibinfo {pages} {116902} (\bibinfo {year} {2023})}\BibitemShut {NoStop}%
\bibitem [{\citenamefont {Udvarhelyi}\ \emph {et~al.}(2023)\citenamefont {Udvarhelyi}, \citenamefont {Clua-Provost}, \citenamefont {Durand}, \citenamefont {Li}, \citenamefont {Edgar}, \citenamefont {Gil}, \citenamefont {Cassabois}, \citenamefont {Jacques},\ and\ \citenamefont {Gali}}]{Udvarhelyi2023}%
  \BibitemOpen
  \bibfield  {author} {\bibinfo {author} {\bibfnamefont {P.}~\bibnamefont {Udvarhelyi}}, \bibinfo {author} {\bibfnamefont {T.}~\bibnamefont {Clua-Provost}}, \bibinfo {author} {\bibfnamefont {A.}~\bibnamefont {Durand}}, \bibinfo {author} {\bibfnamefont {J.}~\bibnamefont {Li}}, \bibinfo {author} {\bibfnamefont {J.~H.}\ \bibnamefont {Edgar}}, \bibinfo {author} {\bibfnamefont {B.}~\bibnamefont {Gil}}, \bibinfo {author} {\bibfnamefont {G.}~\bibnamefont {Cassabois}}, \bibinfo {author} {\bibfnamefont {V.}~\bibnamefont {Jacques}},\ and\ \bibinfo {author} {\bibfnamefont {A.}~\bibnamefont {Gali}},\ }\bibfield  {title} {\bibinfo {title} {A planar defect spin sensor in a two-dimensional material susceptible to strain and electric fields},\ }\href {https://doi.org/10.1038/s41524-023-01111-7} {\bibfield  {journal} {\bibinfo  {journal} {npj Computational Materials}\ }\textbf {\bibinfo {volume} {9}},\ \bibinfo {pages} {150} (\bibinfo {year} {2023})}\BibitemShut {NoStop}%
\bibitem [{\citenamefont {Haykal}\ \emph {et~al.}(2022)\citenamefont {Haykal}, \citenamefont {Tanos}, \citenamefont {Minotto}, \citenamefont {Durand}, \citenamefont {Fabre}, \citenamefont {Li}, \citenamefont {Edgar}, \citenamefont {Iv{\'{a}}dy}, \citenamefont {Gali}, \citenamefont {Michel}, \citenamefont {Dr{\'{e}}au}, \citenamefont {Gil}, \citenamefont {Cassabois},\ and\ \citenamefont {Jacques}}]{Haykal2022}%
  \BibitemOpen
  \bibfield  {author} {\bibinfo {author} {\bibfnamefont {A.}~\bibnamefont {Haykal}}, \bibinfo {author} {\bibfnamefont {R.}~\bibnamefont {Tanos}}, \bibinfo {author} {\bibfnamefont {N.}~\bibnamefont {Minotto}}, \bibinfo {author} {\bibfnamefont {A.}~\bibnamefont {Durand}}, \bibinfo {author} {\bibfnamefont {F.}~\bibnamefont {Fabre}}, \bibinfo {author} {\bibfnamefont {J.}~\bibnamefont {Li}}, \bibinfo {author} {\bibfnamefont {J.~H.}\ \bibnamefont {Edgar}}, \bibinfo {author} {\bibfnamefont {V.}~\bibnamefont {Iv{\'{a}}dy}}, \bibinfo {author} {\bibfnamefont {A.}~\bibnamefont {Gali}}, \bibinfo {author} {\bibfnamefont {T.}~\bibnamefont {Michel}}, \bibinfo {author} {\bibfnamefont {A.}~\bibnamefont {Dr{\'{e}}au}}, \bibinfo {author} {\bibfnamefont {B.}~\bibnamefont {Gil}}, \bibinfo {author} {\bibfnamefont {G.}~\bibnamefont {Cassabois}},\ and\ \bibinfo {author} {\bibfnamefont {V.}~\bibnamefont {Jacques}},\ }\bibfield  {title} {\bibinfo {title} {{Decoherence of {\VB} spin defects in monoisotopic hexagonal boron nitride}},\
  }\href {https://doi.org/10.1038/s41467-022-31743-0} {\bibfield  {journal} {\bibinfo  {journal} {Nature Communications}\ }\textbf {\bibinfo {volume} {13}},\ \bibinfo {pages} {4347} (\bibinfo {year} {2022})},\ \Eprint {https://arxiv.org/abs/2112.10176} {2112.10176} \BibitemShut {NoStop}%
\bibitem [{\citenamefont {Gao}\ \emph {et~al.}(2021{\natexlab{b}})\citenamefont {Gao}, \citenamefont {Pandey}, \citenamefont {Kianinia}, \citenamefont {Ahn}, \citenamefont {Ju}, \citenamefont {Aharonovich}, \citenamefont {Shivaram},\ and\ \citenamefont {Li}}]{Gao2021femto}%
  \BibitemOpen
  \bibfield  {author} {\bibinfo {author} {\bibfnamefont {X.}~\bibnamefont {Gao}}, \bibinfo {author} {\bibfnamefont {S.}~\bibnamefont {Pandey}}, \bibinfo {author} {\bibfnamefont {M.}~\bibnamefont {Kianinia}}, \bibinfo {author} {\bibfnamefont {J.}~\bibnamefont {Ahn}}, \bibinfo {author} {\bibfnamefont {P.}~\bibnamefont {Ju}}, \bibinfo {author} {\bibfnamefont {I.}~\bibnamefont {Aharonovich}}, \bibinfo {author} {\bibfnamefont {N.}~\bibnamefont {Shivaram}},\ and\ \bibinfo {author} {\bibfnamefont {T.}~\bibnamefont {Li}},\ }\bibfield  {title} {\bibinfo {title} {{Femtosecond Laser Writing of Spin Defects in Hexagonal Boron Nitride}},\ }\href {https://doi.org/10.1021/acsphotonics.0c01847} {\bibfield  {journal} {\bibinfo  {journal} {ACS Photonics}\ }\textbf {\bibinfo {volume} {8}},\ \bibinfo {pages} {994} (\bibinfo {year} {2021}{\natexlab{b}})}\BibitemShut {NoStop}%
\bibitem [{\citenamefont {Jin}\ \emph {et~al.}(2009)\citenamefont {Jin}, \citenamefont {Lin}, \citenamefont {Suenaga},\ and\ \citenamefont {Iijima}}]{Jin2009}%
  \BibitemOpen
  \bibfield  {author} {\bibinfo {author} {\bibfnamefont {C.}~\bibnamefont {Jin}}, \bibinfo {author} {\bibfnamefont {F.}~\bibnamefont {Lin}}, \bibinfo {author} {\bibfnamefont {K.}~\bibnamefont {Suenaga}},\ and\ \bibinfo {author} {\bibfnamefont {S.}~\bibnamefont {Iijima}},\ }\bibfield  {title} {\bibinfo {title} {{Fabrication of a freestanding boron nitride single layer and its defect assignments}},\ }\href {https://doi.org/10.1103/PhysRevLett.102.195505} {\bibfield  {journal} {\bibinfo  {journal} {Physical Review Letters}\ }\textbf {\bibinfo {volume} {102}},\ \bibinfo {pages} {195505} (\bibinfo {year} {2009})}\BibitemShut {NoStop}%
\bibitem [{\citenamefont {Mitterreiter}\ \emph {et~al.}(2020)\citenamefont {Mitterreiter}, \citenamefont {Schuler}, \citenamefont {Schuler}, \citenamefont {Cochrane}, \citenamefont {Wurstbauer}, \citenamefont {Wurstbauer}, \citenamefont {Weber-Bargioni}, \citenamefont {Kastl},\ and\ \citenamefont {Holleitner}}]{Mitterreiter2020}%
  \BibitemOpen
  \bibfield  {author} {\bibinfo {author} {\bibfnamefont {E.}~\bibnamefont {Mitterreiter}}, \bibinfo {author} {\bibfnamefont {B.}~\bibnamefont {Schuler}}, \bibinfo {author} {\bibfnamefont {B.}~\bibnamefont {Schuler}}, \bibinfo {author} {\bibfnamefont {K.~A.}\ \bibnamefont {Cochrane}}, \bibinfo {author} {\bibfnamefont {U.}~\bibnamefont {Wurstbauer}}, \bibinfo {author} {\bibfnamefont {U.}~\bibnamefont {Wurstbauer}}, \bibinfo {author} {\bibfnamefont {A.}~\bibnamefont {Weber-Bargioni}}, \bibinfo {author} {\bibfnamefont {C.}~\bibnamefont {Kastl}},\ and\ \bibinfo {author} {\bibfnamefont {A.~W.}\ \bibnamefont {Holleitner}},\ }\bibfield  {title} {\bibinfo {title} {{Atomistic Positioning of Defects in Helium Ion Treated Single-Layer MoS2}},\ }\href {https://doi.org/10.1021/acs.nanolett.0c01222} {\bibfield  {journal} {\bibinfo  {journal} {Nano Letters}\ }\textbf {\bibinfo {volume} {20}},\ \bibinfo {pages} {4437} (\bibinfo {year} {2020})}\BibitemShut {NoStop}%
\bibitem [{\citenamefont {Grzeszczyk}\ \emph {et~al.}(2024)\citenamefont {Grzeszczyk}, \citenamefont {Vaklinova}, \citenamefont {Watanabe}, \citenamefont {Taniguchi}, \citenamefont {Novoselov},\ and\ \citenamefont {Koperski}}]{Grzeszczyk2024}%
  \BibitemOpen
  \bibfield  {author} {\bibinfo {author} {\bibfnamefont {M.}~\bibnamefont {Grzeszczyk}}, \bibinfo {author} {\bibfnamefont {K.}~\bibnamefont {Vaklinova}}, \bibinfo {author} {\bibfnamefont {K.}~\bibnamefont {Watanabe}}, \bibinfo {author} {\bibfnamefont {T.}~\bibnamefont {Taniguchi}}, \bibinfo {author} {\bibfnamefont {K.~S.}\ \bibnamefont {Novoselov}},\ and\ \bibinfo {author} {\bibfnamefont {M.}~\bibnamefont {Koperski}},\ }\bibfield  {title} {\bibinfo {title} {{Electroluminescence from pure resonant states in hBN-based vertical tunneling junctions}},\ }\href {http://dx.doi.org/10.1038/s41377-024-01491-5} {\bibfield  {journal} {\bibinfo  {journal} {Light: Science and Applications}\ }\textbf {\bibinfo {volume} {13}},\ \bibinfo {pages} {155} (\bibinfo {year} {2024})}\BibitemShut {NoStop}%
\bibitem [{\citenamefont {Purdie}\ \emph {et~al.}(2018)\citenamefont {Purdie}, \citenamefont {Pugno}, \citenamefont {Taniguchi}, \citenamefont {Watanabe}, \citenamefont {Ferrari},\ and\ \citenamefont {Lombardo}}]{Purdie2018}%
  \BibitemOpen
  \bibfield  {author} {\bibinfo {author} {\bibfnamefont {D.~G.}\ \bibnamefont {Purdie}}, \bibinfo {author} {\bibfnamefont {N.~M.}\ \bibnamefont {Pugno}}, \bibinfo {author} {\bibfnamefont {T.}~\bibnamefont {Taniguchi}}, \bibinfo {author} {\bibfnamefont {K.}~\bibnamefont {Watanabe}}, \bibinfo {author} {\bibfnamefont {A.~C.}\ \bibnamefont {Ferrari}},\ and\ \bibinfo {author} {\bibfnamefont {A.}~\bibnamefont {Lombardo}},\ }\bibfield  {title} {\bibinfo {title} {Cleaning interfaces in layered materials heterostructures},\ }\href {https://doi.org/10.1038/s41467-018-07558-3} {\bibfield  {journal} {\bibinfo  {journal} {Nature Communications}\ }\textbf {\bibinfo {volume} {9}},\ \bibinfo {pages} {5387} (\bibinfo {year} {2018})}\BibitemShut {NoStop}%
\bibitem [{\citenamefont {Plimpton}(1995)}]{Plimpton1995}%
  \BibitemOpen
  \bibfield  {author} {\bibinfo {author} {\bibfnamefont {S.}~\bibnamefont {Plimpton}},\ }\bibfield  {title} {\bibinfo {title} {{Fast Parallel Algorithms for Short-Range Molecular Dynamics}},\ }\href {https://doi.org/10.1006/jcph.1995.1039} {\bibfield  {journal} {\bibinfo  {journal} {Journal of Computational Physics}\ }\textbf {\bibinfo {volume} {117}},\ \bibinfo {pages} {1} (\bibinfo {year} {1995})}\BibitemShut {NoStop}%
\bibitem [{\citenamefont {Kretschmer}\ \emph {et~al.}(2018)\citenamefont {Kretschmer}, \citenamefont {Maslov}, \citenamefont {Ghaderzadeh}, \citenamefont {Ghorbani-Asl}, \citenamefont {Hlawacek},\ and\ \citenamefont {Krasheninnikov}}]{Kretschmer2018}%
  \BibitemOpen
  \bibfield  {author} {\bibinfo {author} {\bibfnamefont {S.}~\bibnamefont {Kretschmer}}, \bibinfo {author} {\bibfnamefont {M.}~\bibnamefont {Maslov}}, \bibinfo {author} {\bibfnamefont {S.}~\bibnamefont {Ghaderzadeh}}, \bibinfo {author} {\bibfnamefont {M.}~\bibnamefont {Ghorbani-Asl}}, \bibinfo {author} {\bibfnamefont {G.}~\bibnamefont {Hlawacek}},\ and\ \bibinfo {author} {\bibfnamefont {A.~V.}\ \bibnamefont {Krasheninnikov}},\ }\bibfield  {title} {\bibinfo {title} {{Supported Two-Dimensional Materials under Ion Irradiation: The Substrate Governs Defect Production}},\ }\href {https://doi.org/10.1021/acsami.8b08471} {\bibfield  {journal} {\bibinfo  {journal} {ACS Applied Materials and Interfaces}\ }\textbf {\bibinfo {volume} {10}},\ \bibinfo {pages} {30827} (\bibinfo {year} {2018})}\BibitemShut {NoStop}%
\bibitem [{\citenamefont {Ghaderzadeh}\ \emph {et~al.}(2021)\citenamefont {Ghaderzadeh}, \citenamefont {Kretschmer}, \citenamefont {Ghorbani-Asl}, \citenamefont {Hlawacek},\ and\ \citenamefont {Krasheninnikov}}]{Ghaderzadeh2021}%
  \BibitemOpen
  \bibfield  {author} {\bibinfo {author} {\bibfnamefont {S.}~\bibnamefont {Ghaderzadeh}}, \bibinfo {author} {\bibfnamefont {S.}~\bibnamefont {Kretschmer}}, \bibinfo {author} {\bibfnamefont {M.}~\bibnamefont {Ghorbani-Asl}}, \bibinfo {author} {\bibfnamefont {G.}~\bibnamefont {Hlawacek}},\ and\ \bibinfo {author} {\bibfnamefont {A.~V.}\ \bibnamefont {Krasheninnikov}},\ }\bibfield  {title} {\bibinfo {title} {{Atomistic Simulations of Defect Production in Monolayer and Bulk Hexagonal Boron Nitride under Low- and High-Fluence Ion Irradiation}},\ }\href {https://doi.org/10.3390/nano11051214} {\bibfield  {journal} {\bibinfo  {journal} {Nanomaterials}\ }\textbf {\bibinfo {volume} {11}},\ \bibinfo {pages} {1214} (\bibinfo {year} {2021})}\BibitemShut {NoStop}%
\bibitem [{\citenamefont {Lehtinen}\ \emph {et~al.}(2011)\citenamefont {Lehtinen}, \citenamefont {Dumur}, \citenamefont {Kotakoski}, \citenamefont {Krasheninnikov}, \citenamefont {Nordlund},\ and\ \citenamefont {Keinonen}}]{Lehtinen-2011}%
  \BibitemOpen
  \bibfield  {author} {\bibinfo {author} {\bibfnamefont {O.}~\bibnamefont {Lehtinen}}, \bibinfo {author} {\bibfnamefont {E.}~\bibnamefont {Dumur}}, \bibinfo {author} {\bibfnamefont {J.}~\bibnamefont {Kotakoski}}, \bibinfo {author} {\bibfnamefont {A.~V.}\ \bibnamefont {Krasheninnikov}}, \bibinfo {author} {\bibfnamefont {K.}~\bibnamefont {Nordlund}},\ and\ \bibinfo {author} {\bibfnamefont {J.}~\bibnamefont {Keinonen}},\ }\bibfield  {title} {\bibinfo {title} {{Production of defects in hexagonal boron nitride monolayer under ion irradiation}},\ }\href {https://doi.org/10.1016/j.nimb.2010.11.027} {\bibfield  {journal} {\bibinfo  {journal} {Nuclear Instruments and Methods in Physics Research, Section B: Beam Interactions with Materials and Atoms}\ }\textbf {\bibinfo {volume} {269}},\ \bibinfo {pages} {1327} (\bibinfo {year} {2011})}\BibitemShut {NoStop}%
\bibitem [{\citenamefont {Iv{\'a}dy}\ \emph {et~al.}(2020)\citenamefont {Iv{\'a}dy}, \citenamefont {Barcza}, \citenamefont {Thiering}, \citenamefont {Li}, \citenamefont {Hamdi}, \citenamefont {Chou}, \citenamefont {Legeza},\ and\ \citenamefont {Gali}}]{Ivady2020}%
  \BibitemOpen
  \bibfield  {author} {\bibinfo {author} {\bibfnamefont {V.}~\bibnamefont {Iv{\'a}dy}}, \bibinfo {author} {\bibfnamefont {G.}~\bibnamefont {Barcza}}, \bibinfo {author} {\bibfnamefont {G.}~\bibnamefont {Thiering}}, \bibinfo {author} {\bibfnamefont {S.}~\bibnamefont {Li}}, \bibinfo {author} {\bibfnamefont {H.}~\bibnamefont {Hamdi}}, \bibinfo {author} {\bibfnamefont {J.-P.}\ \bibnamefont {Chou}}, \bibinfo {author} {\bibfnamefont {{\"O}.}~\bibnamefont {Legeza}},\ and\ \bibinfo {author} {\bibfnamefont {A.}~\bibnamefont {Gali}},\ }\bibfield  {title} {\bibinfo {title} {Ab initio theory of the negatively charged boron vacancy qubit in hexagonal boron nitride},\ }\href {https://doi.org/10.1038/s41524-020-0305-x} {\bibfield  {journal} {\bibinfo  {journal} {npj Computational Materials}\ }\textbf {\bibinfo {volume} {6}},\ \bibinfo {pages} {41} (\bibinfo {year} {2020})}\BibitemShut {NoStop}%
\bibitem [{\citenamefont {Libbi}\ \emph {et~al.}(2022)\citenamefont {Libbi}, \citenamefont {{De Melo}}, \citenamefont {Zanolli}, \citenamefont {Verstraete},\ and\ \citenamefont {Marzari}}]{Libbi2022}%
  \BibitemOpen
  \bibfield  {author} {\bibinfo {author} {\bibfnamefont {F.}~\bibnamefont {Libbi}}, \bibinfo {author} {\bibfnamefont {P.~M.~M.}\ \bibnamefont {{De Melo}}}, \bibinfo {author} {\bibfnamefont {Z.}~\bibnamefont {Zanolli}}, \bibinfo {author} {\bibfnamefont {M.~J.}\ \bibnamefont {Verstraete}},\ and\ \bibinfo {author} {\bibfnamefont {N.}~\bibnamefont {Marzari}},\ }\bibfield  {title} {\bibinfo {title} {{Phonon-Assisted Luminescence in Defect Centers from Many-Body Perturbation Theory}},\ }\href {https://doi.org/10.1103/PhysRevLett.128.167401} {\bibfield  {journal} {\bibinfo  {journal} {Physical Review Letters}\ }\textbf {\bibinfo {volume} {128}},\ \bibinfo {pages} {167401} (\bibinfo {year} {2022})}\BibitemShut {NoStop}%
\bibitem [{\citenamefont {Reimers}\ \emph {et~al.}(2020)\citenamefont {Reimers}, \citenamefont {Shen}, \citenamefont {Kianinia}, \citenamefont {Bradac}, \citenamefont {Aharonovich}, \citenamefont {Ford},\ and\ \citenamefont {Piecuch}}]{Reimers2020}%
  \BibitemOpen
  \bibfield  {author} {\bibinfo {author} {\bibfnamefont {J.~R.}\ \bibnamefont {Reimers}}, \bibinfo {author} {\bibfnamefont {J.}~\bibnamefont {Shen}}, \bibinfo {author} {\bibfnamefont {M.}~\bibnamefont {Kianinia}}, \bibinfo {author} {\bibfnamefont {C.}~\bibnamefont {Bradac}}, \bibinfo {author} {\bibfnamefont {I.}~\bibnamefont {Aharonovich}}, \bibinfo {author} {\bibfnamefont {M.~J.}\ \bibnamefont {Ford}},\ and\ \bibinfo {author} {\bibfnamefont {P.}~\bibnamefont {Piecuch}},\ }\bibfield  {title} {\bibinfo {title} {{Photoluminescence, photophysics, and photochemistry of the v B- defect in hexagonal boron nitride}},\ }\href {https://doi.org/10.1103/PhysRevB.102.144105} {\bibfield  {journal} {\bibinfo  {journal} {Physical Review B}\ }\textbf {\bibinfo {volume} {102}},\ \bibinfo {pages} {144105} (\bibinfo {year} {2020})}\BibitemShut {NoStop}%
\bibitem [{\citenamefont {Li}\ \emph {et~al.}(2021)\citenamefont {Li}, \citenamefont {Glaser}, \citenamefont {Elias}, \citenamefont {Ye}, \citenamefont {Evans}, \citenamefont {Xue}, \citenamefont {Liu}, \citenamefont {Cassabois}, \citenamefont {Gil}, \citenamefont {Valvin}, \citenamefont {Pelini}, \citenamefont {Yeats}, \citenamefont {He}, \citenamefont {Liu},\ and\ \citenamefont {Edgar}}]{Li2021}%
  \BibitemOpen
  \bibfield  {author} {\bibinfo {author} {\bibfnamefont {J.}~\bibnamefont {Li}}, \bibinfo {author} {\bibfnamefont {E.~R.}\ \bibnamefont {Glaser}}, \bibinfo {author} {\bibfnamefont {C.}~\bibnamefont {Elias}}, \bibinfo {author} {\bibfnamefont {G.}~\bibnamefont {Ye}}, \bibinfo {author} {\bibfnamefont {D.}~\bibnamefont {Evans}}, \bibinfo {author} {\bibfnamefont {L.}~\bibnamefont {Xue}}, \bibinfo {author} {\bibfnamefont {S.}~\bibnamefont {Liu}}, \bibinfo {author} {\bibfnamefont {G.}~\bibnamefont {Cassabois}}, \bibinfo {author} {\bibfnamefont {B.}~\bibnamefont {Gil}}, \bibinfo {author} {\bibfnamefont {P.}~\bibnamefont {Valvin}}, \bibinfo {author} {\bibfnamefont {T.}~\bibnamefont {Pelini}}, \bibinfo {author} {\bibfnamefont {A.~L.}\ \bibnamefont {Yeats}}, \bibinfo {author} {\bibfnamefont {R.}~\bibnamefont {He}}, \bibinfo {author} {\bibfnamefont {B.}~\bibnamefont {Liu}},\ and\ \bibinfo {author} {\bibfnamefont {J.~H.}\ \bibnamefont {Edgar}},\ }\bibfield  {title} {\bibinfo {title} {{Defect Engineering of Monoisotopic
  Hexagonal Boron Nitride Crystals via Neutron Transmutation Doping}},\ }\href {https://doi.org/10.1021/acs.chemmater.1c02849} {\bibfield  {journal} {\bibinfo  {journal} {Chemistry of Materials}\ }\textbf {\bibinfo {volume} {33}},\ \bibinfo {pages} {9231} (\bibinfo {year} {2021})}\BibitemShut {NoStop}%
\bibitem [{\citenamefont {Liu}\ \emph {et~al.}(2021)\citenamefont {Liu}, \citenamefont {Li}, \citenamefont {Yang}, \citenamefont {Yu}, \citenamefont {Meng}, \citenamefont {Wang}, \citenamefont {Li}, \citenamefont {Guo}, \citenamefont {Yan}, \citenamefont {Li}, \citenamefont {Wang}, \citenamefont {Xu}, \citenamefont {Wang}, \citenamefont {Tang}, \citenamefont {Li},\ and\ \citenamefont {Guo}}]{Liu2021}%
  \BibitemOpen
  \bibfield  {author} {\bibinfo {author} {\bibfnamefont {W.}~\bibnamefont {Liu}}, \bibinfo {author} {\bibfnamefont {Z.~P.}\ \bibnamefont {Li}}, \bibinfo {author} {\bibfnamefont {Y.~Z.}\ \bibnamefont {Yang}}, \bibinfo {author} {\bibfnamefont {S.}~\bibnamefont {Yu}}, \bibinfo {author} {\bibfnamefont {Y.}~\bibnamefont {Meng}}, \bibinfo {author} {\bibfnamefont {Z.~A.}\ \bibnamefont {Wang}}, \bibinfo {author} {\bibfnamefont {Z.~C.}\ \bibnamefont {Li}}, \bibinfo {author} {\bibfnamefont {N.~J.}\ \bibnamefont {Guo}}, \bibinfo {author} {\bibfnamefont {F.~F.}\ \bibnamefont {Yan}}, \bibinfo {author} {\bibfnamefont {Q.}~\bibnamefont {Li}}, \bibinfo {author} {\bibfnamefont {J.~F.}\ \bibnamefont {Wang}}, \bibinfo {author} {\bibfnamefont {J.~S.}\ \bibnamefont {Xu}}, \bibinfo {author} {\bibfnamefont {Y.~T.}\ \bibnamefont {Wang}}, \bibinfo {author} {\bibfnamefont {J.~S.}\ \bibnamefont {Tang}}, \bibinfo {author} {\bibfnamefont {C.~F.}\ \bibnamefont {Li}},\ and\ \bibinfo {author} {\bibfnamefont {G.~C.}\ \bibnamefont {Guo}},\
  }\bibfield  {title} {\bibinfo {title} {{Temperature-Dependent Energy-Level Shifts of Spin Defects in Hexagonal Boron Nitride}},\ }\href {https://doi.org/10.1021/acsphotonics.1c00320} {\bibfield  {journal} {\bibinfo  {journal} {ACS Photonics}\ }\textbf {\bibinfo {volume} {8}},\ \bibinfo {pages} {1889} (\bibinfo {year} {2021})}\BibitemShut {NoStop}%
\bibitem [{\citenamefont {Abdi}\ \emph {et~al.}(2018)\citenamefont {Abdi}, \citenamefont {Chou}, \citenamefont {Gali},\ and\ \citenamefont {Plenio}}]{Abdi2018}%
  \BibitemOpen
  \bibfield  {author} {\bibinfo {author} {\bibfnamefont {M.}~\bibnamefont {Abdi}}, \bibinfo {author} {\bibfnamefont {J.-P.}\ \bibnamefont {Chou}}, \bibinfo {author} {\bibfnamefont {A.}~\bibnamefont {Gali}},\ and\ \bibinfo {author} {\bibfnamefont {M.~B.}\ \bibnamefont {Plenio}},\ }\bibfield  {title} {\bibinfo {title} {{Color Centers in Hexagonal Boron Nitride Monolayers: A Group Theory and Ab Initio Analysis}},\ }\href {https://doi.org/10.1021/acsphotonics.7b01442} {\bibfield  {journal} {\bibinfo  {journal} {ACS Photonics}\ }\textbf {\bibinfo {volume} {5}},\ \bibinfo {pages} {1967} (\bibinfo {year} {2018})}\BibitemShut {NoStop}%
\bibitem [{\citenamefont {Linder{\"{a}}lv}\ \emph {et~al.}(2021)\citenamefont {Linder{\"{a}}lv}, \citenamefont {Wieczorek},\ and\ \citenamefont {Erhart}}]{Linderalv2021}%
  \BibitemOpen
  \bibfield  {author} {\bibinfo {author} {\bibfnamefont {C.}~\bibnamefont {Linder{\"{a}}lv}}, \bibinfo {author} {\bibfnamefont {W.}~\bibnamefont {Wieczorek}},\ and\ \bibinfo {author} {\bibfnamefont {P.}~\bibnamefont {Erhart}},\ }\bibfield  {title} {\bibinfo {title} {{Vibrational signatures for the identification of single-photon emitters in hexagonal boron nitride}},\ }\href {https://doi.org/10.1103/PhysRevB.103.115421} {\bibfield  {journal} {\bibinfo  {journal} {Physical Review B}\ }\textbf {\bibinfo {volume} {103}},\ \bibinfo {pages} {115421} (\bibinfo {year} {2021})}\BibitemShut {NoStop}%
\bibitem [{\citenamefont {Weston}\ \emph {et~al.}(2018)\citenamefont {Weston}, \citenamefont {Wickramaratne}, \citenamefont {Mackoit}, \citenamefont {Alkauskas},\ and\ \citenamefont {{Van De Walle}}}]{Weston2018}%
  \BibitemOpen
  \bibfield  {author} {\bibinfo {author} {\bibfnamefont {L.}~\bibnamefont {Weston}}, \bibinfo {author} {\bibfnamefont {D.}~\bibnamefont {Wickramaratne}}, \bibinfo {author} {\bibfnamefont {M.}~\bibnamefont {Mackoit}}, \bibinfo {author} {\bibfnamefont {A.}~\bibnamefont {Alkauskas}},\ and\ \bibinfo {author} {\bibfnamefont {C.~G.}\ \bibnamefont {{Van De Walle}}},\ }\bibfield  {title} {\bibinfo {title} {{Native point defects and impurities in hexagonal boron nitride}},\ }\href {https://doi.org/10.1103/PhysRevB.97.214104} {\bibfield  {journal} {\bibinfo  {journal} {Physical Review B}\ }\textbf {\bibinfo {volume} {97}},\ \bibinfo {pages} {214104} (\bibinfo {year} {2018})}\BibitemShut {NoStop}%
\bibitem [{\citenamefont {Strand}\ \emph {et~al.}(2020)\citenamefont {Strand}, \citenamefont {Larcher},\ and\ \citenamefont {Shluger}}]{Strand2020}%
  \BibitemOpen
  \bibfield  {author} {\bibinfo {author} {\bibfnamefont {J.}~\bibnamefont {Strand}}, \bibinfo {author} {\bibfnamefont {L.}~\bibnamefont {Larcher}},\ and\ \bibinfo {author} {\bibfnamefont {A.~L.}\ \bibnamefont {Shluger}},\ }\bibfield  {title} {\bibinfo {title} {{Properties of intrinsic point defects and dimers in hexagonal boron nitride}},\ }\href {https://doi.org/10.1088/1361-648X/ab4e5d} {\bibfield  {journal} {\bibinfo  {journal} {Journal of Physics: Condensed Matter}\ }\textbf {\bibinfo {volume} {32}},\ \bibinfo {pages} {055706} (\bibinfo {year} {2020})}\BibitemShut {NoStop}%
\bibitem [{\citenamefont {Schu{\'{e}}}\ \emph {et~al.}(2016)\citenamefont {Schu{\'{e}}}, \citenamefont {Stenger}, \citenamefont {Fossard}, \citenamefont {Loiseau},\ and\ \citenamefont {Barjon}}]{Schue2016}%
  \BibitemOpen
  \bibfield  {author} {\bibinfo {author} {\bibfnamefont {L.}~\bibnamefont {Schu{\'{e}}}}, \bibinfo {author} {\bibfnamefont {I.}~\bibnamefont {Stenger}}, \bibinfo {author} {\bibfnamefont {F.}~\bibnamefont {Fossard}}, \bibinfo {author} {\bibfnamefont {A.}~\bibnamefont {Loiseau}},\ and\ \bibinfo {author} {\bibfnamefont {J.}~\bibnamefont {Barjon}},\ }\bibfield  {title} {\bibinfo {title} {{Characterization methods dedicated to nanometer-thick hBN layers}},\ }\href {https://doi.org/10.1088/2053-1583/4/1/015028} {\bibfield  {journal} {\bibinfo  {journal} {2D Materials}\ }\textbf {\bibinfo {volume} {4}},\ \bibinfo {pages} {015028} (\bibinfo {year} {2016})},\ \Eprint {https://arxiv.org/abs/1610.06858} {1610.06858} \BibitemShut {NoStop}%
\bibitem [{\citenamefont {Prasad}\ \emph {et~al.}(2023)\citenamefont {Prasad}, \citenamefont {Al-Ani}, \citenamefont {Goss},\ and\ \citenamefont {Mar}}]{PRM2023}%
  \BibitemOpen
  \bibfield  {author} {\bibinfo {author} {\bibfnamefont {M.~K.}\ \bibnamefont {Prasad}}, \bibinfo {author} {\bibfnamefont {O.~A.}\ \bibnamefont {Al-Ani}}, \bibinfo {author} {\bibfnamefont {J.~P.}\ \bibnamefont {Goss}},\ and\ \bibinfo {author} {\bibfnamefont {J.~D.}\ \bibnamefont {Mar}},\ }\bibfield  {title} {\bibinfo {title} {Charge transfer due to defects in hexagonal boron nitride/graphene heterostructures: An ab initio study},\ }\href {https://doi.org/10.1103/PhysRevMaterials.7.094003} {\bibfield  {journal} {\bibinfo  {journal} {Phys. Rev. Mater.}\ }\textbf {\bibinfo {volume} {7}},\ \bibinfo {pages} {094003} (\bibinfo {year} {2023})}\BibitemShut {NoStop}%
\bibitem [{\citenamefont {Giri}\ \emph {et~al.}(2023)\citenamefont {Giri}, \citenamefont {Jensen}, \citenamefont {Khurana}, \citenamefont {Bocquel}, \citenamefont {Radko}, \citenamefont {Lang}, \citenamefont {Osterkamp}, \citenamefont {Jelezko}, \citenamefont {Berg-S{\o}rensen}, \citenamefont {Andersen},\ and\ \citenamefont {Huck}}]{Giri2023}%
  \BibitemOpen
  \bibfield  {author} {\bibinfo {author} {\bibfnamefont {R.}~\bibnamefont {Giri}}, \bibinfo {author} {\bibfnamefont {R.~H.}\ \bibnamefont {Jensen}}, \bibinfo {author} {\bibfnamefont {D.}~\bibnamefont {Khurana}}, \bibinfo {author} {\bibfnamefont {J.}~\bibnamefont {Bocquel}}, \bibinfo {author} {\bibfnamefont {I.~P.}\ \bibnamefont {Radko}}, \bibinfo {author} {\bibfnamefont {J.}~\bibnamefont {Lang}}, \bibinfo {author} {\bibfnamefont {C.}~\bibnamefont {Osterkamp}}, \bibinfo {author} {\bibfnamefont {F.}~\bibnamefont {Jelezko}}, \bibinfo {author} {\bibfnamefont {K.}~\bibnamefont {Berg-S{\o}rensen}}, \bibinfo {author} {\bibfnamefont {U.~L.}\ \bibnamefont {Andersen}},\ and\ \bibinfo {author} {\bibfnamefont {A.}~\bibnamefont {Huck}},\ }\bibfield  {title} {\bibinfo {title} {{Charge Stability and Charge-State-Based Spin Readout of Shallow Nitrogen-Vacancy Centers in Diamond}},\ }\href {https://doi.org/10.1021/acsaelm.3c01141} {\bibfield  {journal} {\bibinfo  {journal} {ACS Applied Electronic Materials}\ }\textbf
  {\bibinfo {volume} {5}},\ \bibinfo {pages} {6603} (\bibinfo {year} {2023})}\BibitemShut {NoStop}%
\bibitem [{\citenamefont {Frauni{\'{e}}}\ \emph {et~al.}(2025)\citenamefont {Frauni{\'{e}}}, \citenamefont {Clua-Provost}, \citenamefont {Roux}, \citenamefont {Mu}, \citenamefont {Delpoux}, \citenamefont {Seine}, \citenamefont {Lagarde}, \citenamefont {Watanabe}, \citenamefont {Taniguchi}, \citenamefont {Marie}, \citenamefont {Poirier}, \citenamefont {Edgar}, \citenamefont {Grisolia}, \citenamefont {Lassagne}, \citenamefont {Claverie}, \citenamefont {Jacques},\ and\ \citenamefont {Robert}}]{Fraunie2025}%
  \BibitemOpen
  \bibfield  {author} {\bibinfo {author} {\bibfnamefont {J.}~\bibnamefont {Frauni{\'{e}}}}, \bibinfo {author} {\bibfnamefont {T.}~\bibnamefont {Clua-Provost}}, \bibinfo {author} {\bibfnamefont {S.}~\bibnamefont {Roux}}, \bibinfo {author} {\bibfnamefont {Z.}~\bibnamefont {Mu}}, \bibinfo {author} {\bibfnamefont {A.}~\bibnamefont {Delpoux}}, \bibinfo {author} {\bibfnamefont {G.}~\bibnamefont {Seine}}, \bibinfo {author} {\bibfnamefont {D.}~\bibnamefont {Lagarde}}, \bibinfo {author} {\bibfnamefont {K.}~\bibnamefont {Watanabe}}, \bibinfo {author} {\bibfnamefont {T.}~\bibnamefont {Taniguchi}}, \bibinfo {author} {\bibfnamefont {X.}~\bibnamefont {Marie}}, \bibinfo {author} {\bibfnamefont {T.}~\bibnamefont {Poirier}}, \bibinfo {author} {\bibfnamefont {J.~H.}\ \bibnamefont {Edgar}}, \bibinfo {author} {\bibfnamefont {J.}~\bibnamefont {Grisolia}}, \bibinfo {author} {\bibfnamefont {B.}~\bibnamefont {Lassagne}}, \bibinfo {author} {\bibfnamefont {A.}~\bibnamefont {Claverie}}, \bibinfo {author} {\bibfnamefont
  {V.}~\bibnamefont {Jacques}},\ and\ \bibinfo {author} {\bibfnamefont {C.}~\bibnamefont {Robert}},\ }\bibfield  {title} {\bibinfo {title} {{Charge state tuning of spin defects in hexagonal boron nitride}},\ }\href {http://arxiv.org/abs/2501.18206} {\bibfield  {journal} {\bibinfo  {journal} {arXiv}\ ,\ \bibinfo {pages} {11}} (\bibinfo {year} {2025})},\ \Eprint {https://arxiv.org/abs/2501.18206} {arXiv:2501.18206} \BibitemShut {NoStop}%
\bibitem [{\citenamefont {Park}\ \emph {et~al.}(2024)\citenamefont {Park}, \citenamefont {Zhigulin}, \citenamefont {Jung}, \citenamefont {Horder}, \citenamefont {Yamamura}, \citenamefont {Han}, \citenamefont {Cho}, \citenamefont {Jeong}, \citenamefont {Watanabe}, \citenamefont {Taniguchi}, \citenamefont {Oh}, \citenamefont {Lee}, \citenamefont {Jo}, \citenamefont {Aharonovich},\ and\ \citenamefont {Kim}}]{Park2024}%
  \BibitemOpen
  \bibfield  {author} {\bibinfo {author} {\bibfnamefont {G.}~\bibnamefont {Park}}, \bibinfo {author} {\bibfnamefont {I.}~\bibnamefont {Zhigulin}}, \bibinfo {author} {\bibfnamefont {H.}~\bibnamefont {Jung}}, \bibinfo {author} {\bibfnamefont {J.}~\bibnamefont {Horder}}, \bibinfo {author} {\bibfnamefont {K.}~\bibnamefont {Yamamura}}, \bibinfo {author} {\bibfnamefont {Y.}~\bibnamefont {Han}}, \bibinfo {author} {\bibfnamefont {H.}~\bibnamefont {Cho}}, \bibinfo {author} {\bibfnamefont {H.-W.}\ \bibnamefont {Jeong}}, \bibinfo {author} {\bibfnamefont {K.}~\bibnamefont {Watanabe}}, \bibinfo {author} {\bibfnamefont {T.}~\bibnamefont {Taniguchi}}, \bibinfo {author} {\bibfnamefont {M.}~\bibnamefont {Oh}}, \bibinfo {author} {\bibfnamefont {G.-H.}\ \bibnamefont {Lee}}, \bibinfo {author} {\bibfnamefont {M.-H.}\ \bibnamefont {Jo}}, \bibinfo {author} {\bibfnamefont {I.}~\bibnamefont {Aharonovich}},\ and\ \bibinfo {author} {\bibfnamefont {J.}~\bibnamefont {Kim}},\ }\bibfield  {title} {\bibinfo {title} {{Narrowband
  Electroluminescence from Color Centers in Hexagonal Boron Nitride}},\ }\href {https://pubs.acs.org/doi/10.1021/acs.nanolett.4c03824} {\bibfield  {journal} {\bibinfo  {journal} {Nano Letters}\ }\textbf {\bibinfo {volume} {24}},\ \bibinfo {pages} {15268} (\bibinfo {year} {2024})}\BibitemShut {NoStop}%
\bibitem [{\citenamefont {Carbone}\ \emph {et~al.}(2025)\citenamefont {Carbone}, \citenamefont {Breev}, \citenamefont {Figueiredo}, \citenamefont {Kretschmer}, \citenamefont {Geilen}, \citenamefont {Mhenni}, \citenamefont {Arceri}, \citenamefont {Krasheninnikov}, \citenamefont {Wubs}, \citenamefont {Holleitner}, \citenamefont {Huck}, \citenamefont {Kastl},\ and\ \citenamefont {Stenger}}]{Carbone2025}%
  \BibitemOpen
  \bibfield  {author} {\bibinfo {author} {\bibfnamefont {A.}~\bibnamefont {Carbone}}, \bibinfo {author} {\bibfnamefont {I.~D.}\ \bibnamefont {Breev}}, \bibinfo {author} {\bibfnamefont {J.}~\bibnamefont {Figueiredo}}, \bibinfo {author} {\bibfnamefont {S.}~\bibnamefont {Kretschmer}}, \bibinfo {author} {\bibfnamefont {L.}~\bibnamefont {Geilen}}, \bibinfo {author} {\bibfnamefont {A.~B.}\ \bibnamefont {Mhenni}}, \bibinfo {author} {\bibfnamefont {J.}~\bibnamefont {Arceri}}, \bibinfo {author} {\bibfnamefont {A.~V.}\ \bibnamefont {Krasheninnikov}}, \bibinfo {author} {\bibfnamefont {M.}~\bibnamefont {Wubs}}, \bibinfo {author} {\bibfnamefont {A.~W.}\ \bibnamefont {Holleitner}}, \bibinfo {author} {\bibfnamefont {A.}~\bibnamefont {Huck}}, \bibinfo {author} {\bibfnamefont {C.}~\bibnamefont {Kastl}},\ and\ \bibinfo {author} {\bibfnamefont {N.}~\bibnamefont {Stenger}},\ }\href {https://doi.org/10.11583/DTU.28801115.v1} {\bibinfo {title} {Data for: "quantifying the creation of negatively charged boron vacancies in he-ion
  irradiated hexagonal boron nitride"}} (\bibinfo {year} {2025})\BibitemShut {NoStop}%
\bibitem [{\citenamefont {Steffen}\ \emph {et~al.}(2016)\citenamefont {Steffen}, \citenamefont {Sigel},\ and\ \citenamefont {B{\"{o}}rner}}]{Steffen2016}%
  \BibitemOpen
  \bibfield  {author} {\bibinfo {author} {\bibfnamefont {F.~D.}\ \bibnamefont {Steffen}}, \bibinfo {author} {\bibfnamefont {R.~K.}\ \bibnamefont {Sigel}},\ and\ \bibinfo {author} {\bibfnamefont {R.}~\bibnamefont {B{\"{o}}rner}},\ }\bibfield  {title} {\bibinfo {title} {{An atomistic view on carbocyanine photophysics in the realm of RNA}},\ }\href {https://doi.org/10.1039/c6cp04277e} {\bibfield  {journal} {\bibinfo  {journal} {Physical Chemistry Chemical Physics}\ }\textbf {\bibinfo {volume} {18}},\ \bibinfo {pages} {29045} (\bibinfo {year} {2016})}\BibitemShut {NoStop}%
\bibitem [{\citenamefont {Hod}(2012)}]{Hod2012}%
  \BibitemOpen
  \bibfield  {author} {\bibinfo {author} {\bibfnamefont {O.}~\bibnamefont {Hod}},\ }\bibfield  {title} {\bibinfo {title} {{Graphite and Hexagonal Boron-Nitride have the Same Interlayer Distance. Why?}},\ }\href {https://doi.org/10.1021/ct200880m} {\bibfield  {journal} {\bibinfo  {journal} {Journal of Chemical Theory and Computation}\ }\textbf {\bibinfo {volume} {8}},\ \bibinfo {pages} {1360} (\bibinfo {year} {2012})}\BibitemShut {NoStop}%
\bibitem [{\citenamefont {Los}\ \emph {et~al.}(2017)\citenamefont {Los}, \citenamefont {Kroes}, \citenamefont {Albe}, \citenamefont {Gordillo}, \citenamefont {Katsnelson},\ and\ \citenamefont {Fasolino}}]{Los2017}%
  \BibitemOpen
  \bibfield  {author} {\bibinfo {author} {\bibfnamefont {J.~H.}\ \bibnamefont {Los}}, \bibinfo {author} {\bibfnamefont {J.~M.}\ \bibnamefont {Kroes}}, \bibinfo {author} {\bibfnamefont {K.}~\bibnamefont {Albe}}, \bibinfo {author} {\bibfnamefont {R.~M.}\ \bibnamefont {Gordillo}}, \bibinfo {author} {\bibfnamefont {M.~I.}\ \bibnamefont {Katsnelson}},\ and\ \bibinfo {author} {\bibfnamefont {A.}~\bibnamefont {Fasolino}},\ }\bibfield  {title} {\bibinfo {title} {{Extended Tersoff potential for boron nitride: Energetics and elastic properties of pristine and defective h -BN}},\ }\href {https://doi.org/10.1103/PhysRevB.96.184108} {\bibfield  {journal} {\bibinfo  {journal} {Physical Review B}\ }\textbf {\bibinfo {volume} {96}},\ \bibinfo {pages} {184108} (\bibinfo {year} {2017})}\BibitemShut {NoStop}%
\bibitem [{\citenamefont {Ziegler}\ \emph {et~al.}(1985)\citenamefont {Ziegler}, \citenamefont {Biersack},\ and\ \citenamefont {Littmark}}]{Ziegler1985-book}%
  \BibitemOpen
  \bibfield  {author} {\bibinfo {author} {\bibfnamefont {J.~F.}\ \bibnamefont {Ziegler}}, \bibinfo {author} {\bibfnamefont {J.~P.}\ \bibnamefont {Biersack}},\ and\ \bibinfo {author} {\bibfnamefont {U.}~\bibnamefont {Littmark}},\ }\href@noop {} {\emph {\bibinfo {title} {The stopping and range of ions in solids}}}\ (\bibinfo  {publisher} {Pergamon},\ \bibinfo {address} {New York},\ \bibinfo {year} {1985})\BibitemShut {NoStop}%
\bibitem [{\citenamefont {M{\"{o}}ller}\ and\ \citenamefont {Eckstein}(1984)}]{Moller1984}%
  \BibitemOpen
  \bibfield  {author} {\bibinfo {author} {\bibfnamefont {W.}~\bibnamefont {M{\"{o}}ller}}\ and\ \bibinfo {author} {\bibfnamefont {W.}~\bibnamefont {Eckstein}},\ }\bibfield  {title} {\bibinfo {title} {{Tridyn - A TRIM simulation code including dynamic composition changes}},\ }\href {https://doi.org/10.1016/0168-583X(84)90321-5} {\bibfield  {journal} {\bibinfo  {journal} {Nuclear Inst. and Methods in Physics Research, B}\ }\textbf {\bibinfo {volume} {2}},\ \bibinfo {pages} {814} (\bibinfo {year} {1984})}\BibitemShut {NoStop}%
\bibitem [{\citenamefont {Ziegler}\ \emph {et~al.}(2010)\citenamefont {Ziegler}, \citenamefont {Ziegler},\ and\ \citenamefont {Biersack}}]{ZIEGLER20101818}%
  \BibitemOpen
  \bibfield  {author} {\bibinfo {author} {\bibfnamefont {J.~F.}\ \bibnamefont {Ziegler}}, \bibinfo {author} {\bibfnamefont {M.}~\bibnamefont {Ziegler}},\ and\ \bibinfo {author} {\bibfnamefont {J.}~\bibnamefont {Biersack}},\ }\bibfield  {title} {\bibinfo {title} {Srim – the stopping and range of ions in matter (2010)},\ }\href {https://doi.org/https://doi.org/10.1016/j.nimb.2010.02.091} {\bibfield  {journal} {\bibinfo  {journal} {Nuclear Instruments and Methods in Physics Research Section B: Beam Interactions with Materials and Atoms}\ }\textbf {\bibinfo {volume} {268}},\ \bibinfo {pages} {1818} (\bibinfo {year} {2010})}\BibitemShut {NoStop}%
\bibitem [{\citenamefont {Taniguchi}\ and\ \citenamefont {Watanabe}(2007)}]{Taniguchi2007}%
  \BibitemOpen
  \bibfield  {author} {\bibinfo {author} {\bibfnamefont {T.}~\bibnamefont {Taniguchi}}\ and\ \bibinfo {author} {\bibfnamefont {K.}~\bibnamefont {Watanabe}},\ }\bibfield  {title} {\bibinfo {title} {{Synthesis of high-purity boron nitride single crystals under high pressure by using Ba–BN solvent}},\ }\href {https://doi.org/10.1016/j.jcrysgro.2006.12.061} {\bibfield  {journal} {\bibinfo  {journal} {Journal of Crystal Growth}\ }\textbf {\bibinfo {volume} {303}},\ \bibinfo {pages} {525} (\bibinfo {year} {2007})}\BibitemShut {NoStop}%
\bibitem [{\citenamefont {Kubota}\ \emph {et~al.}(2007)\citenamefont {Kubota}, \citenamefont {Watanabe}, \citenamefont {Tsuda},\ and\ \citenamefont {Taniguchi}}]{Kubota2007}%
  \BibitemOpen
  \bibfield  {author} {\bibinfo {author} {\bibfnamefont {Y.}~\bibnamefont {Kubota}}, \bibinfo {author} {\bibfnamefont {K.}~\bibnamefont {Watanabe}}, \bibinfo {author} {\bibfnamefont {O.}~\bibnamefont {Tsuda}},\ and\ \bibinfo {author} {\bibfnamefont {T.}~\bibnamefont {Taniguchi}},\ }\bibfield  {title} {\bibinfo {title} {{Deep Ultraviolet Light-Emitting Hexagonal Boron Nitride Synthesized at Atmospheric Pressure}},\ }\href {https://doi.org/10.1126/science.1144216} {\bibfield  {journal} {\bibinfo  {journal} {Science}\ }\textbf {\bibinfo {volume} {317}},\ \bibinfo {pages} {932} (\bibinfo {year} {2007})}\BibitemShut {NoStop}%
\end{thebibliography}%

\end{document}